\documentclass[fleqn,usenatbib, useAMS]{mnras}

\usepackage{newtxtext,newtxmath}
\usepackage[T1]{fontenc}

\DeclareRobustCommand{\VAN}[3]{#2}
\let\VANthebibliography\thebibliography
\def\thebibliography{\DeclareRobustCommand{\VAN}[3]{##3}\VANthebibliography}

\usepackage{amsmath}	% Advanced maths commands
\usepackage{float}
\usepackage{dsfont}
\usepackage{amsmath}
\usepackage{xcolor}
\usepackage{xparse}
\usepackage{listings}

\usepackage{graphicx} % Required for inserting images
\usepackage{caption}
\usepackage{subcaption}
\usepackage{tabularray}
\usepackage[labelfont=bf]{caption}
\usepackage{gensymb}
\usepackage{setspace}
\usepackage{multirow}
\hypersetup{
  colorlinks=true,
  linkcolor=ForestGreen,
  citecolor=ForestGreen,
  urlcolor=ForestGreen
}
\usepackage[dvipsnames,svgnames,x11names,hyperref]{xcolor}

\usepackage{comment}
\usepackage{textgreek}
\usepackage{xspace}
\usepackage{xargs}

\newcommand{\jwst}{\textit{JWST}\xspace}

\newcommand{\Halpha}{\text{H\textalpha}\xspace}
\newcommand{\Hbeta}{\text{H\textbeta}\xspace}
\newcommand{\Hgamma}{\text{H\textgamma}\xspace}
\newcommand{\Hdelta}{\text{H\textdelta}\xspace}
\newcommandx{\permittedEL}[6][1=O,2=III,3=,4=,5=,6=]{\text{{#1}\,{\sc {#2}}{#3}{#4}{#5}{#6}}\xspace}
\newcommandx{\semiforbiddenEL}[6][1=O,2=III,3=,4=,5=,6=]{\text{{#1}\,{\sc {#2}}]{#3}{#4}{#5}{#6}}\xspace}
\newcommandx{\forbiddenEL}[6][1=O,2=III,3=,4=,5=,6=]{\text{[{#1}\,{\sc{#2}}]{#3}{#4}{#5}{#6}}\xspace}
\newcommand{\kms}{km s$^{-1}$}
\newcommand{\HeIIL}[1][1=1640]{\permittedEL[He][ii][\textlambda][#1]}
\newcommand{\OII}{\forbiddenEL[O][ii]}
\newcommandx{\OIIL}[1][1=3728]{\forbiddenEL[O][ii][\textlambda][#1]}
\newcommand{\OIIall}{\forbiddenEL[O][ii][\textlambda][\textlambda][3727,][3729]}
\newcommand{\OIII}{\forbiddenEL[O][iii]}
\newcommand{\FeII}{\forbiddenEL[Fe][ii]}
\newcommandx{\OIIIL}[1][1=5007]{\forbiddenEL[O][iii][\textlambda][#1]}
\newcommand{\OIIIall}{\forbiddenEL[O][iii][\textlambda][\textlambda][4959,][5007]}
\newcommandx{\NIIL}[1][1=6585]{\forbiddenEL[N][ii][\textlambda][#1]}

\newcommand{\NeIII}{\forbiddenEL[Ne][iii]}
\newcommandx{\NeIIIL}[1][1=3869]{\forbiddenEL[Ne][iii][\textlambda][#1]}

\newcommand{\NeIV}{\forbiddenEL[Ne][iv]}

\newcommand{\Tiii}{\ensuremath{t_3}\xspace}
\newcommand{\Tii}{\ensuremath{t_2}\xspace}

\title[Exploring Low-Mass Galaxies at $z\sim7.6$]{ Exploring Spatially-Resolved Metallicities, Dynamics and Outflows in Low-Mass Galaxies at $z\sim7.6$}
\author[L. R. Ivey et al.]{\parbox[h]{\textwidth}{
L. R.\ Ivey,$^{\! 1,2}$\thanks{E-mail: li247@cam.ac.uk}
J. Scholtz,$^{1, 2}$
A. L. Danhaive,$^{1,2}$
S. Koudmani,$^{1,3,4}$
G. C. Jones,$^{1,2}$
R. Maiolino,$^{1,2,5}$
M. Curti,$^{6}$
F. D'Eugenio,$^{1,2}$
S. Tacchella,$^{1, 2}$
W. M. Baker,$^{7}$
S. Arribas,$^{8}$
S. Charlot,$^{9}$
D. Eisenstein,$^{10}$
Z. Ji$,^{11}$
M. Koller$,^{1, 2}$
N. Laporte,$^{12}$
M. Perna,$^{8}$
D. Puskás,$^{1, 2}$ 
B. Robertson,$^{13}$
D. Sijacki,$^{1,14}$
J. A. A. Trussler,$^{10}$
C. Witten$^{15}$
}\vspace{0.4cm}
\\
$^{1}$Kavli Institute for Cosmology, University of Cambridge, Madingley Road, Cambridge, 
CB3 0HA, UK\\
$^{2}$Cavendish Laboratory, University of Cambridge, 19 JJ Thomson Avenue, Cambridge CB3 0HE, UK\\
$^{3}$St Catharine's College, University of Cambridge, Trumpington Street, Cambridge CB2 1RL, UK\\
$^{4}$Centre for Astrophysics Research, Department of Physics, Astronomy and Mathematics, University of Hertfordshire, College Lane, Hatfield, AL10 9AB, UK\\
$^{5}$Department of Physics and Astronomy, University College London, Gower Street, London WC1E 6BT, UK\\
$^{6}$European Southern Observatory, Karl-Schwarzschild-Strasse 2, 85748 Garching, Germany\\
$^{7}$ DARK, Niels Bohr Institute, University of Copenhagen, Jagtvej 128, DK-2200 Copenhagen, Denmark\\
$^{8}$ Centro de Astrobiolog\'ia (CAB), CSIC–INTA, Cra. de Ajalvir Km.~4, 28850- Torrej\'on de Ardoz, Madrid, Spain\\
$^{9}$ Sorbonne Universit\'e, CNRS, UMR 7095, Institut d'Astrophysique de Paris, 98 bis bd Arago, 75014 Paris, France\\
$^{10}$ Center for Astrophysics $|$ Harvard \& Smithsonian, 60 Garden St., Cambridge MA 02138 USA\\
$^{11}$ Steward Observatory, University of Arizona, 933 N. Cherry Avenue, Tucson, AZ 85721, USA\\
$^{12}$Aix Marseille Univ, CNRS, CNES, LAM, Marseille, France\\
$^{13}$ Department of Astronomy and Astrophysics, University of California, Santa Cruz, 1156 High Street, Santa Cruz, CA 95064, USA\\
$^{14}$Institute of Astronomy, University of Cambridge, Madingley Road, Cambridge, CB3 0HA, UK\\
$^{15}$ Department of Astronomy, University of Geneva, Chemin Pegasi 51, 1290 Versoix, Switzerland\\
}

\date{MNRAS, submitted}

\pubyear{\the\year{}}

\begin{document}
\label{firstpage}
\pagerange{\pageref{firstpage}--\pageref{lastpage}}
\maketitle

\begin{abstract}
A majority of \jwst/NIRSpec/IFU studies at high redshifts to date have focused on UV-bright or massive objects, while our understanding of low-mass galaxies at early cosmic times remains limited. In this work, we present NIRSpec/IFS high-resolution observations of two low-mass ($M_*<10^9 \ M_\odot$), low-metallicity ($[12+\log(\mathrm{O/H})]<8$) galaxies at $z\sim7.66$, one of which we identify as hosting a Type-II AGN. We measure flat strong-line metallicity gradients, suggestive of ISM redistribution by outflows or past merging, but also identify tension with the direct-$T_\mathrm{e}$ metallicity gradient in one galaxy. We measure $v_\mathrm{rot}/\sigma < 1$ in both galaxies, consistent with observations of lower rotational support at early cosmic times. We identify broad kinematical components decoupled from galactic rotation with velocities of $\sim 250-500$ \ \kms \ and argue these components trace outflows, for which we infer outflow rates of $\sim 8-14~M_\odot \ \mathrm{yr}^{-1}$ with $v_\mathrm{out}/v_\mathrm{esc}\sim 1$. We compare our findings to results from the new large-volume \textsc{Aesopica} simulations, which fully incorporate different models of black hole growth and AGN feedback. We find that our observational results of $v_\mathrm{out}/v_\mathrm{esc}$ are consistent with the simulated dwarf AGN population, hinting AGN-driven feedback may contribute to quenching both in our systems and in a wider population of early low-mass galaxies.
This novel study illustrates the necessity of deep IFU observations to decompose the complex kinematics and morphology of high-$z$ galaxies, trace outflows, and constrain the effects of feedback in the early Universe.

\end{abstract}


%%%%%%%%%%%%%%%%%%%%%%%%%%%%%%%%%%%%%%%%%%%%%%%%%%

%%%%%%%%%%%%%%%%% BODY OF PAPER %%%%%%%%%%%%%%%%%%

\section{Introduction}\label{sec:intro}
% quick key overview of jwst for galaxy properties - metallicity gradients included here.
The launch of the \textit{James Webb Space Telescope}  (\textit{JWST}), with its powerful near-infrared spectrograph NIRSpec \citep{Boker22, Jakobsen22}, has enabled observations of rest-frame UV and optical emission above $z>3$, and opened a new window for studying the physical properties of high-$z$ galaxies
(e.g. \citealt{Arrabal_Haro_nature_2023, curtis-lake_2023, Hsiao23,  carniani_z14_2024, castellano_ghz12_2024, d_eugenio_gsz12_2023, maiolino_gnz11_2023,robertson_JOF_LF_2023, Donnan25, Naidu_25_z14}). These observations have shown that high-redshift galaxies have properties fundamentally different to those of local galaxies: lower metallicities \citep{Schaerer22, Langeroodi23,Nakajima23census, Curti_Jades_24}, high ionisation parameters \citep{Cameron23, Sanders23} and electron temperatures \citep{Curti2023}, increasing interstellar medium (ISM) densities \citep{ Isobe23, reddy_ism_2023, Marconcini24} and bursty star-formation histories (SFH; \citealt{dressler24, endsley_sfh_2023, Looser23SFH, Langeroodi24, Witten2025}). 

%  spatially resolved metallicity, but tie it in to kinematics to help with flow
Observations with \jwst NIRSpec/IFU have placed key constraints on spatially-resolved ISM properties, and traced the chemical enrichment of galaxies at high-$z$ \citep{Arribas2024, delPino2024, Scholtz25-COS3018}. For example, radial metallicity gradients, which bear the imprint of gas mixing by various galaxy evolutionary processes, have now been studied out to $z\sim 8$ \citep[e.g.][]{Wang2022, Venturi2024, Tripodi24, Fujimoto2025, Li25metgrad}, finding a significant diversity of negative, flat and positive gradients. Such results are consistent with the diverse gradients measured by other studies up to Cosmic Noon \citep[e.g.][]{Swinbank2012, Troncoso14, Wuyts2016Metal, Wang2017, Curti2020KLEVER, Simons2021, Acharyya2025}; these gradients encode the combination of inflows, outflows and mergers with underlying evolutionary processes and physical conditions in their host galaxies \citep{Kewley2019, Maiolino19}, reflecting the diverse early galaxy population. However, results from simulations across a wide range of numerical methods and input physics suggest increasingly negative gradients at high redshift (e.g. \citealt{Hemler2021, Garcia25}), in tension with observational results. As noted by \cite{Garcia25, Garcia2025_2}, there are currently no simulations which include how unresolved turbulence can redistribute mass and metals through throughout the galaxies, which may drive this tension. An comprehensive understanding of the underlying gas kinematics in high-$z$ galaxies is therefore crucial to accurately modelling their chemical evolution and addressing the current tension between simulations and observations. 

Spatially-resolved spectroscopic surveys have demonstrated that many high-$z$ galaxies have complex kinematics, including mergers or outflows (e.g. \citealt{LeFevre20, Herrera-Camus25}). Additionally, the observed velocity dispersion of disc-like galaxies appears to increase with redshift (e.g. \citealt{Wisnioski2015, Ubler2019, Rizzo24, Lola2025}), with the ionised gas phase typically showing higher dispersion than the cold/molecular gas phase \citep[e.g.][]{Girard2021}. While some evidence for early dynamically cold discs has been found \citep{Rizzo20, Lelli21, Rowland2024, Scholtz+2025_disk}, other NIRSpec/IFU observations have also revealed evidence of small-scale merging (e.g. \citealt{Marconcini24, Jones2025, Scholtz25-COS3018}).  Differences in sample selection, kinematical tracers, and spectral and angular resolution result in significant challenges to reliably comparing kinematics measured by existing studies. To build a full picture of galaxy kinematics in the early Universe, it is therefore crucial to complement existing studies of cold gas kinematics with those of warm, ionised gas, traced with \jwst observations of $\OIII\lambda5007$ or \Halpha emission lines \citep[e.g.][]{deGraaff2024,Lola2025}.

Recent observations with \jwst have found a large fraction of early low-mass ($M_* \lesssim 10^9 \ M_\odot$) galaxies to exhibit bursty star formation \citep{endsley_sfh_2023, Looser23SFH,dressler24}, while others are found to be `mini-quenched' (or napping, e.g. \citealt{Strait23,Looser23a,  Baker25, Covelo-Paz25, Witten25}). High-mass galaxies ($M_* \gtrsim 10^9 \ M_\odot$) have also been found fully-quenched at high redshift \citep{Carnall23a,Carnall23b, Glazebrook24, Nanayakkara24Nature,Nanayakkara25, Baker25quiescent,Baker25Flamingo, Turner25, Weibel25}. Theoretical models argue that massive, fully quenched galaxies cannot result from supernovae feedback alone; feedback driven by active galactic nuclei (AGN) is likely necessary to explain their observed quenching \citep{Gelli24}. Observations of quasar-host galaxies have indeed found results consistent with ionised outflows producing significant feedback even at $z>4.5$ \citep{Parlanti2024, Bischetti2024, Vayner2025, Zhu2025}. However, simulations are able to reproduce quenching in low-mass galaxies, with bursty star formation found to be particularly important in the low-mass regime \citep[][]{Dome24, Dome25, Martin-Alvarez2025}.  Interestingly, \jwst has unveiled an abundance of AGN at high redshift, both Type I \citep{furtak23,greene23, Harikane23, Kocevski23, maiolino_gnz11_2023, Maiolino23JADES, Onoue23,Ubler23,Matthee23, Mazzolaritype1agn24, Juodzbalis25, Taylor25} and Type II \citep{Brinchmann23, MazzolariCEERS24, Scholtz2025_AGN, Tacchella25}, supporting claims that AGN feedback is already operating at early cosmic times. This raises an intriguing question: is AGN feedback contributing to the quenching of early low-mass galaxies? 

Despite the abundance of AGN in the early Universe, finding evidence of AGN-driven feedback (e.g. outflows or hard ionizing radiation) remains difficult due to limited spectral and spatial resolution, as well as the complex morphologies of early galaxies \citep[at $z>7$;][]{Kartaltepe23, Treu23}, all of which limit our ability to accurately distinguish outflows from mergers, inflows or early disks. Efforts to investigate the presence of AGN- and star formation- (SF-) driven outflows in such high-$z$ galaxies have been limited \citep{Carniani24JADES, Xu2025}, even though ionised gas outflows are known to be ubiquitous features of low-$z$ AGN and high-luminosity QSOs \citep{Harrison16,Bischetti2017, Kakkad2020, Cresci2023, Perna23, Perna2025, Marshall24}.

With \jwst, it is now possible to explore ionised gas outflows, traced by \Halpha and $\OIII\lambda5007$ emission, in galaxies at high-$z$, even on spatially-resolved scales. Indeed, numerous studies have found evidence for outflows, primarily by using NIRSpec in its IFS mode, but also with single-slit (MSA) spectroscopy (e.g. \citealt{ Marshall2023, Parlanti2023, Perna23, Ubler23, DEugenio23ifs,Ji2024ganifs,  Jones24, delPino2024, Perez-gonzalez-jekyll, Ubler24, Carniani24JADES, Lamperti2024, Zamora2024, Bertola2025,Scholtz25-COS3018}). However, most of these studies focus on bright targets previously detected by HST and ALMA, resulting in an observational bias towards starburst galaxies or mergers \citep[e.g.,][]{Jones24, Parlanti2025, Scholtz25-COS3018}. In fact, the new population of early intermediate- to low-luminosity AGN found by \jwst appears to show little evidence for AGN-driven feedback in their host galaxies \citep{Deugenio25lrd, Maiolino25xray, Scholtz2025_AGN}. In particular, low-mass star-forming galaxies (SFGs) exhibit fewer outflows than expected, especially considering that \jwst is probing the earliest phases of galaxy formation, when feedback processes were expected to be strongest according to some models \citep{Carniani24JADES, McClymont25_burst}. In contrast to these observations, one possibility that has emerged from recent simulations is that AGN feedback may play an important role in low-mass galaxies \citep{Koudmani21, Sharma23, ArjonaGalvez24}, especially in the early Universe, but that our observations are failing to detect this feedback. It could be that the observed deficit of outflows in low-mass galaxies may arise from MSA spectroscopy-based studies lacking the spatial resolution required to reliably decompose the complex kinematics of high-redshift galaxies and reliably identify outflow signatures. Deep, high-resolution integral field spectroscopy (IFS) is certainly a more adequate tool to investigate the potential weak outflows in low-mass galaxies with low-luminosity AGN.

% explain my targets and briefly introduce the data set
This study focuses on two galaxies observed by \jwst as part of the Early Release Observations (PID: 2736, \citealt{Pontoppidan2022}), as presented in \citet{CarnallSmacs2023} and \citet{Trump2023}. Specifically, we discuss SMACS J0723.3-7327 NIRSpec-ID6355 and -ID10612 (henceforth ID6355 and ID10612, respectively), which belong to a spectroscopically-confirmed protocluster at $z\sim7.66$ (\citealt{Laporte22}, Witten et al. in prep.). Both of these galaxies are weakly-lensed (magnification $\mu \sim 1.2$; derived from the lens model of \citealt{Mahler2023} by \citealt{Curti2023}), with stellar masses of $\log(M_*/M_\odot) = 8.72\pm0.04$ and $8.08\pm0.04$ in ID6355 and ID10612, respectively \citep{Curti2023}, classifying them as low-mass galaxies, as the knee of the galaxy mass function is already $10^{10} \ M_\odot$ at $z \sim 6$ \citep{Weibel24}. Both galaxies were followed up with NIRSpec/IFU observations as part of the GO 2959 programme (PI: Scholtz), obtaining the higher-resolution R2700 spectra, which form the core of this work.

Furthermore, both galaxies discussed in this work are AGN-host candidates. Previous studies using the NIRSpec/MSA R1000 spectra of these targets have detected $\NeIV\lambda\lambda2422,2424$ emission in ID6355 \citep{Brinchmann23,Silcock24}; this high ionisation-energy emission line ($\sim 63 $ eV) is often associated with AGN activity \citep{Feltre16}, and indicates that ID6355 hosts some source of ionisation stronger than star formation alone. While there was no direct detection of this line in ID10612, \citet{Brinchmann23} found by comparison that a majority of SDSS galaxies with a $\OIII\lambda4363$/\Hgamma ratio within $1\sigma$ of that of ID10612 were classified as AGN based on the BPT diagram. This demonstrates there is a strong source of heating in ID10612, hinting at the presence of an AGN. We will further investigate this claim using our higher-resolution NIRSpec/IFU spectra and established AGN diagnostic diagrams.

Furthermore, by exploring the spatially-resolved ISM properties and kinematics of our targets, we will investigate and characterise their outflows (if any). In doing so, we will take an important step towards quantifying the potential impact of AGN-driven feedback on the kinematics and evolution of early low-mass galaxies.
 
This paper is structured as follows: in Sec.~\ref{Data}, we describe the details of the observations and data reduction, then outline our analysis in Sec.~\ref{Methods}. A characterisation of galaxy properties, both integrated and spatially-resolved, is presented in Sec.~\ref{sec:ISMProp}. Then, in Sec.~\ref{sec:kinemresults}, we analyse the kinematics of our targets. Our results will be discussed and situated within in a wider context in Sec.~\ref{Discussion}. Throughout this work, we assume a flat $\Lambda\mathrm{CDM}$ model with $H_0 = 70 \ \mathrm{km} \ \mathrm{s}^{-1} \ \mathrm{Mpc}^{-3}$, $\Omega_{m, 0} = 0.3$ and $\Omega_{\Lambda, 0} = 0.7$. We additionally take solar metallicity as $[12+\log(\mathrm{O/H})]_\odot=8.69$ \citep{Asplund2009}, and use the term low-metallicity to refer to $[12+\log(\mathrm{O/H})]<8$. Finally, we adopt the lensing factors $\mu = 1.23 \pm 0.01$ and $1.34\pm 0.01$ for ID6355 and ID10612, respectively, as derived by \cite{Curti2023}.

\section{Observations and Data Reduction}\label{Data}

\subsection{NIRSpec Data}\label{sec:NIRSpecdata}

Our target galaxies were observed with \jwst/NIRSpec in IFS mode \citep{Jakobsen22,Boker22}. The NIRSpec data were taken on the 29\textsuperscript{th}-30\textsuperscript{th} May 2024, with a medium cycling pattern of eight dither positions and a total integration time of 22.4 ks per galaxy (6.2 h on-source) with the high-resolution grating/filter pair G395H/F290LP, covering the wavelength range $2.87-5.27~\mu$m (spectral resolution $R\sim2000-3500$; \citealp{Jakobsen22}).

Raw data files of these observations were downloaded from the Barbara A.~Mikulski Archive for Space Telescopes (MAST) and then processed with the \jwst Science Calibration pipeline\footnote{\url{https://jwst-pipeline.readthedocs.io/en/stable/jwst/user_documentation/introduction.html}} version 1.15.0 under the Calibration Reference Data System (CRDS) context jwst\_1293.pmap. We use the same modification of the pipeline as described in detail by \citet{Perna23} in order to increase the data quality. We briefly summarise the steps here. Count-rate frames are corrected for $1/f$ noise through a polynomial fit based on an algorithm from \citet{Perna23}. Furthermore, we remove regions affected by failed-open MSA shutters during Stage 2 calibration, along with other artefacts such as snowballs. We also remove regions with strong cosmic ray residuals in several exposures. Any remaining outliers are flagged in individual exposures using an algorithm similar to {\sc lacosmic} \citep{vDokkum01}: we calculate the derivative of the count-rate maps along the dispersion direction, normalise by the local flux (or by three times the root mean square (r.m.s.) noise; whichever is highest), and reject the 95\textsuperscript{th} percentile of the resulting distribution \citep[see][for details]{DEugenio23ifs}. The final cubes are combined using the `drizzle' method. All analysis in this paper is based on the combined cube with a pixel scale of $0.05''$.

We perform background subtraction of our data cubes by first extracting an integrated spectrum from all non-target spaxels, excluding spaxels containing noise spikes via visual inspection. This background spectrum is smoothed with a median filter of 25 pixels to reduce its remaining noise, then subtracted from the spectrum of each spaxel in our data cube. The estimated average background levels are $\sim 9\times 10^{-20}$ and $\sim 6\times 10^{-20} \ \mathrm{erg} \ \mathrm{sec}^{-1} \ \mathrm{cm}^{-2}$ \ \AA$^{-1} \ \mathrm{arcsec}^{-2}$ in the ID6355 and ID10612 cubes, respectively.

\subsection{Error Extension Correction}

\cite{Ubler23} report that uncertainties on flux measurements in the \textsc{err} extension of the IFU data cubes are underestimated compared to the noise estimated from the r.m.s. of the spectrum. However, the \textsc{err} extension still carries crucial information about outliers and the correlated noise between channels. Hence, whenever extracting a spectrum, we also retrieve the uncertainty from the \textsc{err} extension and scale it so that its median uncertainty matches the spectrum's $\sigma$-clipped r.m.s. in emission-line free regions on each detector. The scaling for each detector does not have a wavelength dependence. Utilising the corrected error extension of the data cubes, we also estimate sensitivities of $\sim 5\times 10^{-21}$ and $\sim 3\times 10^{-21} \ \mathrm{erg} \ \mathrm{sec}^{-1} \ \mathrm{cm}^{-2}$ \ \AA$^{-1}$ in the spectral region of \Hbeta \ and \OIIIall \ for the ID6355 and ID10612 cubes, respectively.

\subsection{NIRCam Imaging}\label{sec:NIRCam}

In this work, we also make use of NIRCam images from the original ERO programme of the SMACS J0723.3-7327 cluster (see Sec.~\ref{sec:intro}); the data comprises imaging in six NIRCam bands (F090W, F150W, F200W, F277W, F356W and F444W). A full description of the data reduction, including determination of the astrometry, is presented in \citet{Tacchella23}.

\subsection{Astrometry Realignment}\label{sec:Realignment}

As the field of view (FoV) of the NIRSpec/IFS observations is only 3$\times3''$, we lack the multiple stars or point sources necessary to perform an accurate astrometry calibration. Therefore, we used NIRCam images (Sec.~\ref{sec:NIRCam}) to align the NIRSpec/IFS data to the correct astrometry, allowing us to reliably compare the location of the stellar continuum and emission lines. To do so, we first create a mock F444W image from the NIRSpec cube using the Python package \texttt{sedpy} \citep{SedpyJohnnson}. We identify the brightest pixel in the NIRCam F444W image, and match the WCS coordinates of the brightest mock-NIRCam spaxel to those of the F444W pixel. Later, when using the NIRSpec/IFU data cube for spaxel-by-spaxel fitting, the header of the cube is edited to capture this astrometric realignment.

We adopt an astrometry realignment based on the NIRCam F444W images for two key reasons. Firstly, this filter contains the $\OIII\lambda5007$ emission that is the main focus of our kinematical study, and so this method guarantees alignment between the location of the peak $\OIII\lambda5007$ flux in the NIRSpec and NIRCam data. Secondly, both galaxies appear bright in this filter with good signal-to-noise ratios (SNR). Although the F356W filter has a smaller point spread function (PSF) and could, in theory, yield a more accurate realignment, our goal is to guarantee the best alignment for $\OIII\lambda5007$ emission specifically, so the limitation lies with the IFU data rather than NIRCam.

As all NIRCam filters are aligned with respect to each other, only a single astrometric realignment is required for each galaxy. We additionally check the astrometry corrections based on the F356W filter, as well as those based on a single star in the field of view of ID6355 observations (see Appendix \ref{sec:star}). All methods yield consistent astrometric realignments within the errors.

\section{Data Analysis}\label{Methods}

\subsection{Emission Line Fitting}\label{sec:ELF}

To determine the emission line properties of our sources, we first extract their integrated spectra from a circular aperture ($r\sim0.25''$), centred on the $\OIII\lambda5007$-brightest spaxel in each galaxy. This aperture is chosen as it is sufficiently large to avoid aperture loss from PSF sampling without reducing the SNR of emission lines in the spectrum. We then fit these integrated spectra using a least-squares fitting code, modelling the emission lines as a series of Gaussian profiles. We fit the following emission lines: \Hbeta, \Hgamma, \Hdelta, $\OIII\lambda\lambda4959,5007$, $\OIII\lambda4363$, $\OII\lambda\lambda3727,29$ and $\NeIII\lambda\lambda3869, 3968$. As $\HeIIL[]\lambda$4686 falls right on the edge of the detector gap in the spectra of both galaxies, it has not been included in either full spectral fit. We additionally include a linear component to model the continuum, which is too weak in either galaxy to be reliably fit with a stellar continuum, and which may contain some residual background. As our analysis focuses on the emission lines, modelling the continuum in this way does not impact our results.

Our primary (`narrow-line') model consists of a single Gaussian per emission line; the redshift and intrinsic full width half-maximum (FWHM) of each Gaussian profile are tied together to reduce the number of free parameters, but the flux of each is left free unless a specific line ratio is applied due to underlying atomic physics. The \OIIIall flux ratio is fixed to 2.99 \citep{Dimitrijevic2007} and the $\NeIII\lambda\lambda3869,3968$ flux ratio is fixed to 3.32 \citep[e.g.][]{Jones24}, while the $\OII\lambda\lambda3727,29$ flux ratio is allowed to vary to reflect its dependence on electron density \citep{Sanders15}. For each emission line, the FWHM is convolved with the NIRSpec line spread function (LSF), as retrieved from the JDox.\footnote{Available \href{https://jwst-docs.stsci.edu/jwst-near-infrared-spectrograph/nirspec-instrumentation/nirspec-dispersers-and-filters}{here}.} 

\begin{figure*}

\begin{subfigure}{0.496\textwidth}
   \includegraphics[width=\linewidth]{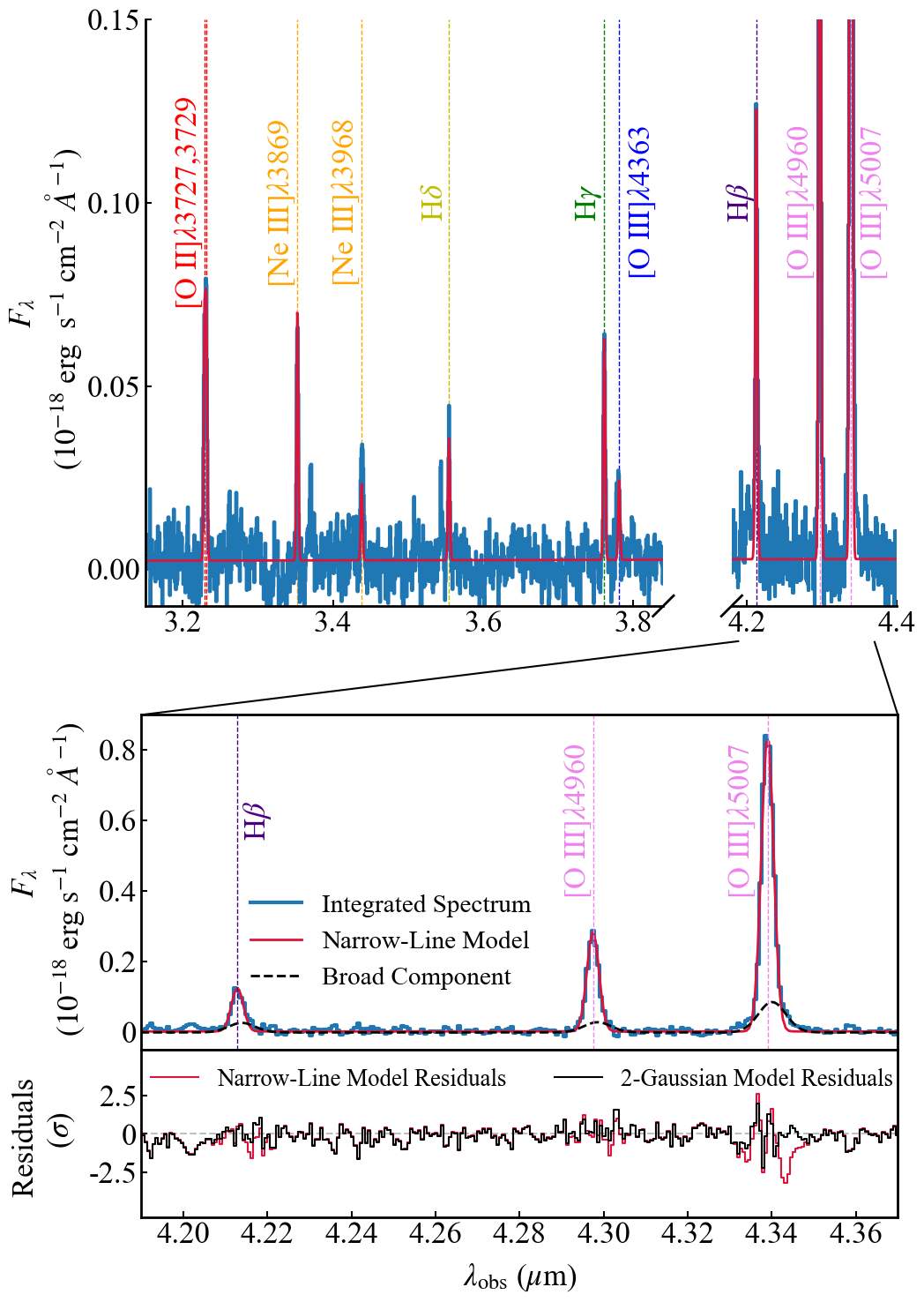}
   \caption{\textbf{ID6355}} \label{fig:6355_sbs}
\end{subfigure}
\hspace*{\fill}
\begin{subfigure}{0.496\textwidth}
   \includegraphics[width=\linewidth]{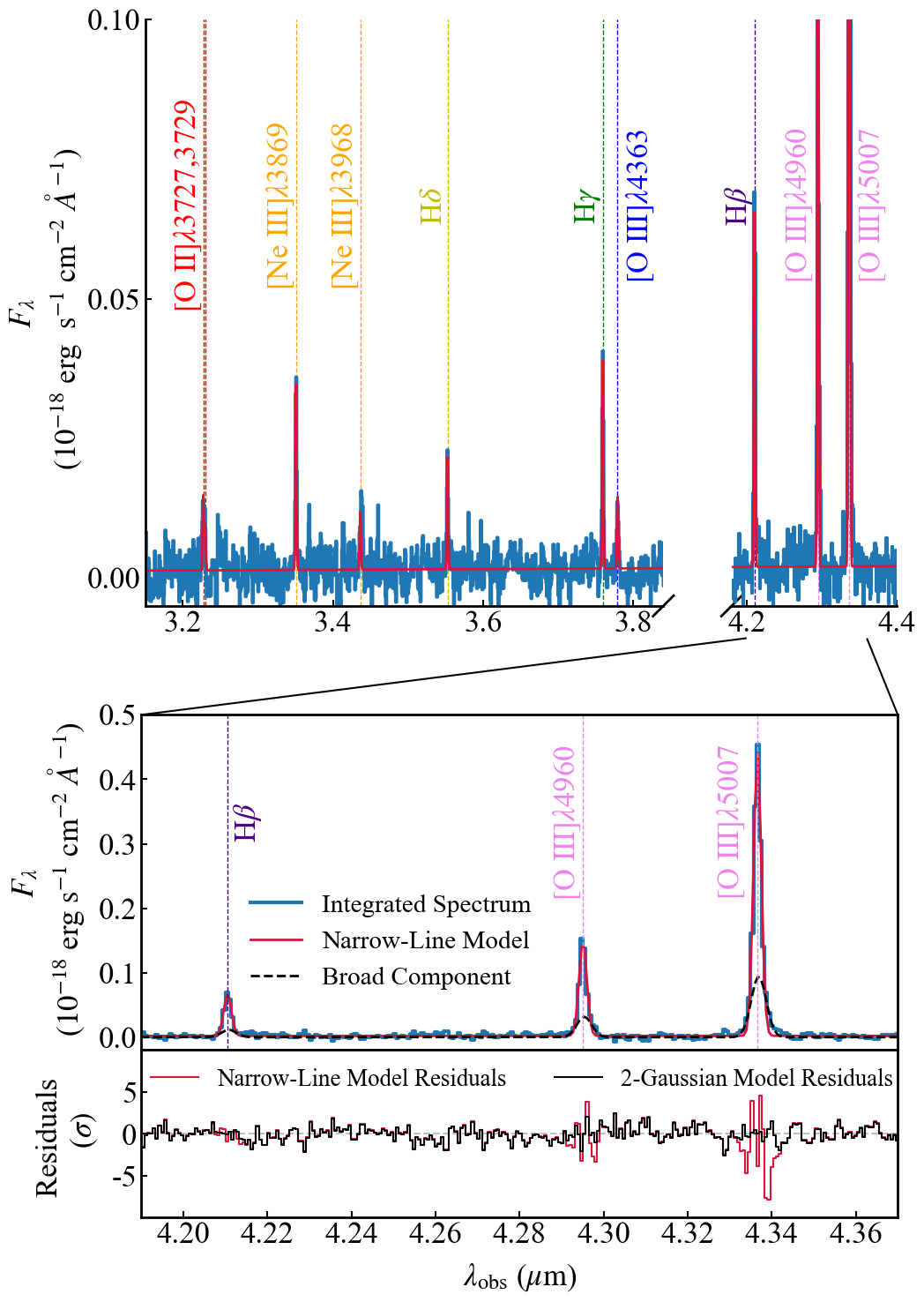}
   \caption{\textbf{ID10612}} \label{fig:10612_sbs}
\end{subfigure}
\hspace*{\fill}
\caption{Summary of background-subtracted integrated spectra, fitted models and model residual comparisons in both target galaxies. \textbf{Top panels}: Full R2700 integrated spectra from the NIRSpec/IFU observations, extracted from circular apertures of radius 5 pixels (0.25"), centred on the $\OIII\lambda5007$ brightest spaxel in each galaxy. The background-subtracted integrated spectrum is plotted in blue, and the fitted narrow-line model (see Sec.~\ref{sec:ELF}) in red. The axes have been cut to eliminate the detector gap in each spectrum from the plots. The spectra shown here have not been continuum-subtracted, as the spectral fitting accounts for the weak continuum with a linear continuum model.  \textbf{Middle panels}: A zoom-in on the \Hbeta and \OIIIall emission lines, showing how these lines are now fitted with the 2-Gaussian model. \textbf{a)} shows the fitted broad component in ID6355 is clearly offset from the narrow line component, suggesting the presence of non-rotational kinematics such as an outflow or merger. The small feature visible blueward of \Hbeta \ in the spectrum of ID6355 is a noise feature close to the detector gap, and our analysis based on SNR and BIC also shows that this feature does not meet our criteria for a detected emission line wing. \textbf{b)} shows the offset between the broad and narrow components is less distinct in ID10612, as shown in .\textbf{Bottom panels}: A comparison of narrow-line and 2-Gaussian model $\chi$ residuals, illustrating the necessity of including broad components for \Hbeta and \OIIIall in our fit.}
\label{fig:fullspec}
\end{figure*}

We also fit each spectrum with a `2-Gaussian model', which is identical to the primary model but additionally includes a broad Gaussian profile in \Hbeta and \OIIIall. This model includes a FWHM threshold at 200 \kms \ as an upper limit on the narrow component, ensuring that the fitted broad component is always broader than the fitted narrow component. Similarly to the primary model, the broad components of the \Hbeta and \OIIIall lines have the redshift and FWHM of their Gaussian profiles tied together, but the broad parameters vary independently from the narrow ones. \Hbeta and \OIIIall are the only lines in our spectra with sufficient SNR to enable detection of a broad component in the line profile.  The minimum SNR threshold considered for the \OIII$\lambda5007$ emission line in a fit is 3. We show the galaxy-integrated spectra and their best fits, with both models, in Fig. \ref{fig:fullspec}.

We repeat the spectral fitting, now spaxel by spaxel, in the region covered by each target, using the same models and overall method as described above. When fitting each spaxel, we first average its spectrum and combine its errors with those of the 8 surrounding spaxels (i.e. within a radius of $0.07''$). As the spaxel size oversamples the PSF of our observations, this increases the SNR of the detected emission lines without reducing the spatial resolution.

Whether the fit to a given spaxel accepts a broad component is statistically determined through two criteria: firstly, when comparing the Bayesian Information Criterion (BIC) of the narrow-line and 2-Gaussian models we require $\Delta\mathrm{BIC}<-10$ (i.e., $\Delta\mathrm{BIC} = \mathrm{BIC}_\mathrm{2-Gaussian} - \mathrm{BIC}_\mathrm{narrow}$); and secondly, the broad component must have velocity-integrated $\mathrm{SNR}>5$. Both criteria must be satisfied to favour the 2-Gaussian model. This procedure prevents overfitting of the data by only allowing an additional broad component to be fit if its inclusion is statistically favoured. The spaxels with evidence for a broad component hence robustly map the spatial extent of this component. For simplicity, the region of spaxels in each galaxy fitting with an additional broad component will henceforth be referred to the `outflow region' of that galaxy. We further discuss the origin of this broad component in Sec.~\ref{sec:otherscenarios}.

\begin{figure*}
        \centering
	\includegraphics[width=0.8\paperwidth]{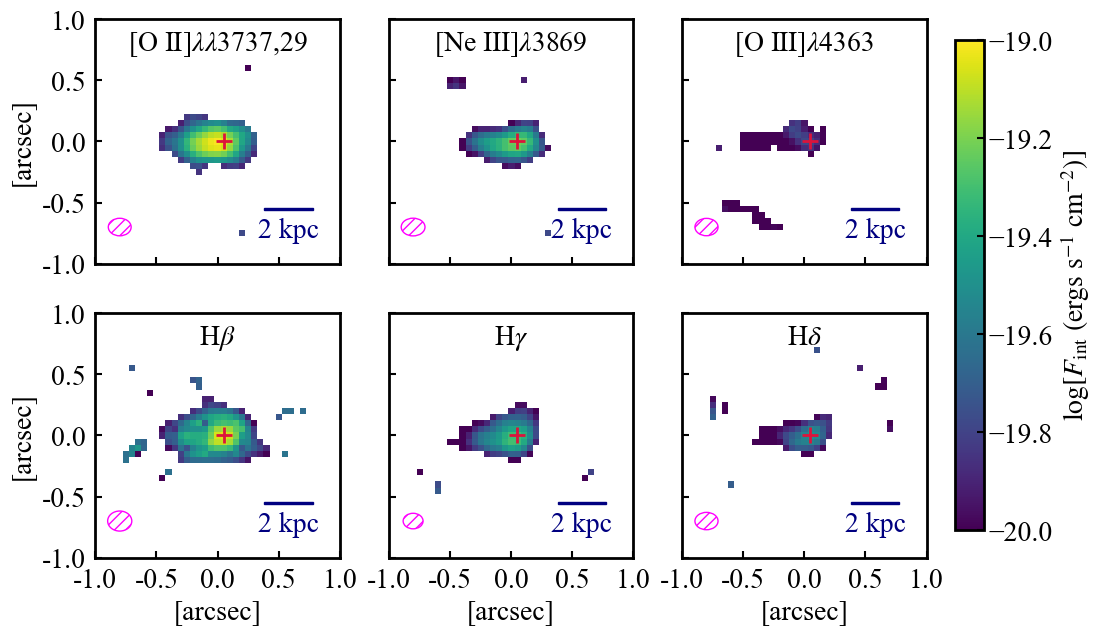}
    \caption{Resolved ($\mathrm{SNR}>3$) emission line maps of ID6355. Integrated fluxes, measured per spaxel, are plotted in $\log$ space to illustrate the relative brightness of each line. The magenta hatched ellipse in the bottom left of each plot illustrates the \jwst/NIRSpec PSF of the relevant emission line. The red cross on each figure corresponds to the centroid of $\OIII\lambda5007$ integrated flux. Fluxes are uncorrected for dust attenuation. \textbf{Top row:} Detected forbidden lines, from left to right: $\OII\lambda\lambda3727,29$, $\NeIII\lambda3869$ and $\OIII\lambda4363$. \textbf{Bottom row:} Detected Balmer lines, from left to right: \Hbeta, \Hgamma \ and \Hdelta. }
    \label{fig:6355intfluxmaps}
\end{figure*}

\begin{figure*}
        \centering
	\includegraphics[width=0.8\paperwidth]{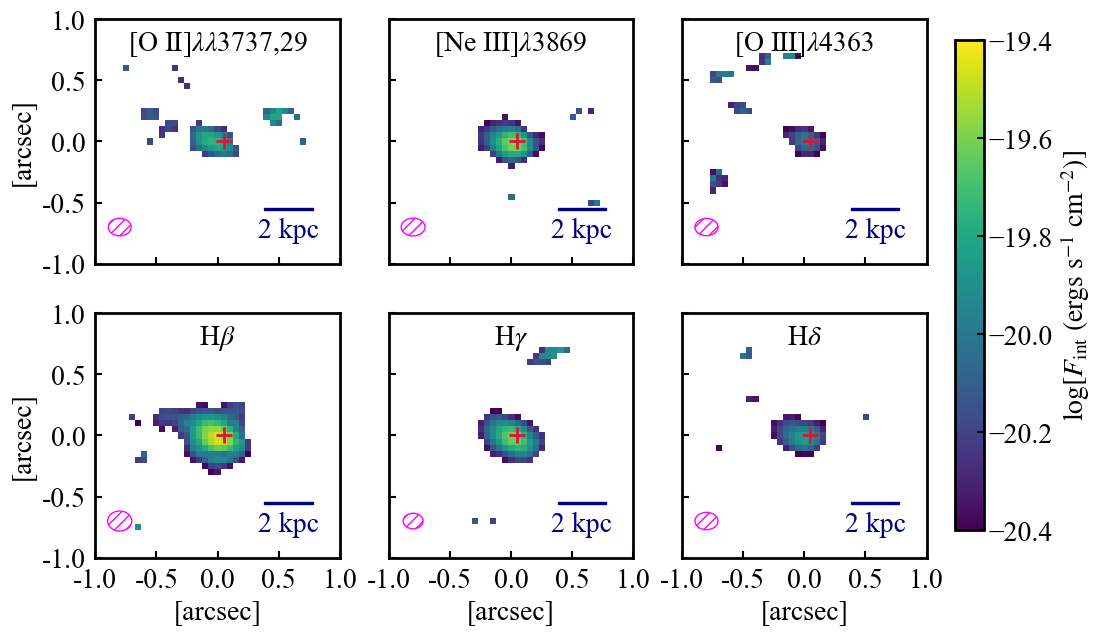}
    \caption{Resolved ($\mathrm{SNR}>3$) emission line maps of ID10612. Integrated fluxes, measured per spaxel, are plotted in $\log$ space to illustrate the relative brightness of each line. The magenta hatched ellipse in the bottom left of each plot illustrates the \jwst/NIRSpec PSF of the relevant emission line. The red cross on each figure corresponds to the centroid of $\OIII\lambda5007$ integrated flux. Fluxes are uncorrected for dust attenuation. \textbf{Top row:} Detected forbidden lines, from left to right: $\OII\lambda\lambda3727,29$, $\NeIII\lambda3869$ and $\OIII\lambda4363$. \textbf{Bottom row:} Detected Balmer lines, from left to right: \Hbeta, \Hgamma \ and \Hdelta.}
    \label{fig:10612intfluxmaps}
\end{figure*}

We present flux maps of the fitted forbidden and Balmer emission lines in Figs.~\ref{fig:6355intfluxmaps} and \ref{fig:10612intfluxmaps}, and note that the narrow and broad components of $\OIII\lambda5007$ integrated flux are mapped in panels i) and ii) of Figs.~\ref{fig:6355_kinem} and \ref{fig:10612_kinem} later in the paper. $\OIII\lambda5007$ and \Hbeta are the brightest detections in both galaxies, with all other emission lines about an order of magnitude fainter. The centroid of each emission line approximately coincides with the centroid of $\OIII\lambda5007$ emission as indicated in Figs.~\ref{fig:6355intfluxmaps} and \ref{fig:10612intfluxmaps}, though \Hbeta emission in particular appears to be extended away from this centroid in both galaxies.

Our resolved detections of $\OIII\lambda4363$ emission in each galaxy are particularly important, as $\OIII\lambda4363$ is a key emission line in AGN diagnostics effective at high-redshift (see Sec.~\ref{sec:AGNdiag}). The extended ($r>0.1''$), spatially-resolved detections of the emission lines shown in Figs.~\ref{fig:6355intfluxmaps} and \ref{fig:10612intfluxmaps} also allow us to investigate the metallicity profiles of our targets (see Sec.~\ref{sec:SpaceISMprop}).  From the results of our spaxel-by-spaxel fitting, we will also present the $\OIII\lambda5007$-derived kinematics maps of our targets later on in Sec.~\ref{sec:kinmaps}.

{\renewcommand{\arraystretch}{1.4}
\begin{table}
\normalsize
\centering
 \begin{tabular}{ c c} 
 \hline
 \hline
  Name &  Emission Line Ratio \\ 
 \hline
 $\mathrm{R}_2$ & $\OIIall$/\Hbeta\\
 $\mathrm{R}_3$ & $\OIII\lambda5007$/\Hbeta\\
 $\hat{\mathrm{R}}$ & $0.47 \ \mathrm{R}_2 + 0.88 \ \mathrm{R}_3$\\
 $\mathrm{O}_3\mathrm{O}_2$ & $\OIII\lambda5007/\OIIall$\\
 $\mathrm{Ne}_3\mathrm{O}_2$ & $\NeIII\lambda3869/\OIIall$\\
 \hline
 \hline
 \end{tabular}
 \caption{Emission line diagnostic ratios used in this work, together with their compact notation. For further discussion of these calibrations, see e.g. \citealt{Curti2020}.}
 \label{table:diagrat}
\end{table}
}

\subsection{Dust Corrections and Gas Metallicity Calibrations}\label{sec:metallicity}

Before estimating metallicity, it is important to first take into account any dust reddening affecting the emission lines. We use the available Balmer emission lines to perform this correction, in particular the \Hgamma/\Hbeta ratio, when both lines are detected with SNR $>3$. We assume the theoretical extinction-free \Hgamma/\Hbeta intrinsic ratio of $0.466$, valid for case-B recombination and an electron temperature of $T_\mathrm{e} \sim 10^4 \ \mathrm{K}$ \citep{OsterbrockFerland2006}, typical of warm ionised gas emitting rest-frame optical lines such as $\OIII\lambda5007$ (and consistent with $T_\mathrm{e}$ measured for these galaxies by \citealt{Curti2023}). To evaluate the dust attenuation $A_V$, we adopt a method similar to the Balmer decrement approach (e.g. \citealt{Calzetti1994, Dominguez13}), replacing the \Halpha/\Hbeta relations with those applicable for \Hgamma/\Hbeta:
\begin{equation}
    A_V = -2.5 \log_{10}\left( \frac{F_\Hgamma/F_\Hbeta}{0.466}\right) \frac{R_V}{k_\Hgamma - k_\Hbeta},
    \label{eq:AVeq}
\end{equation}
where $R_V = 4.05$, and $k_\Hgamma$ and $k_\Hbeta$ correspond to the relevant dust attenuation curve evaluated at the rest-frame wavelengths of \Hgamma and \Hbeta respectively. In this work, we apply the dust attenuation curve presented in \cite{Calzetti2000}, with $R_V = 4.05$, which is suitable for high-$z$, low-metallicity galaxies at the wavelengths of interest here \citep{Shivaei2020}, and similar to the approach adopted in \cite{Venturi2024}.

We note that the Balmer line ratio may also be affected by stellar absorption \citep{Groves2012}, and it is possible to account for this through knowledge of stellar populations obtained from SED fitting (e.g. as done by \citealt{Tacchella23}). However, the difference in spectral resolution between the MSA and IFU observations makes accurate spectral comparison difficult, so we neglect potential stellar absorption effects for the purposes of this work.

In the case of an integrated spectrum, dust correction is based on the integrated \Hgamma/\Hbeta ratio, as the aperture is sufficiently large to render PSF effects negligible. In the case of spaxel-by-spaxel measurements, we first PSF-match our data cubes to the worst PSF, i.e. the PSF at the wavelength of $\OIII\lambda5007$. This is done to mitigate the effects of PSF variation across different wavelengths on spatially-resolved metallicity measurements. We then apply dust corrections to spectra extracted from the PSF-matched cube according to their measured \Hgamma/\Hbeta ratios. After dust correction, we estimate metallicity from the corrected fluxes using two separate methods: strong-line calibrations (e.g. \citealt{Curti2020, Cataldi25}; see Table \ref{table:diagrat}) and the so-called direct-$T_\mathrm{e}$ method. 

We use updated strong-line calibrations, as presented by \citet{Cataldi25} for $z\sim2-3$, to explore the spatial metallicity variation across each galaxy. While these calibrations are based on observations at lower redshifts than our objects, they are still more appropriate to study $z\sim7.6$ galaxies than calibrations based on local-Universe measurements (e.g. \citealt{Curti2020}). In general, we find assuming different metallicity calibrations (e.g. \citealt{Curti2020, Nakajima22, Sanders23, Laseter24}; see \citealt{Maiolino19} for extensive discussion) can introduce systematic uncertainties of $\gtrsim 0.2$ dex. Furthermore, all strong-line calibrations are based on radiation fields typical of star-forming regions; if AGN are present, line ratios can deviate significantly from the SF calibrations due to the hardness of AGN radiation. This can result in inaccurate measurements of overall metallicities and false metallicity gradients. As strong-line measurements are highly uncertain, especially at high redshift, we utilise all five of the strong-line calibrations shown in Table \ref{table:diagrat} to obtain the best possible constraint on strong-line metallicity.

For the direct-$T_\mathrm{e}$ method, we derive the temperature of the \OIII-emitting gas (O$^{++}$) by exploiting the high SNR detection of \OIIIL[5007] and \OIIIL[4363] in the R2700 spectra. We simultaneously derive gas density using the \OIIall doublet ratio, whose lines are detected with SNR>3 in the R2700 integrated spectrum of ID6355. The temperature of the O$^{+}$ emitting region (hereafter \Tii) is assumed to follow the temperature-temperature relation from \cite{Izotov06lowz}, i.e., $t_{2} = 0.693  t_{3} + 2810$, where $t_{3}$ is the temperature of the  O$^{++}$ emitting gas. 

With the derived temperature and density of the O$^{++}$ gas, we then estimate the relative ionic abundances of oxygen and hydrogen using the intensity of each species, taking into account the different temperature- and density-dependent volumetric emissivity of the transitions using \textsc{PyNeb} \citep{PyNeb2015}. With this method, we derive the O$^{++}$/H and O$^{+}$/H ratios from the \OIIIall/\Hbeta (assuming t=\Tiii) and \OII/\Hbeta (assuming t=\Tii) ratios respectively, and compute the total oxygen abundance as O/H = O$^{+}$/H + O$^{++}$/H. The direct-$T_\mathrm{e}$ metallicity is then computed as $[12+\log(\mathrm{O/H})]$.

Using the integrated and spatially-resolved emission line properties, we estimate the metallicity properties of our galaxies on both integrated (Sec.~\ref{sec:teandne}) and spatially-resolved  (Sec.~\ref{sec:SpaceISMprop}) scales. Our main approach to studying spatially-resolved metallicity is to fit spectra extracted from concentric elliptical annuli and obtain radial gradients. We also map strong-line metallicity on a spaxel-by-spaxel basis using the $\OIII\lambda5007$ PSF-matched cubes. 

On a spaxel-by-spaxel basis, it is not always possible to correct a given spaxel for dust extinction, as \Hgamma \ is not as extended as \Hbeta in either galaxy (see Figs.~\ref{fig:6355intfluxmaps} and \ref{fig:10612intfluxmaps}). However, we only require spatially-resolved dust correction for our metallicity analysis. In the case of metallicity gradients, these are derived from spectra integrated across multiple spaxels so each line is detected with sufficient SNR to enable correction. For the strong-line metallicity maps, the metallicity associated to a given spaxel is only calculated when both \Hbeta and \Hgamma \ are detected with SNR$>3$ in its spectrum.

\subsection{Estimating Morphological Parameters}\label{sec:morph-fit}

To obtain the morphological parameters of each source, we fit the NIRCam F444W filter images (see Sec.~\ref{sec:NIRCam}) with a one-component Sérsic model using the Bayesian code \textsc{PySersic} \citep{Pysersic23}. The fitted parameters include ellipticity ($e$), Sérsic index ($n$), effective radius ($R_\mathrm{e}$), and position angle ($\theta$). From these parameters, we then estimate further physical properties of the galaxies, including their inclination ($i$). While performing the morphological fitting assumes the galaxy has a Sérsic brightness profile, no assumptions were made about the location of the photometric or kinematic centre of the galaxy.

The fitted and derived galaxy properties are summarised together in Table \ref{tab:pysersicgalpar}; the models, residuals and posterior corner plots produced from our \textsc{PySersic} fit are presented in Appendix \ref{sec:pysersic}. The fitted morphological centres are illustrated in Fig.~\ref{fig:metmap} later in the paper, and we find the fitted morphological centres are slightly offset from the \OIII$\lambda5007$ flux centroids. Interestingly, we identify half-light radii $R_\mathrm{e} > 1$ kpc in both galaxies, which are unexpectedly large for low-mass galaxies at such high redshifts \citep{Allen25, Miller24, Morishita24, Westcott25, McClymont25reff}.

{\renewcommand{\arraystretch}{1.4}
\begin{table}
\normalsize
\centering
 \begin{tabular}{c c || c c } 
 \hline
 \hline
 Parameter & Unit & \multicolumn{2}{c}{Value} \\ 
 &  & ID6355 & ID10612\\
 \hline
 $e$ & & $0.54 \pm 0.01$ & $0.53\pm0.01$ \\
 $n$ & & $0.95 \pm 0.03$ & $2.63^{+0.13}_{-0.14}$  \\
 $R_\mathrm{e}$ & pix &  $5.75^{+0.05}_{-0.04}$&   $6.45^{+0.19}_{-0.18}$\\
 $R_\mathrm{e}$ & kpc & $1.45 \pm 0.01$& $1.63\pm 0.05$ \\
 $\theta$ & $\degree$ & $93.4\pm0.6$ & $65.3 \pm 1.1$\\
 \hline
 $q$ & &  $0.46\pm0.01$ & $0.47\pm0.01$\\
  $i$ & $\degree$ & $65.0 \pm 0.7$ & $64.3\pm0.7$\\
 \hline
 \hline
  
 \end{tabular}
 \caption{Summary of galaxy parameters obtained by fitting NIRCam F444W images of our targets with \textsc{PySersic}. \textbf{Top segment:} Parameters obtained directly from the \textsc{PySersic} fits. \textbf{Bottom segment:} Derived galaxy parameters. The axis ratio $q$ and inclination $i$ are calculated from the fitted ellipticity; see Eqs. \ref{eq:ellip} and \ref{eq:inclination}. Errors are quoted at the $68\%$ confidence level and are derived from the \textsc{PySersic} fitting posteriors. The uncertainties quoted here exclude systematic errors not accounted for by the modelling and should therefore be taken as a lower limit. }
 \label{tab:pysersicgalpar}
\end{table}
}

The axis ratio $q = b/a$ (ratio of scale height to scale length) is related to the ellipticity by
\begin{equation}
    q = 1-e.
    \label{eq:ellip}
\end{equation}
Galaxy discs are widely thought to be thicker at high redshifts \citep{derWel14}, so the inclination $i$ of the disc is calculated from the measured axis ratio $q$, as demonstrated in \cite{Tully98}:
\begin{equation}
    \cos{i} = \left(\frac{q^2 - q_0^2}{1-q_0^2}\right)^{1/2},
    \label{eq:inclination}
\end{equation}
where $q_0$ is the intrinsic axial ratio of a perfectly edge-on disc galaxy; typically $q_0 \sim 0.2$ for a thin disc. We note that the choice of $q_0$ within its typical range, $0.0-0.2$, does not strongly impact inferred kinematical properties \citep{ForsterSchreiberWuyts20, Price20}.

\section{Galaxy Properties} \label{sec:ISMProp}

\subsection{Auroral Line AGN Diagnostics} \label{sec:AGNdiag}
Previous works \citep{Brinchmann23, Silcock24} have concluded there is evidence to support the presence of AGN in our target galaxies. Here, we further investigate this claim by utilising $\OIII\lambda4363$ auroral line based diagnostics from \cite{Mazzolari24}, which have been shown to reliably identify AGN at high redshift \citep{MazzolariCEERS24, Juodzbalis25}. 

\begin{figure*}
        \centering
	\includegraphics[width=0.85\paperwidth]{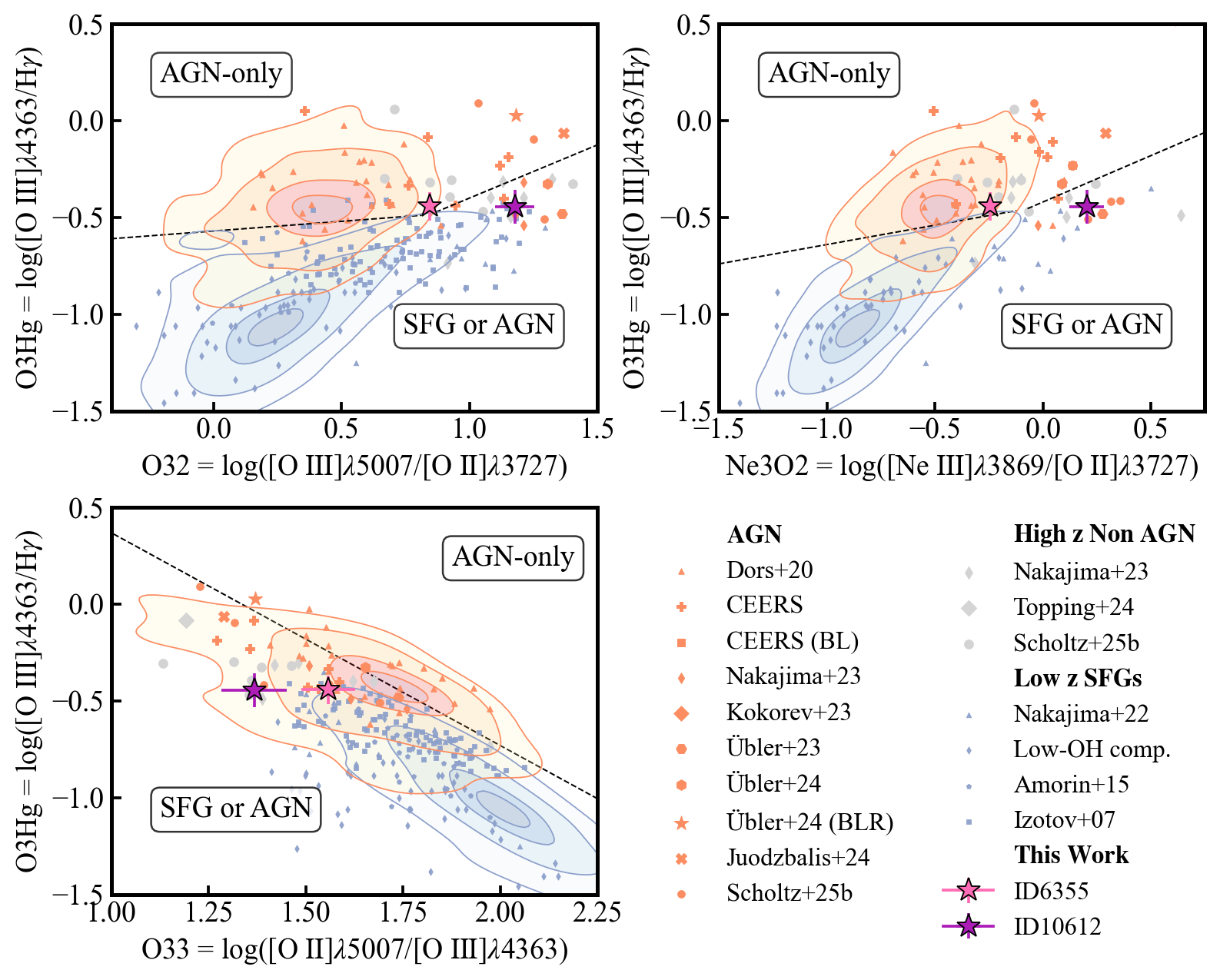}
    \caption{$\OIII\lambda4363$-based AGN diagnostic diagrams \citep{Mazzolari24}. ID6355 and ID10612 are plotted as the pink and purple stars, respectively. Orange points correspond to high-$z$ AGN, blue points to low-$z$ SFGs, and grey points to high-$z$ sources not classified as AGN; the sample shown in these plots is compiled from \citealt{Izotov07, Amorin15,Dors20, Nakajimaempgs22, Nakajima23census, Kokorev23, Ubler23,  Juodzbalis24,Topping24,Ubler24} and  \citealt{Scholtz25-COS3018}. The Low-OH compilation comprises sources from \citealt{Izotov06lowz, Izotov18, Izotov19, Berg12} and \citealt{ Pustilnik20, Pustilnik21}. Orange and blue contours show the distribution of SDSS AGN and SFGs respectively, including the 90\%, 70\%, 30\% and 10\% of the populations. Demarcation lines between AGN-only and SFR/AGN regions are plotted as dashed black lines. Fluxes were not corrected for dust attenuation.}
    \label{fig:mazzodiag}
\end{figure*}

We present the  $\OIII\lambda4363$ diagnostics in Fig. \ref{fig:mazzodiag}, where it can be seen that ID6355 nominally falls in the AGN-only region in two out of three diagrams. Although the points cross the demarcation lines within their uncertainties, given the additional evidence of a high-ionisation source from the detected $\NeIV$ emission in this galaxy \citep{Brinchmann23, Silcock24}, the diagnostics are consistent with the interpretation of ID6355 as hosting an AGN. Furthermore, as the \Hbeta \ emission line does not show evidence for broadening outside of the broad component also associated with the \OIIIall \ emission, this suggests ID6355 hosts a Type-II (narrow-line) AGN.

On the other hand, the line ratios of ID10612 are consistent with either star formation or AGN activity across all three diagrams; therefore, we still lack conclusive evidence for an AGN being the dominant source of ionisation in this galaxy. However, we note that \citealt{Mazzolari24} have emphasised that their demarcation lines are very conservative, and that the AGN-only region likely extends by 0.25--0.5 dex below the lines they provide (to be discussed more extensively in Jones et al., in prep.).

To estimate an upper limit on bolometric luminosity, we utilise the dust-corrected narrow-line \Hbeta \ fluxes from the galaxy-integrated spectra. With the calibrations reported in \cite{Netzer2009}, we estimate $\log_{10}(L_\mathrm{bol} /\mathrm{erg} \ \mathrm{s}^{-1}) \sim  45.7-46.2$ in our targets. However, these calibrations assume the lines used to derive $L_\mathrm{bol}$ are dominated by the AGN emission. This assumption is not necessarily true as we are considering the galaxy-integrated spectra, and therefore the estimated $L_\mathrm{bol}$ represent upper limits. Furthermore, depending on the calibration adopted \citep[e.g.][Hirschmann+in prep.]{Lamastra2009}, uncertainties in the estimated $L_\mathrm{bol}$ can go up to $\sim 1$ dex.

\subsection{Integrated Galaxy Properties}\label{sec:intgalprop}

We report the galaxy-integrated best-fit parameters of the narrow and broad components of a 2-Gaussian model fit to each galaxy (see Sec.~\ref{sec:ELF}) in the upper segment of Table \ref{table:intspecprop}. Importantly, we note that uncertainties in the derived quantities are dominated by systematic uncertainties from the calibrations used in our analysis, rather than from the random flux uncertainties in our measurement. As our quoted uncertainties on the final derived values are determined from the random uncertainties, they should be considered as a lower limit on the true error. In Table \ref{table:intspecprop} we additionally present some derived galaxy properties, as described across the following subsections.

\subsubsection{Dust Attenuation}\label{sec:A_V}
Table \ref{table:intspecprop} shows the observed ratio of integrated \Hgamma/\Hbeta fluxes ($F_\Hgamma/F_\Hbeta$) calculated from our spectral fitting, from which we also calculate the attenuation $A_V$ and its associated error (see Sec.~\ref{sec:metallicity}). We estimate $A_V$ in ID6355 as $0.3^{+0.6}_{-0.3}$, indicative of low dust content, though non-negligible extinction is also possible within the uncertainties. 

We note that the integrated spectrum of ID10612 exhibits a \Hgamma/\Hbeta ratio significantly larger than intrinsic, resulting in an unphysical $A_V$; such anomalous Balmer line ratios in early galaxies have been shown in simulations \citep[e.g.][]{McClymont2025caseb} and can also be seen in stacks of Type-II AGN spectra \citep{Scholtz2025_AGN}. The anomalous line ratios cannot be explained by dust attenuation alone, indicating that density-bound nebulae depart from Case-B physics. Furthermore, Case-B ionisation is highly dependent on particular temperatures and densities, and may only apply to specific regions within a galaxy. Hence, we can attribute the anomalous Balmer ratio in this galaxy to peculiar gas conditions; non-subtracted stellar continuum or other contaminants are unlikely to play a role due to the weakness of the continuum. As we cannot estimate $A_V$ directly in ID10612 from the Balmer line ratios, we assume no dust, consistent with the findings of \citet{Curti2023} and \citet{Tacchella23}. A significant caveat to this approach arises as it does not allow correction for absorption of ionising photons by dust before they ionise hydrogen, but this will have a negligible effect on our results at the low $A_V$ values we find.

\subsubsection{Star Formation Rates}\label{sec:SFRs}

Star-formation rates (SFRs) are typically calculated from \Halpha luminosity, but this emission line is outside of the NIRSpec/IFU wavelength range at the redshifts of our targets. In this work, we instead employ dust-corrected \Hbeta luminosity as a proxy for \Halpha luminosity, based on the case-B intrinsic ratio between the lines ($\Halpha/\Hbeta = 2.86$; \citealt{Osterbrok1989}). This ratio is dependent on both temperature and density, and as mentioned in Sec. \ref{sec:metallicity}, the Balmer-line ratio does not trace the entire dust content of the galaxy. However, due to the low dust attenuation we observe in these galaxies, we deem this approach sufficiently reliable for our purposes. 

To ensure we account for star formation in the full extent of each galaxy, we extract integrated spectra from $0.5''$ and $0.4''$  radius circular apertures for ID6355 and ID10612, respectively (see Fig.~\ref{fig:ifuvmsa} in Appendix~\ref{sec:msavifu}), and calculate the integrated flux of the \Hbeta narrow line component. After correcting for dust attenuation, we estimate $L_{\Hbeta}$ by combining these results with luminosity distance as estimated from our fitted redshifts. We ultimately estimate SFRs by adapting the \Halpha-SFR relation from \citealt{KennicuttEvans2012},
\begin{equation}
    \log_{10}(\mathrm{SFR} \ [M_\odot \ \mathrm{yr}^{-1}]) = \log_{10}(2.86 \times L_{\Hbeta} \ [\mathrm{erg} \ \mathrm{s}^{-1}]) - 41.27.
    \label{eq:sfr}
\end{equation}
We thus identify overall lensing-corrected SFRs of $\log(\mathrm{SFR} \ / \ M_\odot \ \mathrm{yr}^{-1})=1.73 \pm 0.03$ and $1.17 \pm 0.02$ in ID6355 and ID10612, respectively. We note that the Balmer-to-SFR conversion was calibrated for normal SFGs and solar metallicities, meaning that a non-zero ionizing photon escape fraction, or the presence of an AGN (as is likely the case for our targets), will introduce further systematic errors $\gtrsim0.15 \ $ dex in the measured SFRs \citep[i.e.][]{Kennicutt1998, Theios2019}.

 {\renewcommand{\arraystretch}{1.5}
\begin{table*}
\normalsize
\centering
 \begin{tabular}{c c || c c} 
 \hline
 \hline
  &  & ID6355 & ID10612 \\ 
 \hline
 Redshift & & $7.66360\pm0.00003$ & $7.65883\pm0.00002$ \\
 
 $\mathrm{FWHM}_\mathrm{narrow}$ & $\mathrm{km} \ \mathrm{s}^{-1}$ & $189 \pm 4$ & $70 \pm 4$\\
 $\mathrm{FWHM}_\mathrm{broad}$ & $\mathrm{km} \ \mathrm{s}^{-1}$ & $522 \pm 47$ & $275 \pm 12$\\
 $\Delta v$ &  $\mathrm{km} \ \mathrm{s}^{-1}$ & $79 \pm 21$ & $25 \pm 4$\\
 $F_{\mathrm{\OIII\lambda5007, narrow}}$ & $10^{-18} \ \mathrm{erg} \ \mathrm{s}^{-1} \ \mathrm{cm}^{-2}$ & $25.6 \pm 0.8$ & $6.8 \pm 0.3$ \\
 
$F_{\mathrm{\OIII\lambda5007, broad}}$ & $10^{-18} \ \mathrm{erg} \ \mathrm{s}^{-1} \ \mathrm{cm}^{-2}$ & $7.1 \pm 1.5$ & $4.2 \pm 0.4$\\

$F_{\mathrm{\Hbeta}}$ & $10^{-18} \ \mathrm{erg} \ \mathrm{s}^{-1} \ \mathrm{cm}^{-2}$ &  $4.4\pm0.2$ &  $1.5\pm0.1$\\

$F_{\Hgamma}$ & $10^{-18} \ \mathrm{erg} \ \mathrm{s}^{-1} \ \mathrm{cm}^{-2}$ &  $2.0\pm0.1$ & $0.8\pm0.1$\\

$F_{\mathrm{\OIIall}}$ & $10^{-19} \ \mathrm{erg} \ \mathrm{s}^{-1} \ \mathrm{cm}^{-2}$ & $18.0\pm1.1$ & $2.7\pm0.6$\\

$F_{\mathrm{\OIII\lambda4363}}$ & $10^{-19} \ \mathrm{erg} \ \mathrm{s}^{-1} \ \mathrm{cm}^{-2}$ & $7.2\pm1.1$ & $2.9\pm0.5$\\

\hline
\Hgamma/\Hbeta & &  $0.45\pm0.03$ & $0.54 \pm 0.04$\\

$A_V$ & & $0.3^{+0.6}_{-0.3}$& 0 \\
SFR & $\log[M_\odot \ \mathrm{yr}^{-1}]$  & $1.73 \pm 0.03$ & $1.17 \pm 0.02$ \\
$T_\mathrm{e}(\OIII)$ & $10^4$ K & $1.6 \pm 0.2$ & $1.8\pm0.4$\\
$n_\mathrm{e}$ & $\mathrm{cm}^{-3}$ & $790^{+410}_{-250}$ & 1000 (assumed, see Sec.~\ref{sec:teandne})\\
 12+$\log(\mathrm{O}/\mathrm{H})$ & Strong-Line & $\textit{7.81}\pm\textit{0.11}$ (see Sec.~\ref{sec:SpaceISMprop})& $7.41\pm0.15$\\
 12+$\log(\mathrm{O}/\mathrm{H})$ & Direct $T_\mathrm{e}$  & $7.90^{+0.30}_{-0.21}$ & $7.70^{+0.22}_{-0.20}$\\
 \hline
 \hline

 \end{tabular}
 \caption{Summary of galaxy properties, derived from integrated spectra extracted from the aperture described in Sec.~\ref{sec:ELF}, with the exception of SFR, which is determined from an integrated spectrum extracted from a circular aperture of radius $0.5''$ (ID6355) or $0.4''$ (ID10612). Errors were bootstrapped from the  spectral fitting uncertainties and are quoted at the $68\%$ confidence interval. We note that errors will always be dominated by systematics, including systematic uncertainties on the flux calibration of the instrument ($\sim10\%$, \citealt{Scholtz2025DR4}), so the errors quoted here and throughout the rest of this work represent a statistical lower bound on the true error. Quoted fluxes have been corrected for dust correction where possible. \textbf{Upper segment:}  $\Delta v$ represents the velocity offset of the broad component relative to the narrow component. \textbf{Lower segment:} Derived integrated properties of our targets. The strong-line metallicities are estimated using all five diagnostic ratios shown in Table~\ref{table:diagrat}. The strong-line metallicity of ID6355 may be affected by AGN ionisation, as discussed later in Sec.~\ref{sec:SpaceISMprop}. Note that $A_V$ is quoted as 0 for ID10612 as the \Hgamma/\Hbeta ratio indicates no dust in this galaxy.}
 \label{table:intspecprop}
\end{table*}
}

\subsubsection{Electron Temperature, Density and  Integrated Metallicity}\label{sec:teandne}

In ID6355, the individual \OIIall and \OIIIL[4363] emission lines are each detected with SNR \ $>3$, so electron density ($n_\mathrm{e}$) and temperature ($T_\mathrm{e}$) may be measured directly using \textsc{PyNeb}. We estimate a galaxy-integrated electron density of $790^{+410}_{-250} \ \mathrm{cm}^{-3}$ in ID6355, consistent with other densities measured at high-$z$, i.e. $n_\mathrm{e} \sim 300-1000 \ \mathrm{cm}^{-1}$ \citep{Isobe23, Marconcini24, Scholtz25-COS3018}. We also measure an electron temperature of $(1.6\pm0.2)\times10^4$ K.

However, even in the galaxy-integrated spectrum of ID10612, the individual \OIIall lines are not detected with sufficient SNR for direct calculation of electron density to be possible. Instead, we assume $n_\mathrm{e}$ of this galaxy is comparable to ISM electron densities found by studies of other \jwst galaxies at $z\sim 7-9$, i.e.  $n_\mathrm{e}\sim1000 \ \mathrm{cm}^{-3}$  \citep{Isobe23}, leveraging this assumed electron density together with the measured $\OIIIL[4363]/\OIIIL[5007]$ ratio to calculate $T_\mathrm{e}$. Hence, we estimate an electron temperature of  $(1.8\pm0.4)\times10^4$ K in ID10612. We note that measured $T_\mathrm{e}$ does not have a strong dependence on the choice of density estimation provided $n_\mathrm{e} \lesssim 1000 \ \mathrm{cm}^{-3}$, so assuming a lower $n_\mathrm{e}$  has a negligible impact on the measured direct-$T_\mathrm{e}$ metallicity.\footnote{Assuming $n_\mathrm{e} \sim 790 \ \mathrm{cm}^{-3}$, i.e. comparable to the density of ID6355, results in changes of $\sim0.01$ dex for the inferred $T_\mathrm{e}$ and direct-$T_\mathrm{e}$ metallicity of ID10612, insignificant relative to the inferred uncertainties)}

We present the galaxy-integrated metallicities in Table \ref{table:intspecprop}. We measure strong-line metallicities of $12+\log(\mathrm{O/H}) = 7.81\pm0.11$ ($0.13^{+0.04}_{-0.03} \ Z_\odot$) and $7.41 \pm0.15$ ($0.05^{+0.02}_{-0.01} \ Z_\odot$) in the central region ($r\sim1.25$ kpc) of ID6355 and ID10612, respectively. Additionally, we estimate the overall direct-$T_\mathrm{e}$ metallicity of ID6355 as $7.90^{+0.30}_{-0.21}$  ($0.16^{+0.11}_{-0.08}\ Z_\odot$). For ID10612, the assumption of $n_\mathrm{e} \sim 1000 \ \mathrm{cm}^{-3}$ yields a direct-$T_\mathrm{e}$ metallicity of $7.70^{+0.22}_{-0.20}$ ($0.10\pm0.05 \ Z_\odot$). We note that for each galaxy, both methods yield values of metallicity consistent within the associated errors.

\subsection{Metallicity Profiles}\label{sec:SpaceISMprop}

\begin{figure*}
\centering
\begin{subfigure}{\columnwidth}
    \centering
   \includegraphics[width=\linewidth]{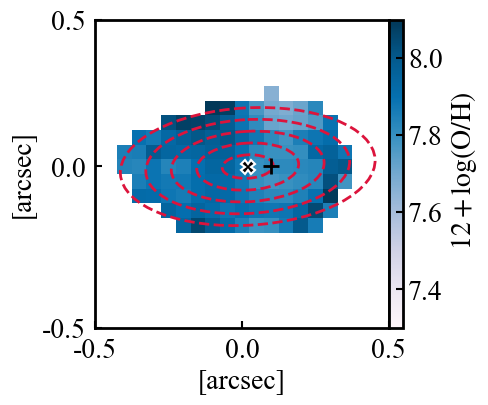}
   \caption{\textbf{ID6355}} \label{fig:6355_metmap}
\end{subfigure}
\hspace*{\fill}
\begin{subfigure}{\columnwidth}
    \centering
   \includegraphics[width=\linewidth]{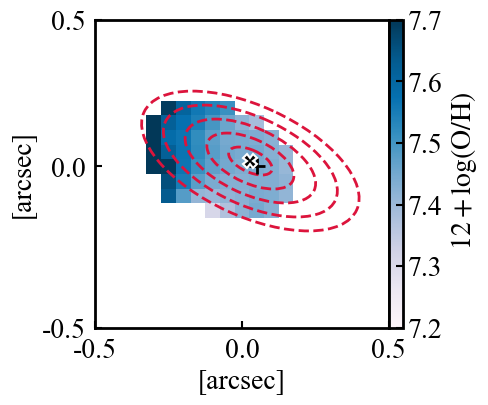}
   \caption{\textbf{ID10612}} \label{fig:10612_metmap}
\end{subfigure}
\hspace*{\fill}
\caption{Strong-line metallicity maps of each galaxy calculated from $\OIII\lambda5007$ PSF-matched cubes, using calibrations from \citet{Cataldi25} and the line ratios from Table~\ref{table:diagrat}, in spaxels where SNR$>3$ for all relevant emission lines. The $\OIII\lambda5007$ flux centroid of each galaxy is indicated by a black cross on the relevant plot. The morphological centre obtained from PySersic fitting (see) is shown as the white-outlined cross on each plot. The annular regions from which the metallicity gradients are derived (Fig. \ref{fig:gradmet}) are highlighted by the red dashed circles on each plot. The average error in metallicity for an individual spaxel is $\sim0.25$ dex for ID6355 and $\sim0.16$ dex for ID10612; hence, both maps are consistent with flat metallicity profiles.}
\label{fig:metmap}
\end{figure*}

\begin{figure}

   \includegraphics[width=0.98\linewidth, trim = {0.5cm 0.5cm 0.5cm 0.5cm}]{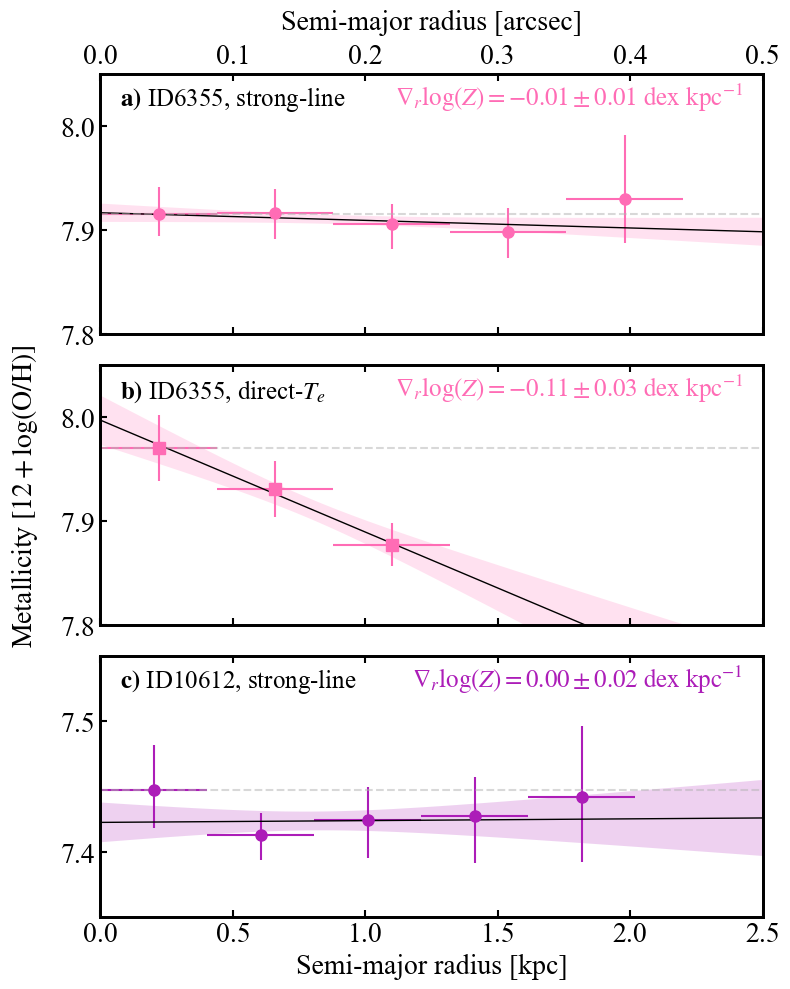}

\caption{Metallicity gradients for our targets, determined from binned spectra extracted from the elliptical apertures shown in Fig.~\ref{fig:metmap}. Top two panels: \textbf{a)} strong-line and \textbf{b)} direct-$T_\mathrm{e}$ gradients in ID6355. Bottom panel, \textbf{c)}: strong-line metallicity gradient for ID10612. Fluxes are corrected for dust attenuation (Sec.~\ref{sec:metallicity}), and strong-line metallicity is calculated using the \citealt{Cataldi25} calibrations with all five diagnostic ratios shown in Table~\ref{table:diagrat}), where $\mathrm{SNR}>3$ for the relevant emission lines. The horizontal gray dashed line on each plot illustrates the corresponding central metallicity value, while the solid black line indicates the best-fit linear profile. The shaded region shows the uncertainty associated to the best fit. The calculated metallicity gradient, along with errors quoted at the 68\% confidence level, are quoted together on each plot.}
\label{fig:gradmet}
\end{figure}

Having measured the integrated chemical properties of the two galaxies, we now utilise our spatially-resolved detection of the emission lines (Figs. \ref{fig:6355intfluxmaps} and \ref{fig:10612intfluxmaps}) to study the spatial variation of their metallicities. 

We first investigate the variation of metallicity on a spaxel-by-spaxel basis, utilising the PSF-matched cubes to present strong-line metallicity maps of ID6355 and ID10612 in Figs.~\ref{fig:6355_metmap} and \ref{fig:10612_metmap}, respectively. The metallicity variation is irregularly distributed in both galaxies (Fig.~\ref{fig:metmap}). At the same time, the error in each metallicity for a given spaxel ($\sim0.2$ dex) is comparable to the magnitude of metallicity variation across the entire galaxy, indicating that both galaxies have flat metallicity profiles overall. We stress that these maps should be interpreted with caution, as the strong-line metallicity calibrations are based on observations  at $z\sim2-3$ \citep{Cataldi25} and may not be reliable at $z\sim7.6$. Furthermore, as previously discussed (see Sec.~\ref{sec:metallicity}), the presence of an AGN in a galaxy will affect its measured line ratios and hence also any inferred strong-line metallicity \citep{Curti2017, Curti2020, Kumari21}, meaning these maps do not necessarily trace metallicity variation alone.

A more robust approach lies in calculating metallicity gradients using radially-binned spectra. We choose to radially bin spaxels in elliptical annuli uniformly-spaced along the semi-major axis (width $\sim0.08''$, see Fig.~\ref{fig:metmap}), obtaining spectra with higher SNR than available from individual spaxels. A different choice of bin (e.g. non-uniformly spaced) does not significantly impact our science results. The elliptical bins were defined using the corresponding position angle and ellipticity of each galaxy, as obtained by morphological fitting (Table~\ref{tab:pysersicgalpar}), and flux uncertainties were appropriately combined during the binning process. Once again, we utilise all five diagnostic emission line ratios from Table~\ref{table:diagrat}, only calculating the metallicity in a particular annulus if all necessary emission lines are detected at SNR $>3$. Metallicities are measured as a function of the distance along the elliptical semi-major axis, and we present these gradients in Fig.~\ref{fig:gradmet}. We measure strong-line metallicity gradients of $-0.01\pm0.01$ and $-0.00 \pm 0.02 \ \mathrm{dex} \ \mathrm{kpc}^{-1}$ in ID6355 and ID10612, respectively (see panels a) and c) of Fig.~\ref{fig:gradmet}), consistent with both galaxies having flat metallicity gradients. However, it must be again be stressed that calculating strong-line metallicities involves the inherent assumption that observed photoionisation is driven by star formation. The AGN present in ID6355 may affect observed line ratios independently of metallicity, with implications for its inferred strong-line metallicity gradient. Hence, direct-$T_\mathrm{e}$ metallicity gradients would offer a better probe of the metallicity variation in ID6355. 

In ID6355, there is sufficiently good detection of \OIIall (SNR > 3 for each line in the doublet) to map electron density in concentric annuli around the centre of the galaxy, though the region we can derive direct-$T_\mathrm{e}$ within is limited by the spatial extent of $\OIII\lambda4363$ detection (see Fig.~\ref{fig:6355intfluxmaps}). When inspecting each annulus-binned spectrum, we identified a small feature blueward of $\OIII\lambda4363$, which grows more prominent and contaminates the $\OIII\lambda4363$ emission at larger radii. We rule out \FeII emission as a cause of this feature, and determine it is most likely caused by a residual outlier that was missed by the data reduction pipeline. Though we masked out the wavelength channels corresponding to this residual feature, it is only possible to avoid its associated $\OIII\lambda4363$ contamination and investigate the direct-$T_\mathrm{e}$ metallicity gradient on $\sim1$ kpc scales. We present the calculated direct-$T_\mathrm{e}$ metallicity profile of ID6355 in Fig. \ref{fig:gradmet}b); we identify a negative gradient of $-0.11 \pm 0.03 \ \mathrm{dex} \ \mathrm{kpc}^{-1}$, in tension with the flat gradient found by strong-line measurements; this tension will be discussed further in Sec.~\ref{sec:metgradimp}.

\section{Galaxy Kinematics}\label{sec:kinemresults}

\subsection{Kinematical Maps}\label{sec:kinmaps}
\begin{figure*}
\begin{subfigure}{0.495\textwidth}
   \includegraphics[width=\linewidth]{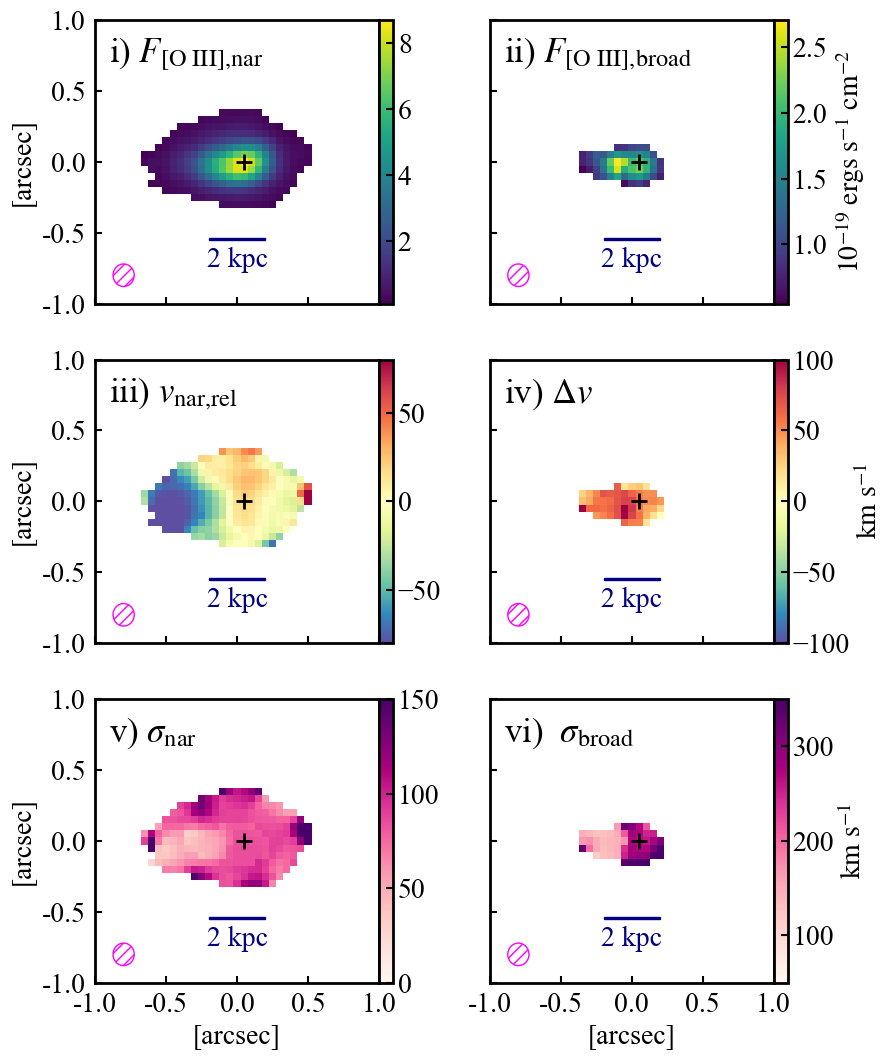}
   \caption{\textbf{ID6355}} \label{fig:6355_kinem}
\end{subfigure}
\hspace*{\fill}
\begin{subfigure}{0.49\textwidth}
   \includegraphics[width=\linewidth]{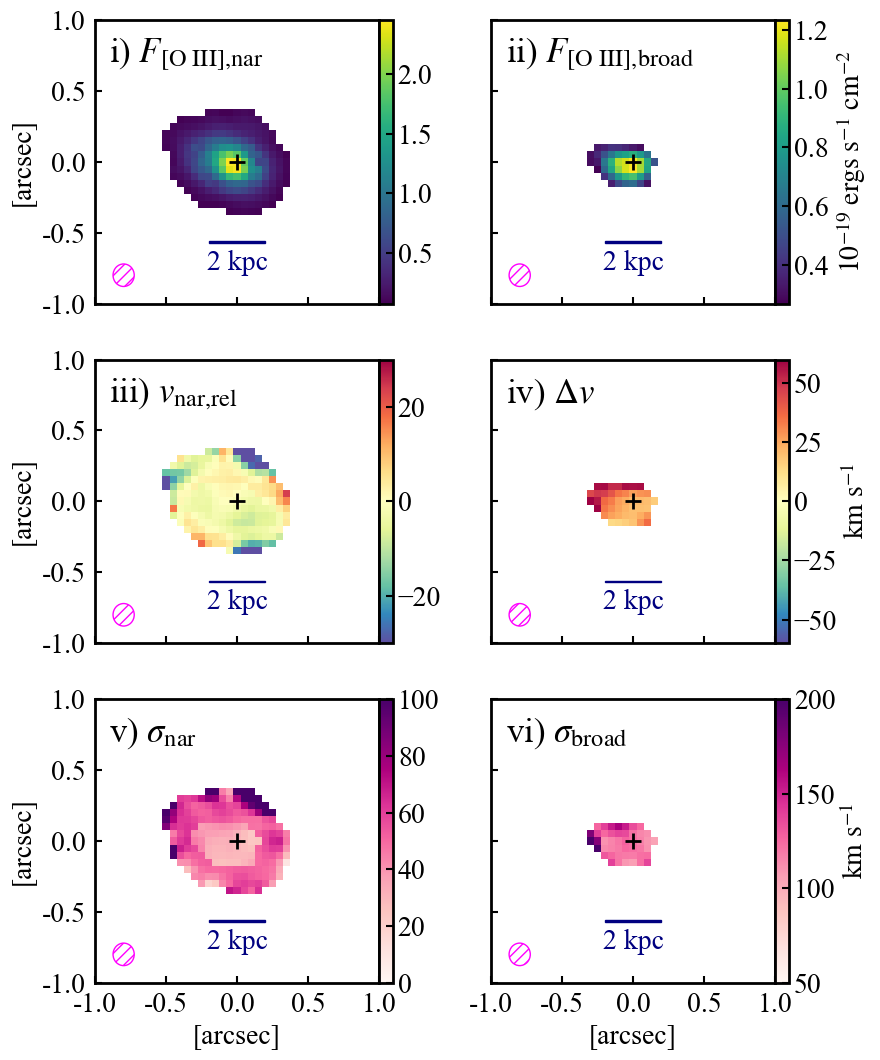}
   \caption{\textbf{ID10612}} \label{fig:10612_kinem}
\end{subfigure}
\hspace*{\fill}
\caption{$\OIII\lambda5007$ kinematic maps of each galaxy, comparing the narrow (\textbf{left column}) and broad (\textbf{right column}) components in each galaxy. These are plotted for all spaxels with $\OIII\lambda5007$ SNR$>$5 for the respective component. The magenta hatched ellipse in the bottom left of each plot indicates the size of the NIRSpec/IFS PSF at the wavelength of the  $\OIII\lambda5007$  emission line, and a $2$ kpc scale bar is also included. The black cross on each
figure corresponds to the centroid of $\OIII\lambda5007$ integrated flux. Individual panels in each figure show the following maps: \textbf{i)} narrow component of $\OIII\lambda5007$ integrated flux; \textbf{ii)} broad component of $\OIII\lambda5007$ integrated flux; \textbf{iii)} $v_\mathrm{nar,rel}$, the narrow line velocity relative to the systematic galaxy velocity; \textbf{iv)} $\Delta v$, the velocity offset between the narrow and broad components; \textbf{v)} $\sigma_\mathrm{nar}$, the intrinsic velocity dispersion of the narrow component; and \textbf{vi)} $\sigma_\mathrm{broad}$, the intrinsic velocity dispersion of the broad component.}
\label{fig:kinem}
\end{figure*}

We present an overview of the narrow- and broad-component $\OIII\lambda5007$ kinematics in Figs. \ref{fig:6355_kinem} and \ref{fig:10612_kinem}. We include integrated $\OIII\lambda5007$ flux and velocity dispersion maps of each component ($\sigma_\mathrm{nar}$ and $\sigma_\mathrm{broad}$, respectively), as well as the narrow-line relative velocity ($v_\mathrm{nar,rel}$) and the velocity offset between the components ($\Delta v$).

While several different lens models for the SMACS J0723.3-7327 cluster have been published \citep{Mahler2023, Pascale2022, Caminha22}, these models yield consistent magnification factors of $\mu \sim 1.2-1.7$ for our targets, and we conclude that overall lensing factors of this magnitude would have a negligible influence on the inferred galaxy kinematics \citep[e.g.][]{Arribas2024, Jonesmerger24}. Hence, an exploration of the effect lensing model choice might have on the systematics introduced to estimated kinematical properties is beyond the scope of this paper. 

Figs.~\ref{fig:6355_kinem})ii) and \ref{fig:10612_kinem})ii) show that the broad component in each galaxy is spatially extended beyond the size of the PSF ($r \sim 1$ kpc) and centrally concentrated. In ID6355, there is a  $\sim0.1''$ ($\sim0.5$ kpc) offset between the narrow and broad $\OIII\lambda5007$ flux centroids, while the two components' flux centroids spatially coincide in ID10612. The velocity offsets, $\Delta v$, between the narrow and broad components are fairly small ($<100$ \kms) in both galaxies. The fitted velocity dispersion is also significantly higher for the broad component than for the narrow component, as can be seen from a comparison of panels v) and vi) in Figs.~\ref{fig:6355_kinem}) and \ref{fig:10612_kinem}).  

From the wavelength resolution of the cubes and the mean $\OIII\lambda5007$ SNR of individual pixels, we estimate the velocity resolution of our observations as $\sim15$ \kms. From Fig.~\ref{fig:6355_kinem})iii), we thus identify a resolved asymmetric velocity gradient in ID6355, but Fig. \ref{fig:10612_kinem})iii) shows there is no clearly resolved rotation in ID10612. Although these galaxies are not rotating axisymmetric discs, we can estimate a constraint on the projected rotational velocities ($v_\mathrm{obs}$) in each galaxy. For ID6355, this is done by calculating the average of the maximum and minimum $v_\mathrm{nar,rel}$, as identified at $R_\mathrm{e}$ in the $v_\mathrm{nar,rel}$ map. As ID10612 lacks a resolved velocity gradient, we instead estimate the upper bound on its projected rotational velocity as the resolution of our observations, i.e. 15 \kms. However, $v_\mathrm{obs}$ must be corrected to account for galaxy inclination, using
\begin{equation}
    v_\mathrm{rot} = v_\mathrm{obs}/\sin(i),
    \label{eq:rotdeproj}
\end{equation}
where $v_\mathrm{rot}$ is the deprojected rotational velocity. We therefore calculated $v_\mathrm{rot}$ at the effective radius in ID6355 and ID10612 as $65 \pm 26$ \kms \ and $<17$ \kms \,  respectively.

To measure the intrinsic velocity dispersion $\sigma$, free of resolved motions, we adopt an approach similar to that of \cite{Wisnioski2015}, measuring the velocity dispersion of the narrow component in the outer regions of the galaxies. This approach reduces the effect of beam smearing, which mainly affects the central region and can be significant when a galaxy has a pronounced velocity gradient (as is the case for ID6355). Using the galaxy-integrated spectra, we measure intrinsic velocity dispersions of $103\pm40$ and $48\pm11$ \kms \ in ID6355 and ID10612, respectively.

We assess the relative importance of rotational and dispersional support in each galaxy through the ratio $v_\mathrm{rot}/\sigma$, measured within the effective radius $R_\mathrm{e}$. We measure $v_\mathrm{rot}/\sigma$ as $0.66\pm0.36$ and $<0.36$ in ID6355 and ID10612, respectively, indicating that ID6355 exhibits a small degree of rotational support, while ID10612 is entirely dispersion-dominated. Our observation that ID6355, the higher-mass of our two galaxies, is also the galaxy exhibiting more rotational support, is as expected from the results of studies at Cosmic Noon \citep[e.g.][]{Wisnioski2015}. 

\begin{figure}
        \centering
	\includegraphics[width=\columnwidth]{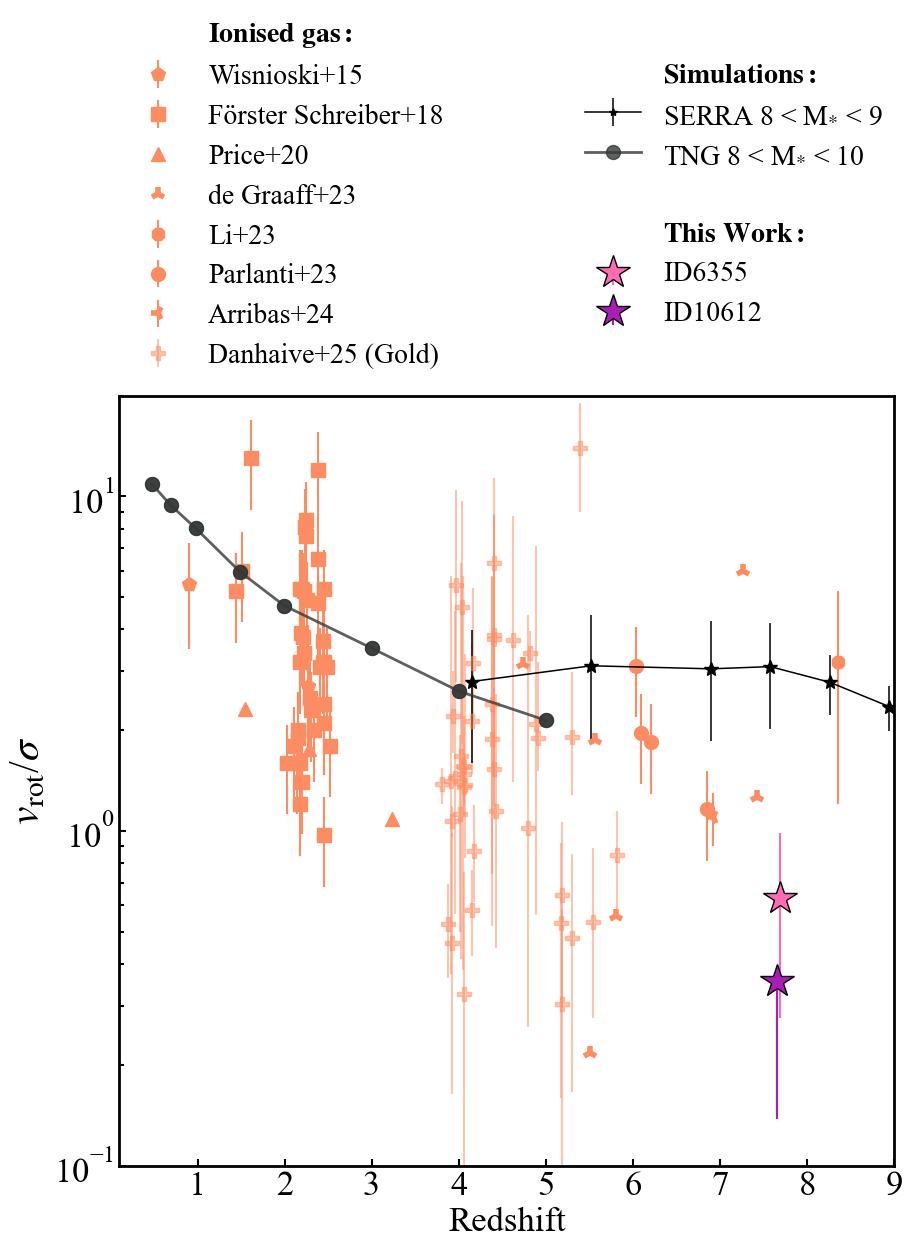}
    \caption{The evolution of $v_\mathrm{rot}/\sigma_\mathrm{v}$, as traced by ionised gas, across cosmic time. We compare ID6355 and ID10612 to a compilation from the literature, including observations of ionised gas: \jwst/NIRCam (\citealt{Livrotsigma23}; gold sample from \citealt{Lola2025}), ground-based IFU \citep{Wisnioski2015, ForsterSchreiber2018}, MOSFIRE  \citep{Price20},  ALMA $z\sim 6$ \citep{Parlanti2023}, NIRSpec \citep{Arribas2024, deGraaff2024}; and simulations: Illustris-TNG \citep{Pillepich2019} and SERRA \citep{Kohandel2024}. ID10612 only has a lower error bar in this plot, as its estimated rotational velocity is an upper limit.}
    \label{fig:vrotsigma}
\end{figure}

In Fig. \ref{fig:vrotsigma}, we compare our $v_\mathrm{rot}/\sigma$ to those from other ionised gas studies across cosmic time. Previous studies of ionised \citep{Parlanti2023, Arribas2024, deGraaff2024, Lola2025} and cold \citep{Rowland2024} gas at high redshift have indicated that highly turbulent systems such as our targets are not uncommon at early cosmic times, though we note that cold gas kinematics are known to be different from those of ionised gas. Our results are in tension with SERRA simulations of ionised gas for galaxies in a stellar mass range similar to that of our targets \citep{Kohandel2024}, which predict higher $v_\mathrm{rot}/\sigma$ across early cosmic times. We instead find our results are consistent with an expected increase in $\sigma$ and decrease in rotational support towards early cosmic times, as predicted by some other simulations such as TNG \citep{Pillepich2019} and now observed in many galaxies at $z\gtrsim4$ \citep{Lola2025}. 

\subsection{Dynamical Masses} \label{sec:dynmass}
To constrain the dynamical mass of our targets, we follow the same approach as other works (e.g. \citealt{Ubler23, Carniani25}), using the relation from \cite{vanderwel22}\footnote{We note that this approach was calibrated for the stellar kinematics of massive ($M_* \sim 10^9 -10^{10} \ M_\odot$) galaxies at $z\sim0.8$, but other calibrations provide similar answers, within $\sim 0.3$ dex \citep{Marconcini24}.}:
\begin{equation}
    M_\mathrm{dyn} = \beta(n)K(q)\frac{\sigma_*^2R_\mathrm{e}}{G},
    \label{eq:dynmass}
\end{equation}
where $\sigma_*$ is the stellar velocity dispersion. $\beta(n) = 8.87-0.831n+0.0241n^2$ and $K(q) = \left[0.87+0.38e^{-3.71(1-q)}\right]^2$ are correction factors for galaxy profile and inclination, following \cite{Cappellari2006} and \cite{vanderwel22}, respectively. $\sigma_*$ is inferred from $\sigma$ using the best-fit relations from \cite{Bezanson18}, $\sigma_* \simeq 1.26 \times \sigma$. $M_\mathrm{dyn}$ represents twice the mass inside the sphere of radius $R_\mathrm{e}$ \citep{Cappellari2006}; in the following analysis, we treat $M_\mathrm{dyn}$ as the total mass, even though it represents an extrapolation and the true total mass can thus be much larger.

{\renewcommand{\arraystretch}{1.4}
\begin{table}
\normalsize
\centering
 \begin{tabular}{c c || c c } 
 \hline
 \hline
 Property & Unit & \multicolumn{2}{c}{Value} \\ 
 & & ID6355 & ID10612\\
 \hline
 $v_\mathrm{rot}$ & $\mathrm{km} \ \mathrm{s}^{-1}$ & $65\pm26$ & $<17$ \\
 $\sigma$ & \kms & $103\pm40$ & $48\pm11$ \\
 $\sigma_*$ & $\mathrm{km} \ \mathrm{s}^{-1}$  & $130 \pm 50$ & $61\pm14$ \\
 $v_\mathrm{rot}/\sigma$ & & $0.63\pm0.36$ & $<0.36$\\
 $\log_{10}(M_\mathrm{dyn}/M_\odot)$ &  & $10.6^{+0.2}_{-0.6}$&$9.9^{+0.2}_{-0.3}$ \\
 $M_*/M_\mathrm{dyn}$ &  & $0.013 \pm 0.010$ & $0.015\pm0.007$ \\
 $v_\mathrm{esc}$ & $\mathrm{km} \ \mathrm{s}^{-1}$  & $480\pm190$ & $210 \pm 50$ \\
 \hline
 \hline
  
 \end{tabular}
 \caption{Summary of derived dynamical properties of each galaxy, along with bootstrapped errors quoted at the $68\%$ confidence interval. $v_\mathrm{rot}$ been corrected for galaxy inclination; see Eq. \ref{eq:rotdeproj}. For the comparison of dynamical to stellar mass, we note that \citealt{Curti2023} measure stellar masses of $\log(M_*/M_\odot) = 8.72 \pm 0.04$ and $8.08\pm0.04$ in ID6355 and ID10612, respectively.}
 \label{tab:dynpar}
\end{table}
}

We report measured dynamical masses of ID6355 and ID10612 as $\log_{10}(M_\mathrm{dyn}/M_\odot) = 10.6^{+0.2}_{-0.6}$ and $9.9^{+0.2}_{-0.3}$, respectively (see Table \ref{tab:dynpar}). It is also interesting to investigate the $M_*/M_\mathrm{dyn}$ ratio, as this gives an indication of the gas (and dark matter) fraction in a galaxy. Taking $M_*$ as measured by \citealt{Curti2023} (see caption of Table \ref{tab:dynpar}), we find $M_*/M_\mathrm{dyn}=0.013 \pm 0.010$ and $0.015 \pm 0.007$ in ID6355 and ID10612, respectively. These low ratios of stellar to dynamical mass indicate both galaxies have high gas fractions and/or high dark matter fractions in their central regions, which have also been found for other distant, low-mass galaxies (\citealt{deGraaff2024}, Danhaive et al. in prep.). In galaxies with low stellar masses (i.e. $\log(M_*/M_\odot) \leq 9.5-10$), high gas fractions such as these  combine with shallow potential wells to promote central starbursts. Gas inflows to the central regions \citep{DekelBurker14, El-Badry16, Tacchella16b, Hopkins23}, and outflows triggered by star formation \citep[e.g.][]{Carniani24JADES}, disrupt disc settling and inject turbulence into the ISM. Low-mass galaxies are also more unstable to disruption from gravitational instabilities. The interplay of these disruptive mechanisms can help explain the low degree of rotational support measured in our two galaxies and other galaxies in the early Universe \citep{Lola2025}.

\subsection{Identification of High-Velocity Gas}\label{sec:disentangle}

By mapping the kinematics of the $\OIII\lambda5007$ emission line in each galaxy, we have identified a resolved velocity gradient in ID6355 and a lack of resolved velocity gradient in ID10612 (see Figs.~\ref{fig:6355_kinem})iii) and \ref{fig:10612_kinem})iii), respectively). We have also found broad components in both ID6355 and ID10612, which are red-shifted relative to their narrow line counterparts. We now seek to disentangle the broad components in each galaxy from systemic disc emission, to show clearly that the broad components are not associated with galactic rotation but rather have a separate kinematical origin, such as an outflow or merger.

\begin{figure*}
        \centering
	\includegraphics[width=0.85\paperwidth]{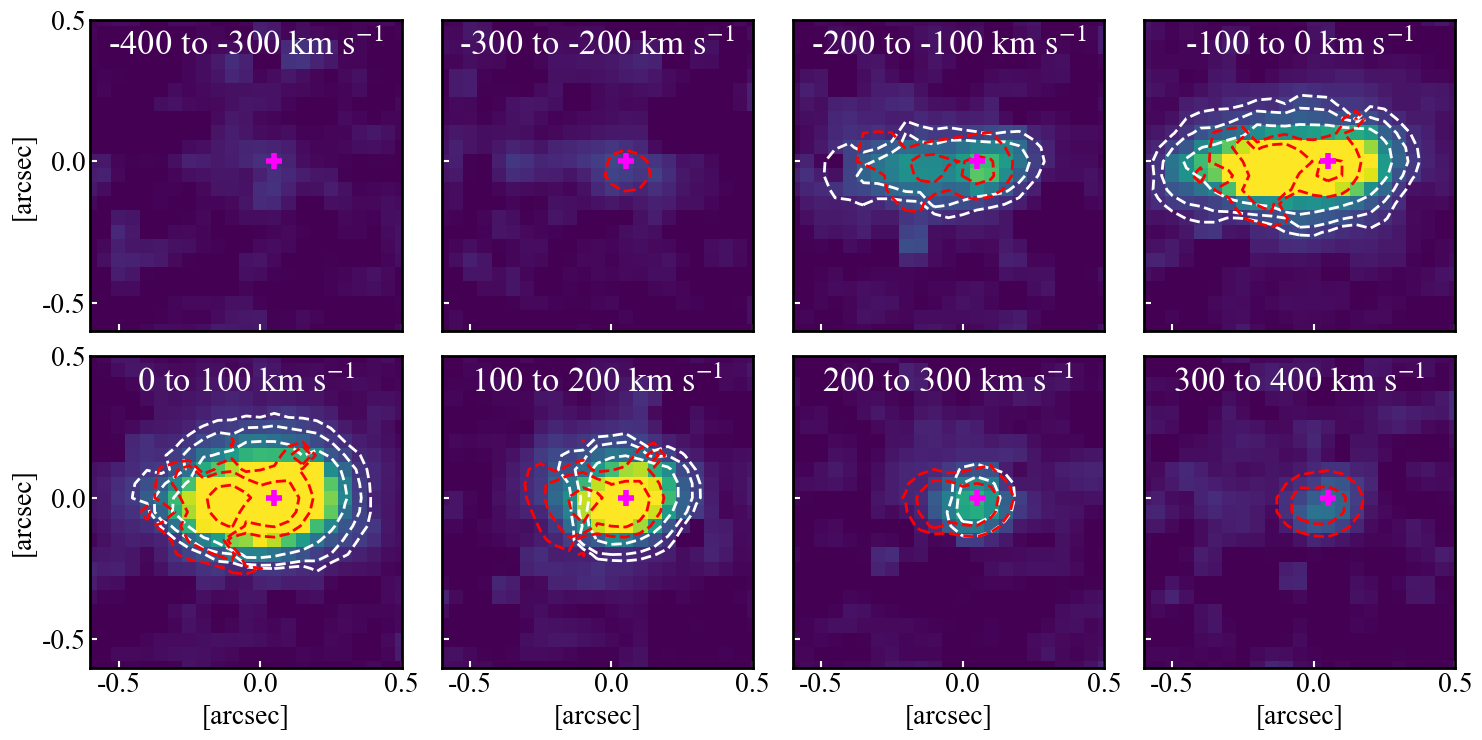}
    \caption{Velocity channel maps of ID6355, created by collapsing the spectral flux into 100 \kms \ relative velocity channels centred on the observed peak of the $\OIII\lambda5007$ emission line. The $\OIII\lambda5007$ flux centroid is shown as a magenta cross in each subplot. The dashed contours in white and red correspond to the 3, 5, and 10$\sigma$ flux levels of the fitted narrow and broad components, respectively, in each channel. }
    \label{fig:6355velmap}
\end{figure*}

\begin{figure*}
        \centering
	\includegraphics[width=0.85\paperwidth]{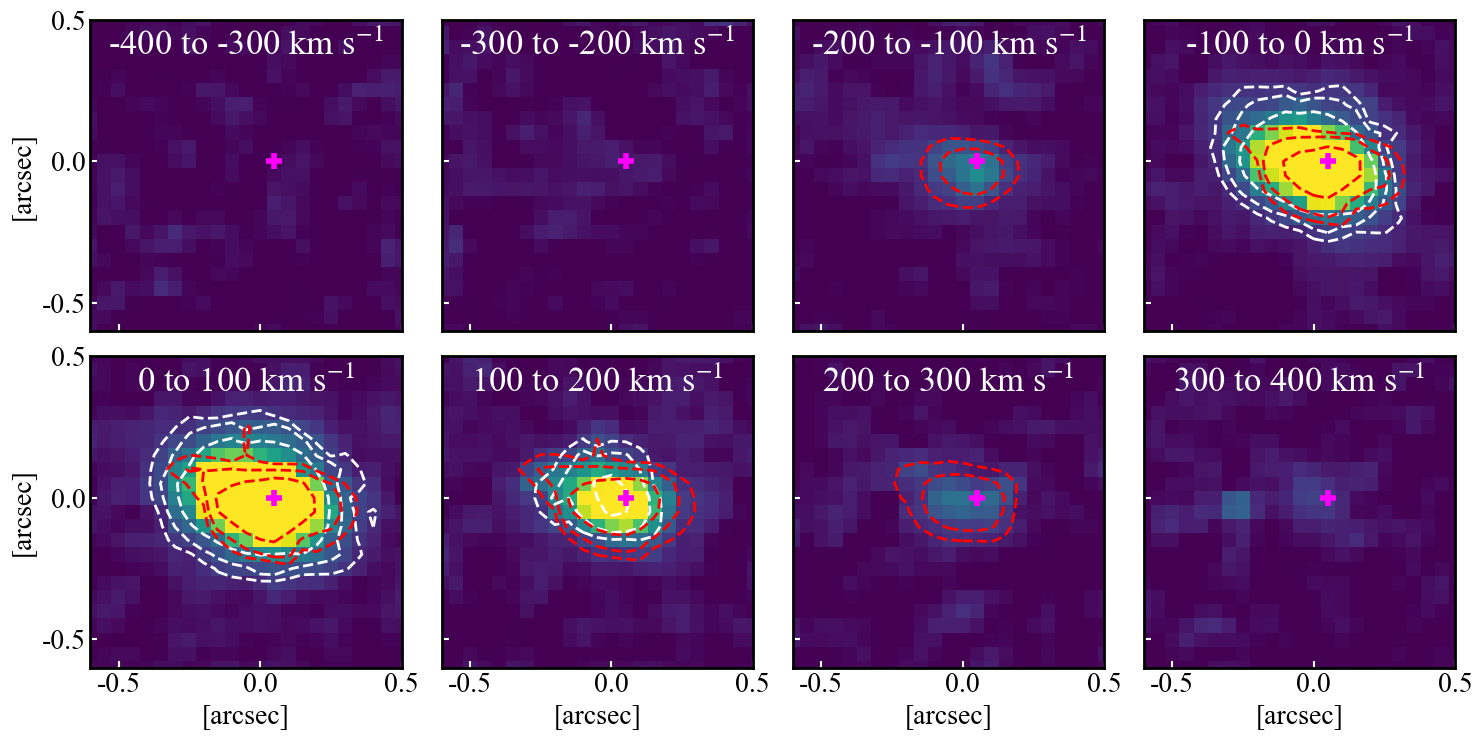}
    \caption{Velocity channel maps of ID10612, created by collapsing the spectral flux into 100 \kms relative velocity channels centred on the observed peak of the $\OIII\lambda5007$ emission line. The $\OIII\lambda5007$ flux centroid is shown as a magenta cross in each subplot. The dashed contours in white and red correspond to the 3, 5, and 10$\sigma$ flux levels of the fitted narrow and broad components, respectively, in each channel. }
    \label{fig:10612velmap}
\end{figure*}

To more clearly illustrate the motion of gas, we create velocity channel maps for each galaxy (see Figs. \ref{fig:6355velmap} and \ref{fig:10612velmap}) by collapsing the IFU cubes into velocity channels of 100 \kms, centred around the observed peak of $\OIII\lambda5007$ emission. If these galaxies exhibit pure rotation, we would expect the peak flux to be symmetric across the $0 \ \mathrm{km} \ \mathrm{s}^{-1}$ boundary in Figs.~\ref{fig:6355velmap} and \ref{fig:10612velmap}. However, we instead observe a faint flux in the high velocity ($\sim 200-400 \ $\kms) channels of both galaxies, which is not noticeably present in the corresponding low velocity channel, indicating a discrepancy. The additional broad kinematical component in these galaxies can manifest as an asymmetry in the channels such as we observe here, as any broad component such as outflow or merger is generally complex and affected by attenuation by ISM in the disc. 

In Figs. \ref{fig:6355velmap} and \ref{fig:10612velmap}, we additionally show the contours of fitted narrow and broad component flux in each velocity channel, plotted at the 1, 3, and 5$\sigma$ confidence levels. In each galaxy, the broad component of the $\OIII\lambda5007$ line is asymmetrically distributed about the central wavelength, contributing significantly to the high velocity ($\sim 200-400 \ $\kms), red-shifted component. As the broad flux falls well outside what we can reasonably attribute to the galactic rotation, we conclude that these broad components arise due to a second kinematical component in each galaxy.

\subsection{Origin of the Broad Component}\label{sec:otherscenarios}
Across the previous sections, we have presented evidence for an additional high-velocity ($>$200 \kms), non-rotating kinematical component in the $\OIII\lambda5007$ emission line-based kinematics of both galaxies (see Figs.
\ref{fig:fullspec}, \ref{fig:kinem}, \ref{fig:6355velmap} and \ref{fig:10612velmap}). We will now discuss the possible origin of these broad components.

\cite{Tacchella23} analysed the NIRCam photometry of these galaxies, finding that both have shown a recent increase in SFR. They further identified both galaxies as having complex morphologies consisting of multiple components, suggesting that their recent increases in SFR are driven by in-situ gravitational instabilities or mergers. In fact, enhanced merger rates are expected in protocluster environments like the one which hosts our target galaxies \citep{Marcelin25}. Our measured broad components are also consistent with a merger, as there are small spectral ($<100 \ $\kms) and spatial ($<0.2''$) separations between the narrow and broad components in each object. However, it should be pointed out that the galaxies were shown to be clumpy in the UV NIRCam filters, suggesting bursty star formation rather than a merger origin. Additionally, we do not see a multiply-peaked $\OIII\lambda5007$ or \Hbeta flux distribution in our spatial maps as is typically found in mergers (e.g. \citealt{Jonesmerger24}). 

Interestingly, the broad component in both galaxies is red-shifted relative to the narrow component; this indicates the gas is moving along the line of sight away from us, suggesting an inflow of gas. However, there are two significant problems with this explanation. Firstly, we have determined both galaxies to have low dust content (based on measurements of \Hgamma/\Hbeta; see Sec.~\ref{sec:A_V}), rendering it unlikely that a significant \OIIIall blue wing is present but dust-obscured. Secondly, in the case of an inflow, we would expect to see the broad component correspond to nearly pristine gas with $\OIII\lambda5007/\Hbeta <1$ \citep{Vanzella23, Hsiao25, Maiolino25BH, Nakajima25}. However, we note that our measured broad components have $\OIII\lambda5007$/H$\beta$ ratios $4.0\pm0.9$ and $8.6\pm2.3$ in ID6355 and ID10612, respectively, significantly higher than expected for pristine inflowing gas. 

Another more likely explanation for the observed broad component is an outflow. Indeed, \cite{Xu2025} utilised the MSA spectra of ID6355 and ID10612 to identify their broad components and characterised them as outflows. The broad component velocities we now infer are high ($\gtrsim250$ \ \kms), as are the velocity dispersions of both the narrow and broad components (Table \ref{tab:dynpar}), more suggestive of gas that is highly turbulent and re-energised in some way, possibly from feedback processes. Furthermore, Fig.~\ref{fig:6355velmap} indicates the broad component in ID6355 does contribute to high-velocity gas at $-300$ to $-200$ \ \kms, as we might traditionally expect for an outflow. The red-shifted nature of the broad component could arise due to outflow orientation, but due to our small sample size, we cannot distinguish whether red-shifted outflow components are a systemic property of early galaxies like our targets, or whether we have simply managed to identify this population by chance. Regardless, purely red-shifted outflows have already been observed as part of lower redshift studies such as the KASHz survey at Cosmic Noon, where $\sim16\%$ of outflows are found to be red-shifted rather than blue-shifted \citep{Harrison16, Scholtz2020}. We will require a larger sample of similar targets to distinguish whether red-shifted outflows are common at high-$z$.

Robustly identifying the origin of a broad component is not straightforward at such high redshifts, and is complicated further by the low masses of our targets. With multi-phase or higher-resolution observations, it may be possible to conclusively distinguish between the different scenarios, but currently we work within the limitations of our data. Given the high outflow velocity ($\sim500$ \kms) and central concentration of the broad kinematical component in ID6355, we conclude that this component is not part of an inflow or a merger. As such, taking into consideration the caveats discussed above, we attribute the broad component to an outflow in ID6355. On the other hand, interpreting the broad component in ID10612 is more challenging. The lower outflow velocity of 255 \kms and lack of settled kinematics (see Sec.~\ref{sec:kinemresults}) in this galaxy make the outflow interpretation difficult. Hence, the broad component could be an artifact of the complex kinematics of this object; we can note the obvious `footprint' of the broad component visible in Fig. \ref{fig:10612_kinem})v) as evidence of the difficulties of disentangling the kinematical components in this galaxy. Therefore, from now on we tentatively attribute the broad component of ID10612 to an outflow.

\subsection{Outflow Properties}\label{sec:outflowcalc}
Having identified evidence that the broad component in each galaxy arises from an ionised gas outflow, we now estimate the associated outflow properties. We first extract an integrated spectrum from the `outflow region' of each galaxy, as identified in Figs.~\ref{fig:6355_kinem})ii) and \ref{fig:10612_kinem})ii), where the broad component of $\OIII\lambda5007$ emission is detected with SNR$\ >5$. To determine the velocity of the outflowing gas ($v_\mathrm{out}$), we use the prescription from \cite{Rupke2005}:
\begin{equation}
v_\mathrm{out} = |v_\mathrm{broad} -  v_\mathrm{narrow}|+2\sigma_\mathrm{broad, deconv},
\label{eq:outvel}
\end{equation}
where $|v_\mathrm{broad} -  v_\mathrm{narrow}| = \Delta v$ is the velocity offset between the peaks of the broad and narrow components. $\sigma_\mathrm{broad,deconv}$ is the velocity dispersion of the broad component (measured as $210\pm20$ and $120\pm10$ \kms \ in ID6355 and ID10612, respectively; Table \ref{table:outflowprop}), which has been deconvolved with the instrumental linespread function $\sigma_\mathrm{LSF}$ (i.e. $\sigma_\mathrm{broad,deconv}^2 = \sigma_\mathrm{broad}^2 - \sigma_\mathrm{LSF}^2$). We adopt this method as it guarantees outflow velocities are not significantly dependent on the inclination of the outflow cone with respect to the line of sight \citep{Rupke2005, Fiore2017}. We estimate outflow velocities of $512^{+83}_{-84}$ \kms \ and $255^{+18}_{-27}$ \kms \ for ID6355 and ID10612, respectively, which are slow compared to the outflow velocities one might typically expect for AGN-driven outflows, $\sim400-3500 \ $ \kms \ (e.g. \citealt{Davies2020}). We additionally note that $v_\mathrm{out}$ should be treated as an upper limit, as only a small fraction of outflowing material would be moving with this high velocity.

\begin{figure*}
        \centering
	\includegraphics[width=0.85\paperwidth]{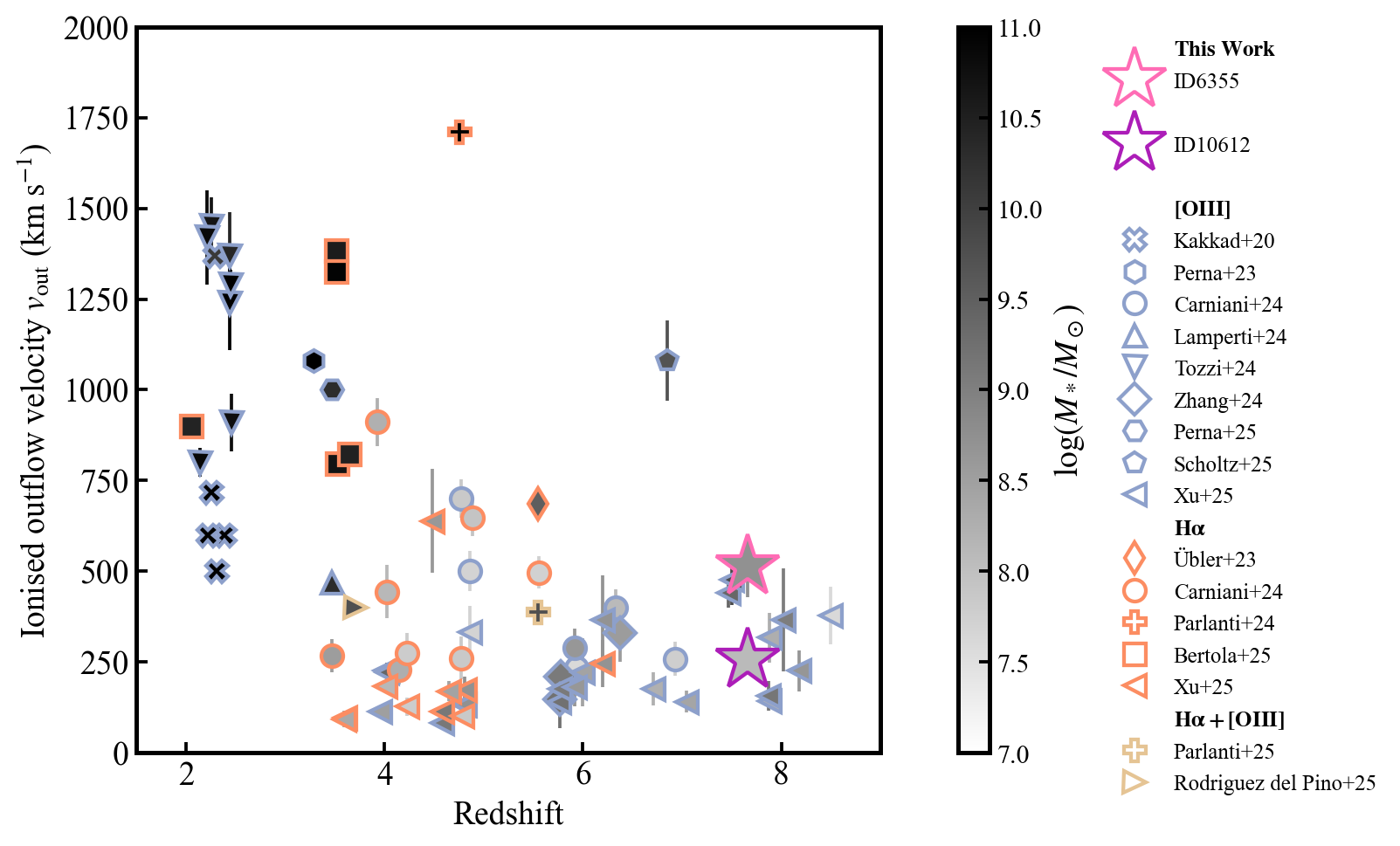}
    \caption{Cosmic evolution of ionised gas outflow velocities measured from rest-frame optical emission lines, with dependence on stellar mass also illustrated. ID6355 and ID10612 are plotted as the pink- and purple-outlined stars, respectively. Points in blue, orange, and yellow outlines correspond to outflows detected in H$\alpha$, \OIII, or both, respectively. The comparison samples shown in this plot include outflow velocities measured in: low-mass, high-redshift AGN and SFGs \citep{Carniani24JADES,Zhang2024, Xu2025}; higher-mass galaxies and quasars from the GA-NIFS literature \citep{Bertola2025, Scholtz25-COS3018, Parlanti2025, Perna2025, Lamperti2024, delPino2024, Parlanti2024, Perna23, Ubler23}; and massive AGN-host galaxies at $z\sim 2$ \citep{Kakkad2020, Tozzi2024}.}
    \label{fig:voutvsz}
\end{figure*}

We compare the outflow velocity of our sources to those of other \jwst spectroscopically-identified outflows (traced by H$\alpha$ or $\OIII\lambda5007$, i.e. ionised gas) across cosmic time in Fig. \ref{fig:voutvsz}; our results are consistent with other \OIII-detected outflows at $z\sim7-8$ in having outflow velocities $<1000$ \kms. This figure also includes 3 QSOs observed with NIRSpec/IFS at high-$z$ (e.g. \citealt{Marshall2023}), showing that high-luminosity or massive sources can show dramatically faster outflows even at early cosmic times \citep{Brazzini25}. This is in contrast to the lower-mass regime probed by our targets, where we expect to see lower-velocity outflows. Our results show outflow velocities consistent with those in SFGs, hinting at bursty star formation as a potential driver for the outflows. 

\begin{figure*}
        \centering
	\includegraphics[width=0.85\paperwidth]{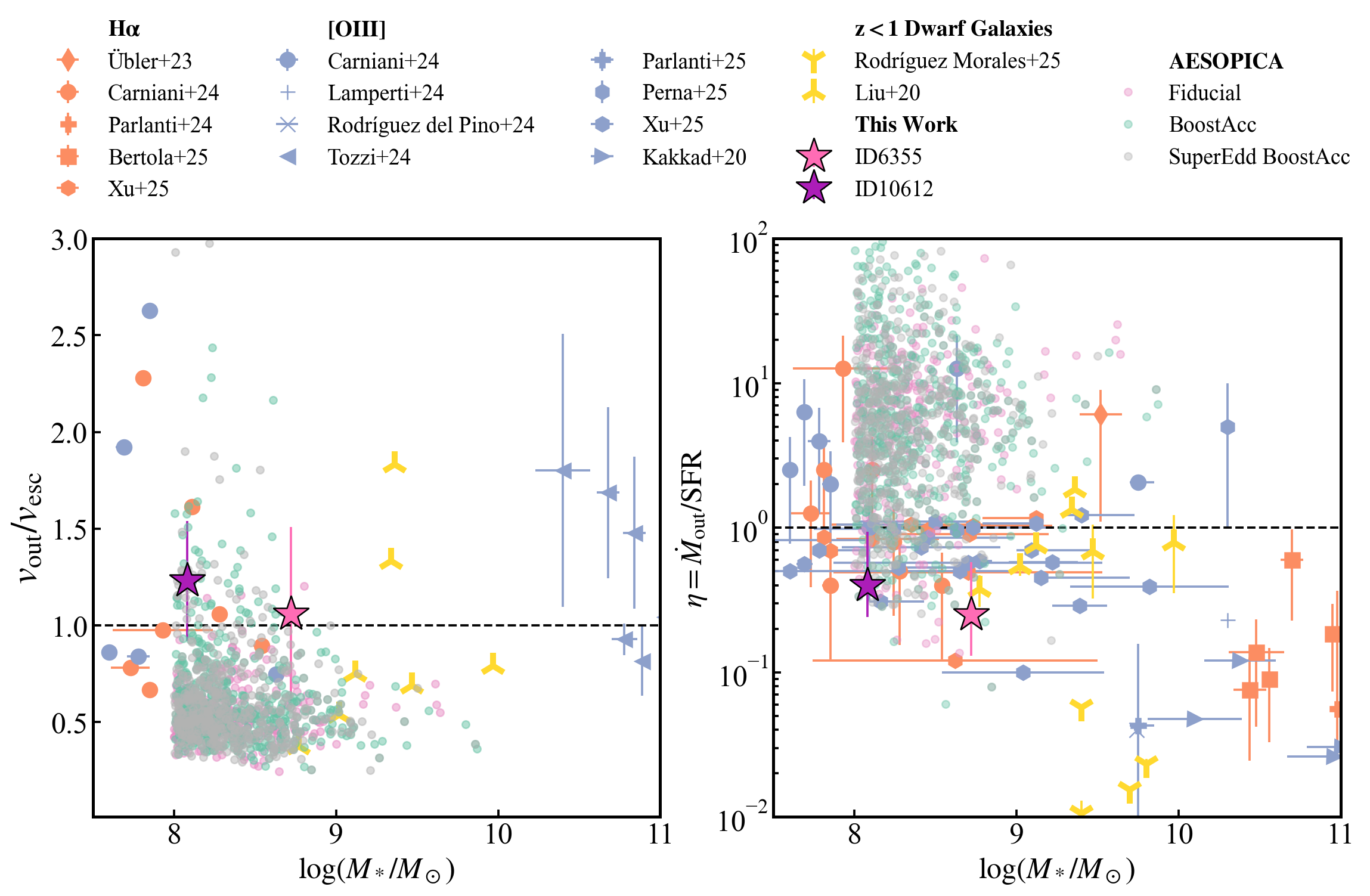}
    \caption{A summary of ionised gas outflow properties across observations and simulations. We compare our results for ID6355 and ID10612 (the pink and purple stars, respectively, in each plot) to outflows in $z>2$ AGN-host and star-forming galaxies (a subsample of the literature shown in Fig. \ref{fig:voutvsz}), as well as local ($z<1$) AGN-host dwarf galaxies from \citealt{Liu2020dwarfAGN} and \citealt{RM25dwarfagn}. We additionally compare our results to three different modes of the \textsc{Aesopica} simulations (Koudmani et al, in prep) of outflows in $z\sim7.6$ dwarf galaxy AGN: \textit{Fiducial}, boosted accretion (\textit{BoostAcc}), and super-Eddington with boosted accretion (\textit{SuperEdd+BoostAcc}). 
    \textbf{Left panel:} Ratio of the ionised gas outflow velocity to escape velocity ($v_\mathrm{out}/v_\mathrm{esc}$) as a function of stellar mass. $v_\mathrm{esc}$ for our targets have been estimated via calculated dynamical mass as described in Sec.~\ref{sec:dynmass}. The black dashed line indicates $v_\mathrm{out}/v_\mathrm{esc} = 1$. 
    \textbf{Right panel:} Mass loading factor ($\eta=\dot{M}_\mathrm{out}/\mathrm{SFR}$) as a function of stellar mass. The black dashed line indicates $\eta = 1$. }
    \label{fig:etavmstar}
\end{figure*}

Over time, outflows can have a significant effect on a galaxy's evolution, either by evacuating the gas into the circumgalactic medium (CGM) or by heating the CGM and thereby inhibiting future gas accretion \citep{Brownson19, Jones23, Bennett2024}. We assess the long-term impact of our detected outflows by comparing the measured outflow velocity to the escape velocity of the host galaxy. From our estimate of $M_\mathrm{dyn}$ (see Sec.~\ref{sec:dynmass}), we calculate the escape velocity $v_\mathrm{esc}$ as
\begin{equation}
    v_\mathrm{esc} = \sqrt{\frac{2GM_\mathrm{dyn}}{R_\mathrm{e}}}.
    \label{eq:escvel}
\end{equation} 
We find v$_{\rm out}$/v$_{\rm esc}$ ratios of $1.06\pm0.45$ and $1.23 \pm 0.31$ for ID6355 and ID10612, respectively, indicating that $v_\mathrm{out} \gtrsim v_\mathrm{esc}$ for the highest-velocity outflowing gas in both of our targets. While only part of the outflowing gas may escape the gravitational potential of the halo, this may enrich the CGM with metals and additionally suppress star formation over long timescales, as gas is gradually depleted from the galaxy. We compare the relationship of the $v_\mathrm{out}/v_\mathrm{esc}$ ratio to stellar mass both in our targets and in samples selected from the literature in the left-hand panel of Fig.~\ref{fig:etavmstar} (see Sec.~\ref{sec:Comparison} for further discussion).

We estimate the mass of outflowing gas, $M_\mathrm{out}$, directly from the dust-corrected luminosity $L^\mathrm{corr}_{\OIII}$ of the broad component of $\OIII\lambda5007$ (see e.g. \citealt{Carniani2015}), using
\begin{equation}
    M_\mathrm{out} = 0.8 \times 10^8 \left(\frac{L^\mathrm{corr}_\mathrm{\OIII}}{10^{44} \ \mathrm{erg} \ \mathrm{s}^{-1}}\right)\left(\frac{Z_\mathrm{out}}{Z_\odot}\right)^{-1} \left(\frac{n_\mathrm{out}}{500\ \mathrm{cm}^{-3}}\right)^{-1} \ \mathrm{M}_\odot,
\label{eq:outmass}
\end{equation}
where $Z_\mathrm{out}$ and $n_\mathrm{out}$ are the metallicity and electron density of the outflowing gas, respectively. However, diagnostic emission lines necessary to measure $n_\mathrm{e}$ and direct-$T_\mathrm{e}$ metallicity are too faint to be detected in our observed outflows, and assumptions must be made. Some studies of outflows at high-$z$ \citep[e.g.][]{Concas22, Carniani24JADES} assume an outflow electron density of $380~\mathrm{cm}^{-3}$, the typical value estimated from deep observations of $z\sim 2$ SF-driven outflows \citep{ForsterSchrieber2019}. However, we opt for a more conservative approach, setting the outflow density for both galaxies as $n_\mathrm{e} = 1000 \ \mathrm{cm}^{-3}$, the median ISM density of $z \sim 7-9$ \jwst galaxies as found by \cite{Isobe23}. Additionally, we set $Z_\mathrm{out}$ as the direct-$T_\mathrm{e}$ gas-phase metallicity of each galaxy\footnote{We note as a caveat that outflows are known to be more metal enriched than narrow line regions (e.g. \citealt{Pu2014}).}, i.e $0.16^{+0.11}_{-0.08} \ Z_\odot$ and  $0.10^{+0.06}_{-0.04} \ Z_\odot$ in ID6355 and ID10612, respectively (see Sec.~\ref{sec:teandne}). 

To estimate mass outflow rates, we adopt the same approach as other high-redshift outflow studies \citep[e.g.][]{Carniani24JADES, Cooper2025}, assuming a uniformly-filled conical outflow: 
\begin{equation}
\dot{M}_\mathrm{out} = 3 M_\mathrm{out} v_\mathrm{out}/r_\mathrm{out},
\label{eq:outrate}
\end{equation}
where $r_\mathrm{out}$ is the extension of the outflow (e.g. \citealt{Maiolino2012, GonzalezAlfonso17}). This approach also assumes that the outflow rate is constant with time (e.g. \citealt{Lutz2020}). We estimate the outflow extension $r_\mathrm{out}$ by measuring the area of the outflow region, approximating it as circular, and calculating the corresponding radius; this yields $r_\mathrm{out}\sim1$ kpc in each galaxy, comparable with the measured effective radii (see Table \ref{tab:pysersicgalpar}), and therefore representative of extended outflows. 

We estimate the mass outflow rates of our sources to be $14^{+18}_{-6}$ and $8^{+11}_{-3} M_\odot \mathrm{yr}^{-1}$ in ID6355 and ID10612, respectively, and report the properties of our detected outflows in Table \ref{table:outflowprop}. We will further discuss the impact of the outflows in Sec.~\ref{sec:impact}. It should be noted that the measured outflow masses and rates are highly sensitive to the choice of assumptions related to metallicity, electron density and outflow geometry.  For example, we find that assuming the outflow metallicity is closer to solar would decrease the mass outflow rate by a factor of $\sim$7-10 in both galaxies. Furthermore, assuming a shell rather than filled multiconical outflow geometry would reduce the measured mass outflow rate by a factor of three \citep{Lutz2020}. As an additional caveat, we note that \OIII-based energetics are typically underestimated with respect to \Hbeta-based ones \citep{Carniani2015, Fiore2017, Perna2019}, which could also reduce our measured mass outflow rate.

{\renewcommand{\arraystretch}{1.4}
\begin{table}
\normalsize
\centering
 \begin{tabular}{c c || c c} 
 \hline
 \hline
 Property & Unit &  ID6355 & ID10612 \\ 
 \hline
 $\sigma_\mathrm{broad,deconv}$ & \kms & $210 \pm 20$ & $120\pm10$\\
 $\Delta v_\mathrm{outflow}$  & \kms & $92\pm23$ & $30\pm10$\\
 
 $r_\mathrm{out}$ & kpc & $0.95^{+0.16}_{-0.20}$ & $0.73^{+0.14}_{-0.18}$ \\
 $v_\mathrm{out}$ & $\mathrm{km} \ \mathrm{s}^{-1}$& $512^{+83}_{-84}$ & $255^{+18}_{-27}$\\
 $v_\mathrm{out}/v_\mathrm{esc}$ & & $1.06 \pm 0.45$ & $1.23 \pm 0.31$\\
 $M_\mathrm{out}$ & $10^6\ M_\odot$&  $8.2^{+10.4}_{-3.5}$  & $7.4^{+7.0}_{-2.9}$\\
 $\dot{M}_\mathrm{out}$ & $\dot{M}_\odot \ \mathrm{yr}^{-1}$ & $14^{+18}_{-6}$ & $8^{+11}_{-3}$\\
 $\eta$ &  & $0.3^{+0.3}_{-0.1}$ & $0.4^{+0.5}_{-0.2}$\\
 \hline
 \hline
 \end{tabular}
 \caption{Summary of outflow properties, estimated by fitting the 2-Gaussian model (Sec.~\ref{sec:ELF}) to an integrated spectrum extracted from the `outflow region' of each galaxy, along with bootstrapped errors quoted at the $68\%$ confidence interval. $M_\mathrm{out}$ is calculated based on the direct-$T_\mathrm{e}$ metallicity of each galaxy (see Table \ref{table:intspecprop}). $r_\mathrm{out}$ and $M_\mathrm{out}$ have been corrected for lensing.}
 \label{table:outflowprop}
\end{table}
}

A further way of evaluating outflow impact is through estimating the mass loading factor ($\eta$) of the outflow, \begin{equation}
\eta = \dot{M}_\mathrm{out}/\mathrm{SFR},
\label{eq:eta}
\end{equation}
which indicates the primary source of gas consumption in the system. We estimate $\eta = 0.3^{+0.3}_{-0.1}$  and $0.4^{+0.5}_{-0.2}$ in ID6355 and ID10612, respectively, showing that outflows are sub-dominant to star formation in both galaxies. We must be conservative in interpreting $\eta$, as the assumptions involved in calculating mass-loading factor, including the caveats discussed above, can have systematic uncertainties of up to $\sim 1 \ \mathrm{dex}$ \citep{Kakkad2020}. Furthermore, we note that $\eta$ is an instantaneous quantity that can change on timescales of a few Myr (e.g. \citealt{Qiu21}), and hence is not representative of the long-term effect of feedback. We investigate the relationship of $\eta$ to stellar mass in our targets through comparison to a literature sample in the right-hand panel of Fig.~\ref{fig:etavmstar} (see Sec.~\ref{sec:Comparison} for discussion).

\section{Discussion} \label{Discussion}

\subsection{Comparison to NIRSpec/MSA Integrated Measurements}\label{sec:Comparison}

A number of existing studies (e.g. \citealt{Schaerer22, Brinchmann23, CarnallSmacs2023, Curti2023, Tacchella23, Trump2023,  Xu2025}) have utilised NIRSpec/MSA data from the ERO programme to characterise the two galaxies presented in this study. In this section, we present a brief overview of how our results, obtained from the higher-resolution NIRSpec/IFU observations, compare to those from previous studies.  

First of all, we discuss star formation rates. Previous studies all present measured SFRs that are significantly smaller than our values across both galaxies \citep{Brinchmann23, Curti2023, Nakajima23census, Tacchella23}. Across these previous measurements there is a significant spread in values depending on the methods used in calculating SFR ($\log(\mathrm{SFR} / M_\odot \ \mathrm{yr}^{-1})=1.29-1.49$ and $0.70-1.23$ for ID6355 and ID10612, respectively), indicating the assumptions involved play a nontrivial role in the final SFR estimates. As \citet{Brinchmann23, Curti2023, Tacchella23} estimate SFR from SED fitting, \citet{Nakajima23census} offers the best comparison to our results, as their estimated SFRs are also derived from dust-corrected \Hbeta fluxes. They measure $\log(\mathrm{SFR} / M_\odot \ \mathrm{yr}^{-1}) =1.41\pm0.01$ and $1.14^{+0.01}_{-0.02}$ for ID6355 and ID10612, respectively. The SFR we derived for ID10612, $\log(\mathrm{SFR} / M_\odot\ \mathrm{yr}^{-1}) = 1.17\pm0.02$, is slightly higher than their value while remaining marginally consistent within the quoted uncertainties. In contrast, for ID6355 we derived a much higher and more discrepant SFR, $\log(\mathrm{SFR} / M_\odot\ \mathrm{yr}^{-1}) =1.73\pm0.03$. We note \cite{Nakajima23census} have adopted the \cite{Chabrier2003} IMF in their SFR estimation, while we have applied SFR calibrations assuming a \cite{Kroupa2003} IMF, but \cite{KennicuttEvans2012} comment that these IMFs should yield similar results. Hence, the discrepancy in measured SFR instead most likely arises because our integrated spectra are extracted from a larger area compared to the slit of the NIRSpec/MSA observations (see Fig.~\ref{fig:ifuvmsa}); hence, we obtain a more complete estimate of the total SFR of the galaxies using IFU spectra. The discrepancy is therefore also more significant for ID6355 is it more extended than ID10612 and therefore not as well covered by the MSA slit.

The difference in aperture size between the MSA and IFU could also drive discrepancies in measured ISM properties, particularly if key emission lines (e.g. \OIII$\lambda4363$) are poorly captured by the MSA slit. For instance, we now turn to electron temperature. Comparing to results from \citealt{Curti2023}, we first note that we have measured a consistent electron temperature for ID10612, finding $(1.8\pm0.4)\times10^{4}\ \mathrm{K}$ compared to their measurement of $(1.75\pm0.16)\times10^{4}\ \mathrm{K}$. However, we find a significantly higher value for ID6355 compared to this earlier study (our value $(1.6\pm0.2)\times10^{4}\ \mathrm{K}$ compared to their value $(1.20\pm0.07)\times10^{4}\ \mathrm{K}$). However, this analysis is based on the entire \OIIIall and \Hbeta line profiles. When recalculating $T_\mathrm{e}$, now deconvolving the broad component detected in each galaxy, we instead find yet higher electron temperatures of $(1.8\pm0.4)\times10^4 \ \mathrm{K}$ and $(2.2\pm0.4)\times10^4 \ \mathrm{K}$, revealing discrepantly higher $T_\mathrm{e}$ for both galaxies, which naturally has implications for the direct-$T_\mathrm{e}$ metallicities we infer. 

Compared to previous studies \citep{Curti2023, Nakajima23census, Tacchella23}, our measured direct-$T_\mathrm{e}$ metallicities are significantly lower. In particular, in comparison to \citet{Curti2023}, the direct-$T_\mathrm{e}$ metallicity we have identified for ID6355 ($7.90^{+0.30}_{-0.21}$) is significantly lower than their quoted value of $8.24\pm0.07$, and the two results do not agree within their margins of error. Furthermore, when recalculating the direct-$T_\mathrm{e}$ metallicity after removing the broad component, we measure values of $7.48^{+0.43}_{-0.27}$ and $7.12^{+0.24}_{-0.10}$ in ID6355 and ID10612, respectively; these values are significantly lower than our earlier estimate that included both narrow and broad components ($7.90^{+0.30}_{-0.21}$ and $7.70^{+0.22}_{-0.20}$), and discrepant to the values found by \citet{Curti2023}, $8.24\pm0.07$ and $7.73 \pm 0.12$. This difference suggests the narrow and broad kinematical components in each galaxy have different metallicities and therefore distinct physical properties, which affects galaxy properties inferred from an overall line profile. Hence, when working with high-resolution spectra where separate broad components can be resolved, it is important to account for the presence of a broad component when making calculations of ISM properties. 

Finally, \cite{Xu2025} measure outflow properties in our target galaxies using the R1000 spectral data, finding a lower outflow velocity for ID6355 ($386\pm33$ \kms \ compared to our value of $512^{+83}_{-84}$ \kms), but obtaining a comparable result for ID10612 ($254\pm35$ \kms \ compared to $255^{+18}_{-27}$ \kms) within the respective errors. Furthermore, they have obtained mass loading factors of $0.59$ and $0.70$ for ID6355 and ID10612, which are consistent with our measured values of $\eta = 0.3^{+0.3}_{-0.1}$ and $0.4^{+0.5}_{-0.2}$ within the uncertainties. With our IFS observations, we were able to directly estimate the outflow size, unlike MSA observation-based studies such as \cite{Xu2025} which must rely on assumed sizes. Furthermore, the spatial resolution of our observations enabled us to trace outflows and SFRs more completely than would be possible with the MSA observations. These significant improvements of IFU over MSA, combined with the inherently uncertain nature of $\eta$,  therefore do not allow for meaningful comparison between our results and those of \cite{Xu2025}.

\subsection{Metallicity Profile Implications}\label{sec:metgradimp}

Metallicity gradients bear the imprint of underlying processes such as gas accretion, star formation, outflows, and mergers, as well as the interplay of these processes \citep{Maiolino19}. Overall, negative (radially decreasing) gradients are often interpreted as arising from inside-out galaxy formation, with stars forming earlier in the inner parts of a galaxy and having more time to chemically enrich the inner regions than the outer ones \citep{Garcia25}. Positive (radially increasing) gradients can be produced by accretion of external pristine gas towards the centre of the galaxy, or by metal-rich outflows from the galaxy centre driving metals outward \citep{Cresci10, Troncoso14, Chisholm2018}. Flat gradients can arise from the radial mixing of gas or redistribution of metals on galaxy scales, as induced by SN winds or outflows \citep{Baker23metfundep}; galaxy mergers can also produce flat gradients. Although we do not favour the merger scenario for the objects presented in this study (see Sec.~\ref{sec:otherscenarios}), it is worth mentioning that, based on recent \jwst observations, the galaxy major merger rate is $2-8\ \mathrm{Gyr}^{-1}$ \citep{Puskas25}. This is significantly higher than at low-$z$, meaning that it is likely that high-$z$ galaxies have undergone one or more major mergers in their lifetimes, which will have had an impact on their observed metallicity profiles.

Studies of spatially-resolved metallicity gradients are currently very limited at $z>5$, and primarily present strong-line metallicity gradients. \citet{Venturi2024} measured metallicity gradients in a sample of three galaxies at $z\sim 6-8$, which are comparable to our targets in terms of their stellar mass and bulk gas-phase metallicity. They identify metallicity gradients of $\sim - 0.05 \ \mathrm{to} \ + 0.14 \ \mathrm{dex} \ \mathrm{kpc}^{-1}$, flat within their uncertainties and consistent with processes mixing gas and metals on galaxy scales (i.e. mergers, outflows, or SN winds). \citet{Vallini} measure metallicity gradients in five UV-luminous Lyman-break galaxies at $z\sim7$ to be mildly negative, also compatible with flat gradients within the uncertainties.
\citet{Tripodi24} infer slight positive gradients by analysing 2D slit spectra stacked along the slit; such inverted gradients could be tracing near-pristine gas inflows toward the centre; again, the gradients they measure are also consistent with being flat within the uncertainties. In contrast to these results, \citet{Li25metgrad} present an investigation of metallicity gradients around the redshift of our sources based on the stacking of grism spectra and find evidence of steep negative gradients, which they ascribe to inside-out growth with little feedback. They suggest that the flat gradients found by \citet{Venturi2024} might be due to the sample comprising of merging systems. However, we note that the stacking technique is subject to potential issues associated with combining galaxies with different sizes and properties. Finally, \cite{Arribas2024} have identified a gradient of $\sim0.1$ dex $\mathrm{kpc}^{-1}$ across the whole structure of SPT0311-58-E, which they conclude is explained by accretion of pristine gas from the IGM. Overall, the flat gradients we measure in this study are therefore consistent with current observational picture of diverse metallicity gradients in the early Universe. 

While the flat strong-line gradients and irregular metallicity distributions we observe in our targets may be due to some merger activity (see Sec.~\ref{sec:otherscenarios}), our favoured outflow scenario would suggest that that the negative gradients suggested by the analysis of \citet{Li25metgrad} are not ubiquitous among isolated galaxies at this cosmic epoch, and that metal redistribution via feedback processes might be at work in low-mass galaxies at these early times. \cite{Garcia2025_2} also suggest that the flat gradients often measured at high-$z$ may arise due to feedback associated with bursty star formation, though this association is less clear at the low-mass end.

In contrast to the strong-line results, the negative direct-$T_\mathrm{e}$ gradient inferred for ID6355 may point to inside-out structure formation and rotational support playing some role in its metallicity profile. The tension between our strong-line and direct-$T_\mathrm{e}$ results also highlights some of the challenges currently facing high-redshift metallicity gradient analyses. For example, the tension may arise in part due to strong-line calibrations being inappropriate for studying metallicity in the inner region of ID6355, as AGN ionisation could artificially flatten the observed strong-line gradient. On the other hand, direct-$T_\mathrm{e}$ metallicity measurements assume a constant density applies to the aperture \citep[e.g.][]{Martinez2025}, which may not be a valid assumption in the case of clumpy high-$z$ galaxies. A detailed study of how such factors affect measured strong-line and direct-$T_\mathrm{e}$ metallicity gradients, and of how to resolve this tension, is beyond the scope of this paper.

\subsection{Comparison to Outflow Properties in Local Analogues and Simulations}\label{sec:impact}

In Sec.~\ref{sec:disentangle}, we argued that the broad component seen in the \OIIIall emission line of ID6355 arises due to an outflow; we also suggested a tentative outflow detection in ID10612. In this section, we now discuss the impact of the detected outflows on their host galaxies. 

The results from Sec.~\ref{sec:otherscenarios} and Sec.~\ref{sec:outflowcalc} are consistent with findings at Cosmic Noon and local observations, which have shown the evidence for AGN feedback is subtle and elusive. Indeed, \citet{Piotrowska22} and \citet{Bluck2022} have shown that the effect of AGN feedback is cumulative over many AGN episodes; hence, we do not expect to observe any strong evidence of AGN feedback on a short timescale \citep{Scholtz2020, Scholtz24, Lamperti21}, so the low mass-loading factors calculated for our targets do not reveal the overall impact of the outflows. Furthermore, the lower velocities of our outflows are consistent with the findings of \cite{Harrison16} and \cite{ ForsterSchrieber2019} of a strong correlation between outflow velocity and stellar mass, and also with the trend hinted at in Fig. \ref{fig:voutvsz} of outflow velocity decreasing toward earlier cosmic times in normal $M_* <10^9 \  M_\odot$ AGN-host and SF galaxies. This evidence points towards slower outflows in early low-mass galaxies, increasing the difficulty of conclusively identifying outflows at high-$z$. A systematic, large-sample study accounting for both $L_\mathrm{bol}$ and $M_*$ will be necessary to confirm this trend, but such an analysis is beyond the scope of this paper. Furthermore, both of our sources hosting red-shifted outflow components indicates that outflow morphology may be complex or different to expectations at in the early Universe, further complicating the matter of identifying outflows.

In Fig. \ref{fig:etavmstar}, we compare our detection of outflows in our high-$z$ targets to the best local analogues we could identify in the literature: local AGN-host dwarf galaxies at $z<1$ (from \citealt{Liu2020dwarfAGN} and \citealt{RM25dwarfagn}). In terms of low mass-loading factor, our targets are consistent with some of these local dwarf galaxies; however, we note that although these dwarf galaxies are considered to be `low-mass' \textit{locally}, most of them have a stellar mass almost 0.5 dex higher than those of our targets.  To properly evaluate how local dwarf galaxies compare to our high-redshift sample and investigate how AGN feedback evolves across cosmic time in low-mass systems, we would require a larger sample with lower-mass local AGN-host dwarf galaxies. We also find that our observations are consistent with other \jwst/NIRSpec samples from the literature, particularly those probing lower stellar mass \citep{Carniani24JADES}, which ultimately show low mass-loading factors and $v_\mathrm{out}/v_\mathrm{esc}\sim1$.

To assist with the further interpretation of our results, we make a comparison of our observations to the results of \textsc{Aesopica} simulations. \textsc{Aesopica} is a new suite of large-volume cosmological simulations (Koudmani et al., in prep.) built upon the \textsc{Fable} galaxy formation model \citep{Henden18}, with targeted updates for modelling the growth of infant SMBHs in the early Universe. \textsc{Aesopica} explores three key modifications to fiducial galaxy formation models: enabling efficient accretion in the low-mass regime \citep{Koudmani22}, incorporating super-Eddington accretion, and examining a broad range of seed masses ($10^{2}~\mathrm{M_\odot}$ to $10^{5}~\mathrm{M_\odot}$) following seed evolution from early cosmic epochs ($z \sim 20$). From the simulations, we select simulated AGN-host galaxies at $z\sim7.6$, i.e. the redshift of our sources. The simulated galaxies roughly match the dynamical and stellar mass of our targets. We utilise three different simulation modes: (i) fiducial accretion following \textsc{Fable} (`\textit{Fiducial}'), (ii) boosted accretion allowing efficient AGN activity in low-mass galaxies following \citet{Koudmani22} (`\textit{BoostAcc}'), and (iii) boosted accretion with super-Eddington accretion (up to ten times the Eddington limit; `\textit{SuperEdd+BoostAcc}'). 

In the right-hand panel of Fig. \ref{fig:etavmstar}, we note that outflows observed with \jwst have a wide range of measured mass loading factors scattered around $\eta \sim1$. The results from the \textsc{Aesopica} simulations on average have larger mass loading factors, but display a considerable scatter at a fixed stellar mass with some simulated galaxies in good agreement with our observational estimates. We however caution that estimating the mass outflow rate in observations includes major systematic uncertainties with regards to the assumed density and metallicity of the outflow, resulting in systematic uncertainties of up to 1 dex (as discussed in Sec.~\ref{sec:outflowcalc}). Furthermore, there are significant uncertainties in the modelling of high-redshift AGN in simulations and, crucially, in measuring the simulated outflow properties in an analogous manner to the observations \citep[for a recent detailed discussion on this topic, see][]{Martin-Alvarez2025}. Importantly, the mass-weighted outflow velocities calculated from simulations are significantly biased low compared to outflow velocities inferred from mock observations based on the same simulations. We therefore utilise the 95th percentile of the (mass-weighted) gas velocity distribution as a more reliable tracer of the broad-line based outflow velocity \citep[see discussion in Appendix B of][]{Martin-Alvarez2025}. Finally, we use Eq.~\ref{eq:outrate} to calculate the outflow rate, integrating over all outflowing mass between the observational resolution limit (conservatively $\sim 40~\mathrm{km \, s^{-1}}$) and the outflow velocity based on the 95th percentile. 

Although it is reassuring that the \textsc{Aesopica} simulations contain some systems that resemble our observations (see Fig.~\ref{fig:etavmstar}), we ultimately cannot make any firm conclusions on the nature of the observed outflows. For a true apple-to-apple comparison, we would need to produce full mock observations of the \textsc{Aesopica} outputs; however, this is beyond the scope of this exploratory work, and we defer utilising this approach until future studies with larger samples of similar outflow observations. We stress that accounting for observational effects would likely increase the scatter in the simulated outflow distributions. Furthermore, outflows with low mass loading factors and high outflow velocities are more likely to arise from simulations with a resolved multi-phase ISM, see e.g. \citet{Koudmani19, Steinwandel22}.

While we conclude the broad components in each galaxy correspond to outflows, we must also comment on what may be driving them. In simulations of dwarf galaxies without efficient AGN activity (our \textit{Fiducial} model), the high outflow velocities observed for our targets are quite rare (see left-hand panel of Fig. \ref{fig:etavmstar}). High outflow velocities are much more common in the \textsc{Aesopica} simulations with boosted AGN accretion efficiency (i.e. the \textit{BoostAcc} or \textit{SuperEdd+BoostAcc} models). This is in line with previous results from the \textsc{Fable} simulations \citep{Koudmani21} as well as high-resolution simulations of idealised galaxies \citep{Koudmani19}. Interestingly, we found from our analysis of the simulations that high outflow velocities are generally associated with overmassive BHs (see Sec.~\ref{sec:simchoice}); this is also a common feature in AGN observed with \jwst \citep[e.g.][]{maiolino_gnz11_2023, Ubler23, Juodzbalis25}. Combining with the tentative detection of an AGN in ID6355, we propose that this may be an example of an AGN-driven outflow operating in a low-mass galaxy in the early Universe. On the other hand, as we have not conclusively identified ID10612 as hosting an AGN, we cannot currently distinguish whether an outflow in this galaxy is driven by AGN or star formation.

Our analysis of these galaxies has demonstrated that searching for outflows at the Epoch of Reionisation is significantly more challenging than at lower redshifts. Low outflow velocities can result in outflows being confused with galactic rotation, merging galaxies or tidal interactions \citep{Scholtz25-COS3018, Concas22, Jones2025}, requiring deep, high spectral resolution IFS observation to disentangle the different kinematical components.

\section{Conclusions} \label{sec:conclusions}

In this work, we have presented a detailed analysis of new, high spectral resolution NIRSpec/IFS observations of two low-mass, potential AGN-host galaxies at $z\sim7.6$. Through our integrated and spatially-resolved investigations of the ISM properties and kinematics in our targets, we find that:

\begin{enumerate}
    \item Based on established AGN diagnostics based on the $\OIII\lambda4363$ auroral line, our observations of ID6355 are consistent with AGN ionisation, while ID10612 is consistent with either AGN or SF activity (see Fig.~\ref{fig:mazzodiag}). Combining with earlier results including the detection of high ionisation lines ($\NeIV\lambda\lambda2422,2424$; \citealt{Brinchmann23}) and full spectral fitting \citep{Silcock24}, we conclude ID6355 hosts a Type-II AGN, while ID10612 remains a tentative AGN-host candidate.

    \item Our results show marginal discrepancies in measured SFR and direct-$T_\mathrm{e}$ metallicity when compared to the results of earlier studies using the NIRSpec/MSA observations of these galaxies; we measure higher SFRs and lower metallicities overall, though almost all findings are consistent with earlier results within the associated uncertainties. The discrepancies can primarily be explained by difference in aperture size between the IFU and MSA observations.

    \item Taking advantage of spatially-resolved observations, we find evidence for flat strong-line metallicity gradients of $-0.01 \pm 0.01$ and $0.00 \pm 0.02\ \mathrm{dex} \ \mathrm{kpc}^{-1}$ in ID6355 and ID10612 respectiely (Fig. \ref{fig:gradmet}). When compared to the limited previous studies of metallicity gradients in galaxies at comparable redshift \citep{Arribas2024,Tripodi24, Venturi2024,Vallini, Li25metgrad},
    our results suggest that a redistribution of metals is occurring on a galactic scale in our targets. We note that these gradients are not radially symmetric, and that while maps of strong-line metallicity (see Fig. \ref{fig:metmap}) are not particularly reliable, they indicate the flat strong-line gradients may arise due to an irregular metallicity distribution in these galaxies.

    \item We identify a negative direct-$T_\mathrm{e}$ metallicity gradient of $-0.11\pm 0.03 \ \mathrm{dex} \ \mathrm{kpc}^{-1}$, in significant tension with the strong-line result. This tension could arise from problems with the strong-line calibrations or the constant-density assumptions of the direct-$T_\mathrm{e}$ calculation. As we cannot distinguish between these scenarios without further study on these gradients in larger sample of galaxies, we conclude that ID6355 exhibits a flat-to-negative gradient, consistent with scenarios ranging from inside-out structure formation to gas mixing on kpc scales within the galaxy.

    \item Both galaxies exhibit complex kinematics with multiple components (see Sec.~\ref{sec:kinemresults}). ID6355 exhibits evidence for some rotational support with high dispersion (Fig.~\ref{fig:6355_kinem}), while ID10612 is completely dominated by dispersive motions (Fig. \ref{fig:10612_kinem}). We identify $v_\mathrm{rot}/\sigma$ ratios of $0.63\pm0.36$ and $<0.36$ in the respective galaxies, consistent with the emerging theoretical and observational picture of decreasing rotational support towards early cosmic times (Fig. \ref{fig:vrotsigma}).  

    \item By combining morphological modelling with our kinematics maps, we estimate intrinsic rotational velocities and constrain the dynamical masses of our targets. We find $M_*/M_\mathrm{dyn}$ ratios of $0.013\pm0.010$ and $0.015\pm0.007$ in ID6355 and ID10612, respectively, indicating high gas and/or dark matter fractions, consistent with findings for other early low-mass galaxies \citep[e.g.][]{deGraaff2024}.

    \item We find strong evidence for an additional broad kinematical component in each galaxy, with $FWHM = 522\pm47$ and $275\pm12$ \ \kms \ in ID6355 and ID10612, respectively (Figs. \ref{fig:6355velmap} and \ref{fig:10612velmap}). We evaluate inflow, outflow, and merger scenarios as possible origins for this component, ultimately adopting the outflow interpretation in our analysis.

    \item Assuming the broad component arises from an outflow in each galaxy, we constrain the associated outflow properties. We identify ionised outflows with $v_\mathrm{out}$ of 512$^{+83}_{-84}$ and 255$_{-27}^{18}$ \kms in ID6355 and ID10612 respectively, comparable to that measured for other high-redshift outflows  (Fig. \ref{fig:voutvsz}), but considerably lower than those associated with AGN at lower redshifts. We estimate outflow radii of $\sim 1$ kpc in each galaxy, indicating the outflows are spatially extended. We estimate mass outflow rates of $14^{+18}_{-6}$ and $8^{+11}_{-3} \ M_\odot \ \mathrm{yr}^{-1}$ in ID6355 and ID10612, respectively. Both galaxies have mass loading factors $\eta < 1$ (see right panel of Fig. \ref{fig:etavmstar}), illustrating that at this point in cosmic time, the outflows cannot completely quench star formation in their galaxies. However, we additionally estimate the escape velocity of each galaxy based on estimated dynamical masses, finding that the measured outflows are capable of driving gas out of the galaxies into the CGM (see left panel of Fig. \ref{fig:etavmstar}) and could contribute to the suppression of star formation over longer timescales. We also show that the outflow properties of our systems and of high-$z$ AGN observed with \jwst are consistent within systematic uncertainties.

    \item To investigate the outflow origins, we compare the outflow properties of our targets with those from the new \textsc{Aesopica} simulations (Koudmani et al. in prep.) of $z \sim 7.6$ AGN-host dwarf galaxies in Sec.~\ref{sec:impact} and Fig. \ref{fig:etavmstar}. We find $v_\mathrm{out}/v_\mathrm{esc}$ ratios that are more commonly associated with the simulations including efficient AGN accretion at early cosmic times. We also find that our measured mass-loading factors can be reproduced by simulated systems of similar stellar and dynamical mass. These observations together hint that AGN activity could be the driver of the detected outflows. However, we caution that the majority of simulated galaxies have larger $\eta$ values, and stress that $\eta$ is a largely uncertain outflow parameter across both simulations and observations. Mock observations, when combined with a larger future observational sample, will be necessary to tackle the larger issue of cosmological simulations tending to find larger mass-loading factors than observations. Hence, we cannot conclusively state whether the measured outflows are driven by AGN or by star formation.
\end{enumerate}

Overall, this novel study demonstrates the importance of obtaining deep NIRSpec/IFS observations of low-mass galaxies at $z>7$, as such observations enable detailed, spatially-resolved analyses of the galaxy properties and kinematics in this challenging, poorly-understood demographic. The high spatial-resolution observations possible with NIRSpec/IFS are particularly crucial for disentangling any detected broad components from galactic rotation, and for evaluating the origin of these broad components. The galaxies discussed in this work represent just a small fraction of a wider high-$z$ population of low-mass galaxies within which AGN feedback may be operating. Future spatially-resolved studies based on deep NIRSpec/IFU observations, similar to that presented in this work, will therefore be crucial to our wider understanding of how outflows influence the kinematics, chemical enrichment, and overall evolution of low-mass galaxies at this epoch.

\section*{Acknowledgements}
We thank the anonymous referee for their thoughtful comments, which helped improve the quality of this paper. Double check affiliations
Reorder the acknowledgements. We thank Giovanni Mazzolari for providing the base code and data compilation for AGN diagnostic diagrams shown in this work.

LRI, JS, GCJ, FDE and RM acknowledge support by the Science and Technology Facilities Council (STFC), ERC Advanced Grant 695671 ``QUENCH" and the UKRI Frontier Research grant RISEandFALL. 
ALD thanks the University of Cambridge Harding Distinguished Postgraduate Scholars Programme and Technology Facilities Council (STFC) Center for Doctoral Training (CDT) in Data Intensive Science at the University of Cambridge (STFC grant number 2742605) for a PhD studentship, and gratefully acknowledges support by the Royal Society Research Grant G125142.
SK has been supported by a Junior Research Fellowship from St Catharine's College, Cambridge and a Research Fellowship from the Royal Commission for the Exhibition of 1851. RM also acknowledges funding from a research professorship from the Royal Society.
MC acknowledges support from ESO via the ESO Fellowship Europe.
ST acknowledges support by the Royal Society Research Grant G125142. 
WMB gratefully acknowledges support from DARK via the DARK fellowship. This work was supported by a research grant (VIL54489) from VILLUM FONDEN.
SA acknowledges grant PID2021-127718NB-I00 funded by the Spanish Ministry of Science and Innovation/State Agency of Research (MICIN/AEI/ 10.13039/501100011033).
DJE is supported as a Simons Investigator and by JWST/NIRCam contract to the University of Arizona, NAS5-02015 and by NASA through a grant from the Space Telescope Science Institute, which is operated by the Association of Universities for Research in Astronomy, Inc., under NASA contract NAS 5-03127.
ZJ acknowledges JWST/NIRCam contract to the University of Arizona, NAS5-02015.
MK thanks the University of Cambridge Harding Distinguished Postgraduate Scholars Programme, UK STFC CDT in Data Intensive Science, and Girton College Cambridge for a PhD studentship.
MP acknowledges support through the grants PID2021-127718NB-I00 and RYC2023-044853-I, funded by the Spain Ministry of Science and Innovation/State Agency of Research MCIN/AEI/10.13039/501100011033 and El Fondo Social Europeo Plus FSE+.
DP acknowledges support by the Huo Family Foundation through a P.C. Ho PhD Studentship.
BER acknowledges support from the NIRCam Science Team contract to the University of Arizona, NAS5-02015, and JWST Program 3215.
DS acknowledges support from the STFC, grant code ST/W000997/1.
JAAT acknowledges support from the Simons Foundation and JWST program 3215. Support for program 3215 was provided by NASA through a grant from the Space Telescope Science Institute, which is operated by the Association of Universities for Research in Astronomy, Inc., under NASA contract NAS 5-03127.
CW acknowledges support from the Swiss State Secretariat for Education, Research and Innovation (SERI) under contract number MB22.00072.

%%%%%%%%%%%%%%%%%%%%%%%%%%%%%%%%%%%%%%%%%%%%%%%%%%
\section*{Data Availability}
The datasets were derived from sources in the public domain. JWST/NIRSpec and NIRCam data can be downloaded from the MAST (\url{https://mast.stsci.edu/portal/Mashup/Clients/Mast/Portal.html}) under PIDs 2959 and 2736, respectively.



\bibliographystyle{mnras}
\bibliography{biblio} % if your bibtex file is called example.bib

@ARTICLE{McClymont25_burst,
       author = {{McClymont}, William and {Tacchella}, Sandro and {Smith}, Aaron and {Kannan}, Rahul and {Puchwein}, Ewald and {Borrow}, Josh and {Garaldi}, Enrico and {Keating}, Laura and {Vogelsberger}, Mark and {Zier}, Oliver and {Shen}, Xuejian and {Popovic}, Filip and {Simmonds}, Charlotte},
        title = "{The THESAN-ZOOM project: burst, quench, repeat {\textendash} unveiling the evolution of high-redshift galaxies along the star-forming main sequence}",
      journal = {\mnras},
         year = 2025,
        month = nov,
       volume = {544},
       number = {1},
        pages = {513-534},
          doi = {10.1093/mnras/staf1660},
archivePrefix = {arXiv},
       eprint = {2503.00106},
 primaryClass = {astro-ph.GA},
       adsurl = {https://ui.adsabs.harvard.edu/abs/2025MNRAS.544..513M}
}

@ARTICLE{Tripodi24,
       author = {{Tripodi}, Roberta and {D'Eugenio}, Francesco and {Maiolino}, Roberto and {Curti}, Mirko and {Scholtz}, Jan and {Tacchella}, Sandro and {Marconcini}, Cosimo and {Bunker}, Andrew J. and {Trussler}, James A.~A. and {Cameron}, Alex J. and {Arribas}, Santiago and {Baker}, William M. and {Brada{\v{c}}}, Maru{\v{s}}a and {Carniani}, Stefano and {Charlot}, St{\'e}phane and {Ji}, Xihan and {Ji}, Zhiyuan and {Robertson}, Brant and {{\"U}bler}, Hannah and {Venturi}, Giacomo and {Willmer}, Christopher N.~A. and {Witstok}, Joris},
        title = "{Spatially resolved emission lines in galaxies at 4 {\ensuremath{\leq}} z < 10 from the JADES survey: Evidence for enhanced central star formation}",
      journal = {\aap},
         year = 2024,
        month = dec,
       volume = {692},
          eid = {A184},
        pages = {A184},
          doi = {10.1051/0004-6361/202449980},
archivePrefix = {arXiv},
       eprint = {2403.08431},
 primaryClass = {astro-ph.GA},
       adsurl = {https://ui.adsabs.harvard.edu/abs/2024A&A...692A.184T}
}

@ARTICLE{Donnan25,
       author = {{Donnan}, Callum T. and {Dickinson}, Mark and {Taylor}, Anthony J. and {Arrabal Haro}, Pablo and {Finkelstein}, Steven L. and {Stanton}, Thomas M. and {Jung}, Intae and {Papovich}, Casey and {Akins}, Hollis B. and {Koekemoer}, Anton M. and {McLeod}, Derek J. and {Napolitano}, Lorenzo and {Amor{\'\i}n}, Ricardo O. and {Begley}, Ryan and {Burgarella}, Denis and {Carnall}, Adam C. and {Casey}, Caitlin M. and {Calabr{\`o}}, Antonello and {Cullen}, Fergus and {Dunlop}, James S. and {Ellis}, Richard S. and {Fern{\'a}ndez}, Vital and {Giavalisco}, Mauro and {Hirschmann}, Michaela and {Hu}, Weida and {Illingworth}, Garth and {Kartaltepe}, Jeyhan S. and {Kocevski}, Dale D. and {Kokorev}, Vasily and {Leung}, Ho-Hin and {Lucas}, Ray A. and {Morales}, Alexa M. and {McLure}, Ross and {Pentericci}, Laura and {P{\'e}rez-Gonz{\'a}lez}, Pablo G. and {Somerville}, Rachel S. and {Stevenson}, Struan and {Trump}, Jonathan R. and {Yung}, L.~Y. Aaron and {Zavala}, Jorge A.},
        title = "{Very bright, very blue, and very red: JWST CAPERS analysis of highly luminous galaxies with extreme UV slopes at $\mathbf{z = 10}$}",
      journal = {arXiv e-prints},
         year = 2025,
        month = jul,
          eid = {arXiv:2507.10518},
        pages = {arXiv:2507.10518},
          doi = {10.48550/arXiv.2507.10518},
archivePrefix = {arXiv},
       eprint = {2507.10518},
 primaryClass = {astro-ph.GA},
       adsurl = {https://ui.adsabs.harvard.edu/abs/2025arXiv250710518D}
}

@ARTICLE{Naidu_25_z14,
       author = {{Naidu}, Rohan P. and {Oesch}, Pascal A. and {Brammer}, Gabriel and {Weibel}, Andrea and {Li}, Yijia and {Matthee}, Jorryt and {Chisholm}, John and {Pollock}, Clara L. and {Heintz}, Kasper E. and {Johnson}, Benjamin D. and {Shen}, Xuejian and {Hviding}, Raphael E. and {Leja}, Joel and {Tacchella}, Sandro and {Ganguly}, Arpita and {Witten}, Callum and {Atek}, Hakim and {Belli}, Sirio and {Bose}, Sownak and {Bouwens}, Rychard and {Dayal}, Pratika and {Decarli}, Roberto and {de Graaff}, Anna and {Fudamoto}, Yoshinobu and {Giovinazzo}, Emma and {Greene}, Jenny E. and {Illingworth}, Garth and {Inoue}, Akio K. and {Kane}, Sarah G. and {Labbe}, Ivo and {Leonova}, Ecaterina and {Marques-Chaves}, Rui and {Meyer}, Romain A. and {Nelson}, Erica J. and {Roberts-Borsani}, Guido and {Schaerer}, Daniel and {Simcoe}, Robert A. and {Stefanon}, Mauro and {Sugahara}, Yuma and {Toft}, Sune and {van der Wel}, Arjen and {van Dokkum}, Pieter and {Walter}, Fabian and {Watson}, Darach and {Weaver}, John R. and {Whitaker}, Katherine E.},
        title = "{A Cosmic Miracle: A Remarkably Luminous Galaxy at $z_{\rm{spec}}=14.44$ Confirmed with JWST}",
      journal = {arXiv e-prints},
         year = 2025,
        month = may,
          eid = {arXiv:2505.11263},
        pages = {arXiv:2505.11263},
          doi = {10.48550/arXiv.2505.11263},
archivePrefix = {arXiv},
       eprint = {2505.11263},
 primaryClass = {astro-ph.GA},
       adsurl = {https://ui.adsabs.harvard.edu/abs/2025arXiv250511263N}
}

@ARTICLE{Koudmani19,
       author = {{Koudmani}, Sophie and {Sijacki}, Debora and {Bourne}, Martin A. and {Smith}, Matthew C.},
        title = "{Fast and energetic AGN-driven outflows in simulated dwarf galaxies}",
      journal = {\mnras},
         year = 2019,
        month = apr,
       volume = {484},
       number = {2},
        pages = {2047-2066},
          doi = {10.1093/mnras/stz097},
archivePrefix = {arXiv},
       eprint = {1812.04629},
 primaryClass = {astro-ph.GA},
       adsurl = {https://ui.adsabs.harvard.edu/abs/2019MNRAS.484.2047K}
}

@ARTICLE{Martin-Alvarez2025,
       author = {{Martin-Alvarez}, Sergio and {Sijacki}, Debora and {Haehnelt}, Martin G. and {Concas}, Alice and {Yuan}, Yuxuan and {Maiolino}, Roberto and {Wechsler}, Risa H. and {Rodr{\'\i}guez Montero}, Francisco and {Farcy}, Marion and {Sanati}, Mahsa and {Dubois}, Yohan and {Rosdahl}, Joki and {Lopez-Rodriguez}, Enrique and {Clark}, Susan E.},
        title = "{The Pandora project. II: how non-thermal physics drives bursty star formation and temperate mass-loaded outflows in dwarf galaxies}",
      journal = {arXiv e-prints},
         year = 2025,
        month = jun,
          eid = {arXiv:2506.03245},
        pages = {arXiv:2506.03245},
          doi = {10.48550/arXiv.2506.03245},
archivePrefix = {arXiv},
       eprint = {2506.03245},
 primaryClass = {astro-ph.GA},
       adsurl = {https://ui.adsabs.harvard.edu/abs/2025arXiv250603245M}
}

@ARTICLE{Mazzolari24,
       author = {{Mazzolari}, Giovanni and {{\"U}bler}, Hannah and {Maiolino}, Roberto and {Ji}, Xihan and {Nakajima}, Kimihiko and {Feltre}, Anna and {Scholtz}, Jan and {D'Eugenio}, Francesco and {Curti}, Mirko and {Mignoli}, Marco and {Marconi}, Alessandro},
        title = "{New AGN diagnostic diagrams based on the [OIII]{\ensuremath{\lambda}}4363 auroral line}",
      journal = {\aap},
         year = 2024,
        month = nov,
       volume = {691},
          eid = {A345},
        pages = {A345},
          doi = {10.1051/0004-6361/202450407},
archivePrefix = {arXiv},
       eprint = {2404.10811},
 primaryClass = {astro-ph.GA},
       adsurl = {https://ui.adsabs.harvard.edu/abs/2024A&A...691A.345M}
}

@ARTICLE{Sanders15,
       author = {{Sanders}, Ryan L. and {Shapley}, Alice E. and {Kriek}, Mariska and {Reddy}, Naveen A. and {Freeman}, William R. and {Coil}, Alison L. and {Siana}, Brian and {Mobasher}, Bahram and {Shivaei}, Irene and {Price}, Sedona H. and {de Groot}, Laura},
        title = "{The MOSDEF Survey: Mass, Metallicity, and Star-formation Rate at z \raisebox{-0.5ex}\textasciitilde 2.3}",
      journal = {\apj},
         year = 2015,
        month = feb,
       volume = {799},
       number = {2},
          eid = {138},
        pages = {138},
          doi = {10.1088/0004-637X/799/2/138},
archivePrefix = {arXiv},
       eprint = {1408.2521},
 primaryClass = {astro-ph.GA},
       adsurl = {https://ui.adsabs.harvard.edu/abs/2015ApJ...799..138S}
}

@ARTICLE{Dimitrijevic2007,
       author = {{Dimitrijevi{\'c}}, M.~S. and {Popovi{\'c}}, L. {\v{C}}. and {Kova{\v{c}}evi{\'c}}, J. and {Da{\v{c}}i{\'c}}, M. and {Ili{\'c}}, D.},
        title = "{The flux ratio of the [OIII] {\ensuremath{\lambda}}{\ensuremath{\lambda}}5007, 4959 lines in AGN: comparison with theoretical calculations}",
      journal = {\mnras},
         year = 2007,
        month = jan,
       volume = {374},
       number = {3},
        pages = {1181-1184},
          doi = {10.1111/j.1365-2966.2006.11238.x},
archivePrefix = {arXiv},
       eprint = {astro-ph/0610848},
 primaryClass = {astro-ph},
       adsurl = {https://ui.adsabs.harvard.edu/abs/2007MNRAS.374.1181D}
}

@ARTICLE{Carniani24JADES,
       author = {{Carniani}, Stefano and {Venturi}, Giacomo and {Parlanti}, Eleonora and {de Graaff}, Anna and {Maiolino}, Roberto and {Arribas}, Santiago and {Bonaventura}, Nina and {Boyett}, Kristan and {Bunker}, Andrew J. and {Cameron}, Alex J. and {Charlot}, Stephane and {Chevallard}, Jacopo and {Curti}, Mirko and {Curtis-Lake}, Emma and {Eisenstein}, Daniel J. and {Giardino}, Giovanna and {Hausen}, Ryan and {Kumari}, Nimisha and {Maseda}, Michael V. and {Nelson}, Erica and {Perna}, Michele and {Rix}, Hans-Walter and {Robertson}, Brant and {Del Pino}, Bruno Rodr{\'\i}guez and {Sandles}, Lester and {Scholtz}, Jan and {Simmonds}, Charlotte and {Smit}, Renske and {Tacchella}, Sandro and {{\"U}bler}, Hannah and {Williams}, Christina C. and {Willott}, Chris and {Witstok}, Joris},
        title = "{JADES: The incidence rate and properties of galactic outflows in low-mass galaxies across 3 < z < 9}",
      journal = {\aap},
         year = 2024,
        month = may,
       volume = {685},
          eid = {A99},
        pages = {A99},
          doi = {10.1051/0004-6361/202347230},
archivePrefix = {arXiv},
       eprint = {2306.11801},
 primaryClass = {astro-ph.GA},
       adsurl = {https://ui.adsabs.harvard.edu/abs/2024A&A...685A..99C}
}

@ARTICLE{Tacchella23,
       author = {{Tacchella}, Sandro and {Johnson}, Benjamin D. and {Robertson}, Brant E. and {Carniani}, Stefano and {D'Eugenio}, Francesco and {Kumari}, Nimisha and {Maiolino}, Roberto and {Nelson}, Erica J. and {Suess}, Katherine A. and {{\"U}bler}, Hannah and {Williams}, Christina C. and {Adebusola}, Alabi and {Alberts}, Stacey and {Arribas}, Santiago and {Bhatawdekar}, Rachana and {Bonaventura}, Nina and {Bowler}, Rebecca A.~A. and {Bunker}, Andrew J. and {Cameron}, Alex J. and {Curti}, Mirko and {Egami}, Eiichi and {Eisenstein}, Daniel J. and {Frye}, Brenda and {Hainline}, Kevin and {Helton}, Jakob M. and {Ji}, Zhiyuan and {Looser}, Tobias J. and {Lyu}, Jianwei and {Perna}, Michele and {Rawle}, Timothy and {Rieke}, George and {Rieke}, Marcia and {Saxena}, Aayush and {Sandles}, Lester and {Shivaei}, Irene and {Simmonds}, Charlotte and {Sun}, Fengwu and {Willmer}, Christopher N.~A. and {Willott}, Chris J. and {Witstok}, Joris},
        title = "{JWST NIRCam + NIRSpec: interstellar medium and stellar populations of young galaxies with rising star formation and evolving gas reservoirs}",
      journal = {\mnras},
         year = 2023,
        month = jul,
       volume = {522},
       number = {4},
        pages = {6236-6249},
          doi = {10.1093/mnras/stad1408},
archivePrefix = {arXiv},
       eprint = {2208.03281},
 primaryClass = {astro-ph.GA},
       adsurl = {https://ui.adsabs.harvard.edu/abs/2023MNRAS.522.6236T}
}

@ARTICLE{Mahler2023,
       author = {{Mahler}, Guillaume and {Jauzac}, Mathilde and {Richard}, Johan and {Beauchesne}, Benjamin and {Ebeling}, Harald and {Lagattuta}, David and {Natarajan}, Priyamvada and {Sharon}, Keren and {Atek}, Hakim and {Claeyssens}, Ad{\'e}la{\"\i}de and {Cl{\'e}ment}, Benjamin and {Eckert}, Dominique and {Edge}, Alastair and {Kneib}, Jean-Paul and {Niemiec}, Anna},
        title = "{Precision Modeling of JWST's First Cluster Lens SMACS J0723.3-7327}",
      journal = {\apj},
         year = 2023,
        month = mar,
       volume = {945},
       number = {1},
          eid = {49},
        pages = {49},
          doi = {10.3847/1538-4357/acaea9},
archivePrefix = {arXiv},
       eprint = {2207.07101},
 primaryClass = {astro-ph.GA},
       adsurl = {https://ui.adsabs.harvard.edu/abs/2023ApJ...945...49M}
}

@ARTICLE{Calzetti2000,
       author = {{Calzetti}, Daniela and {Armus}, Lee and {Bohlin}, Ralph C. and {Kinney}, Anne L. and {Koornneef}, Jan and {Storchi-Bergmann}, Thaisa},
        title = "{The Dust Content and Opacity of Actively Star-forming Galaxies}",
      journal = {\apj},
         year = 2000,
        month = apr,
       volume = {533},
       number = {2},
        pages = {682-695},
          doi = {10.1086/308692},
archivePrefix = {arXiv},
       eprint = {astro-ph/9911459},
 primaryClass = {astro-ph},
       adsurl = {https://ui.adsabs.harvard.edu/abs/2000ApJ...533..682C}
}

@ARTICLE{Rupke2005,
       author = {{Rupke}, David S. and {Veilleux}, Sylvain and {Sanders}, D.~B.},
        title = "{Outflows in Infrared-Luminous Starbursts at z < 0.5. II. Analysis and Discussion}",
      journal = {\apjs},
         year = 2005,
        month = sep,
       volume = {160},
       number = {1},
        pages = {115-148},
          doi = {10.1086/432889},
archivePrefix = {arXiv},
       eprint = {astro-ph/0506611},
 primaryClass = {astro-ph},
       adsurl = {https://ui.adsabs.harvard.edu/abs/2005ApJS..160..115R}
}

@ARTICLE{Carniani2015,
       author = {{Carniani}, S. and {Marconi}, A. and {Maiolino}, R. and {Balmaverde}, B. and {Brusa}, M. and {Cano-D{\'\i}az}, M. and {Cicone}, C. and {Comastri}, A. and {Cresci}, G. and {Fiore}, F. and {Feruglio}, C. and {La Franca}, F. and {Mainieri}, V. and {Mannucci}, F. and {Nagao}, T. and {Netzer}, H. and {Piconcelli}, E. and {Risaliti}, G. and {Schneider}, R. and {Shemmer}, O.},
        title = "{Ionised outflows in z \raisebox{-0.5ex}\textasciitilde 2.4 quasar host galaxies}",
      journal = {\aap},
         year = 2015,
        month = aug,
       volume = {580},
          eid = {A102},
        pages = {A102},
          doi = {10.1051/0004-6361/201526557},
archivePrefix = {arXiv},
       eprint = {1506.03096},
 primaryClass = {astro-ph.GA},
       adsurl = {https://ui.adsabs.harvard.edu/abs/2015A&A...580A.102C}
}

@ARTICLE{Fiore2017,
       author = {{Fiore}, F. and {Feruglio}, C. and {Shankar}, F. and {Bischetti}, M. and {Bongiorno}, A. and {Brusa}, M. and {Carniani}, S. and {Cicone}, C. and {Duras}, F. and {Lamastra}, A. and {Mainieri}, V. and {Marconi}, A. and {Menci}, N. and {Maiolino}, R. and {Piconcelli}, E. and {Vietri}, G. and {Zappacosta}, L.},
        title = "{AGN wind scaling relations and the co-evolution of black holes and galaxies}",
      journal = {\aap},
         year = 2017,
        month = may,
       volume = {601},
          eid = {A143},
        pages = {A143},
          doi = {10.1051/0004-6361/201629478},
archivePrefix = {arXiv},
       eprint = {1702.04507},
 primaryClass = {astro-ph.GA},
       adsurl = {https://ui.adsabs.harvard.edu/abs/2017A&A...601A.143F}
}

@ARTICLE{Maiolino2012,
       author = {{Maiolino}, R. and {Gallerani}, S. and {Neri}, R. and {Cicone}, C. and {Ferrara}, A. and {Genzel}, R. and {Lutz}, D. and {Sturm}, E. and {Tacconi}, L.~J. and {Walter}, F. and {Feruglio}, C. and {Fiore}, F. and {Piconcelli}, E.},
        title = "{Evidence of strong quasar feedback in the early Universe}",
      journal = {\mnras},
         year = 2012,
        month = sep,
       volume = {425},
       number = {1},
        pages = {L66-L70},
          doi = {10.1111/j.1745-3933.2012.01303.x},
archivePrefix = {arXiv},
       eprint = {1204.2904},
 primaryClass = {astro-ph.CO},
       adsurl = {https://ui.adsabs.harvard.edu/abs/2012MNRAS.425L..66M}
}

@ARTICLE{GonzalezAlfonso17,
       author = {{Gonz{\'a}lez-Alfonso}, E. and {Fischer}, J. and {Spoon}, H.~W.~W. and {Stewart}, K.~P. and {Ashby}, M.~L.~N. and {Veilleux}, S. and {Smith}, H.~A. and {Sturm}, E. and {Farrah}, D. and {Falstad}, N. and {Mel{\'e}ndez}, M. and {Graci{\'a}-Carpio}, J. and {Janssen}, A.~W. and {Lebouteiller}, V.},
        title = "{Molecular Outflows in Local ULIRGs: Energetics from Multitransition OH Analysis}",
      journal = {\apj},
         year = 2017,
        month = feb,
       volume = {836},
       number = {1},
          eid = {11},
        pages = {11},
          doi = {10.3847/1538-4357/836/1/11},
archivePrefix = {arXiv},
       eprint = {1612.08181},
 primaryClass = {astro-ph.GA},
       adsurl = {https://ui.adsabs.harvard.edu/abs/2017ApJ...836...11G}
}

@ARTICLE{Tully98,
       author = {{Tully}, R. Brent and {Pierce}, Michael J. and {Huang}, Jia-Sheng and {Saunders}, Will and {Verheijen}, Marc A.~W. and {Witchalls}, Peter L.},
        title = "{Global Extinction in Spiral Galaxies}",
      journal = {\aj},
         year = 1998,
        month = jun,
       volume = {115},
       number = {6},
        pages = {2264-2272},
          doi = {10.1086/300379},
archivePrefix = {arXiv},
       eprint = {astro-ph/9802247},
 primaryClass = {astro-ph},
       adsurl = {https://ui.adsabs.harvard.edu/abs/1998AJ....115.2264T}
}

@ARTICLE{Jones24,
       author = {{Jones}, Gareth C. and {Bowler}, Rebecca and {Bunker}, Andrew J. and {Arribas}, Santiago and {Carniani}, Stefano and {Charlot}, Stephane and {Perna}, Michele and {Rodr{\'\i}guez Del Pino}, Bruno and {{\"U}bler}, Hannah and {Willott}, Chris J. and {Chevallard}, Jacopo and {Cresci}, Giovanni and {Parlanti}, Eleonora and {Scholtz}, Jan and {Venturi}, Giacomo},
        title = "{GA-NIFS: interstellar medium properties and tidal interactions in the evolved massive merging system B14-65666 at z = 7.152}",
      journal = {arXiv e-prints},
         year = 2024,
        month = dec,
          eid = {arXiv:2412.15027},
        pages = {arXiv:2412.15027},
          doi = {10.48550/arXiv.2412.15027},
archivePrefix = {arXiv},
       eprint = {2412.15027},
 primaryClass = {astro-ph.GA},
       adsurl = {https://ui.adsabs.harvard.edu/abs/2024arXiv241215027J}
}

@ARTICLE{Pysersic23,
       author = {{Pasha}, Imad and {Miller}, Tim B.},
        title = "{pysersic: A Python package for determining galaxy structural properties via Bayesian inference, accelerated with jax}",
      journal = {The Journal of Open Source Software},
         year = 2023,
        month = sep,
       volume = {8},
       number = {89},
          eid = {5703},
        pages = {5703},
          doi = {10.21105/joss.05703},
archivePrefix = {arXiv},
       eprint = {2306.05454},
 primaryClass = {astro-ph.GA},
       adsurl = {https://ui.adsabs.harvard.edu/abs/2023JOSS....8.5703P}
}

@ARTICLE{ForsterSchreiberWuyts20,
       author = {{F{\"o}rster Schreiber}, Natascha M. and {Wuyts}, Stijn},
        title = "{Star-Forming Galaxies at Cosmic Noon}",
      journal = {\araa},
         year = 2020,
        month = aug,
       volume = {58},
        pages = {661-725},
          doi = {10.1146/annurev-astro-032620-021910},
archivePrefix = {arXiv},
       eprint = {2010.10171},
 primaryClass = {astro-ph.GA},
       adsurl = {https://ui.adsabs.harvard.edu/abs/2020ARA&A..58..661F}
}

@ARTICLE{Price20,
       author = {{Price}, Sedona H. and {Kriek}, Mariska and {Barro}, Guillermo and {Shapley}, Alice E. and {Reddy}, Naveen A. and {Freeman}, William R. and {Coil}, Alison L. and {Shivaei}, Irene and {Azadi}, Mojegan and {de Groot}, Laura and {Siana}, Brian and {Mobasher}, Bahram and {Sanders}, Ryan L. and {Leung}, Gene C.~K. and {Fetherolf}, Tara and {Zick}, Tom O. and {{\"U}bler}, Hannah and {F{\"o}rster Schreiber}, Natascha M.},
        title = "{The MOSDEF Survey: Kinematic and Structural Evolution of Star-forming Galaxies at 1.4 {\ensuremath{\leq}} z {\ensuremath{\leq}} 3.8}",
      journal = {\apj},
         year = 2020,
        month = may,
       volume = {894},
       number = {2},
          eid = {91},
        pages = {91},
          doi = {10.3847/1538-4357/ab7990},
archivePrefix = {arXiv},
       eprint = {1902.09554},
 primaryClass = {astro-ph.GA},
       adsurl = {https://ui.adsabs.harvard.edu/abs/2020ApJ...894...91P}
}

@ARTICLE{derWel14,
       author = {{van der Wel}, A. and {Chang}, Yu-Yen and {Bell}, E.~F. and {Holden}, B.~P. and {Ferguson}, H.~C. and {Giavalisco}, M. and {Rix}, H. -W. and {Skelton}, R. and {Whitaker}, K. and {Momcheva}, I. and {Brammer}, G. and {Kassin}, S.~A. and {Martig}, M. and {Dekel}, A. and {Ceverino}, D. and {Koo}, D.~C. and {Mozena}, M. and {van Dokkum}, P.~G. and {Franx}, M. and {Faber}, S.~M. and {Primack}, J.},
        title = "{Geometry of Star-forming Galaxies from SDSS, 3D-HST, and CANDELS}",
      journal = {\apjl},
         year = 2014,
        month = sep,
       volume = {792},
       number = {1},
          eid = {L6},
        pages = {L6},
          doi = {10.1088/2041-8205/792/1/L6},
archivePrefix = {arXiv},
       eprint = {1407.4233},
 primaryClass = {astro-ph.GA},
       adsurl = {https://ui.adsabs.harvard.edu/abs/2014ApJ...792L...6V}
}

@ARTICLE{Lola2025,
       author = {{Danhaive}, A. Lola and {Tacchella}, Sandro and {{\"U}bler}, Hannah and {de Graaff}, Anna and {Egami}, Eiichi and {Johnson}, Benjamin D. and {Sun}, Fengwu and {Arribas}, Santiago and {Bunker}, Andrew J. and {Carniani}, Stefano and {Jones}, Gareth C. and {Maiolino}, Roberto and {McClymont}, William and {Parlanti}, Eleonora and {Simmonds}, Charlotte and {Villanueva}, Natalia C. and {Baker}, William M. and {Jaffe}, Daniel T. and {Eisenstein}, Daniel and {Hainline}, Kevin and {Helton}, Jakob M. and {Ji}, Zhiyuan and {Lin}, Xiaojing and {Liu}, Yichen and {Pusk{\'a}s}, D{\'a}vid and {Rieke}, Marcia and {Rinaldi}, Pierluigi and {Robertson}, Brant and {Scholz}, Jan and {Williams}, Christina C. and {Willmer}, Christopher N.~A.},
        title = "{The dawn of discs: unveiling the turbulent ionized gas kinematics of the galaxy population at z {\ensuremath{\sim}} 4{\textendash}6 with JWST/NIRCam grism spectroscopy}",
      journal = {\mnras},
         year = 2025,
        month = nov,
       volume = {543},
       number = {4},
        pages = {3249-3302},
          doi = {10.1093/mnras/staf1540},
archivePrefix = {arXiv},
       eprint = {2503.21863},
 primaryClass = {astro-ph.GA},
       adsurl = {https://ui.adsabs.harvard.edu/abs/2025MNRAS.543.3249D}
}

@ARTICLE{Curti2023,
       author = {{Curti}, Mirko and {D'Eugenio}, Francesco and {Carniani}, Stefano and {Maiolino}, Roberto and {Sandles}, Lester and {Witstok}, Joris and {Baker}, William M. and {Bennett}, Jake S. and {Piotrowska}, Joanna M. and {Tacchella}, Sandro and {Charlot}, Stephane and {Nakajima}, Kimihiko and {Maheson}, Gabriel and {Mannucci}, Filippo and {Amiri}, Amirnezam and {Arribas}, Santiago and {Belfiore}, Francesco and {Bonaventura}, Nina R. and {Bunker}, Andrew J. and {Chevallard}, Jacopo and {Cresci}, Giovanni and {Curtis-Lake}, Emma and {Hayden-Pawson}, Connor and {Jones}, Gareth C. and {Kumari}, Nimisha and {Laseter}, Isaac and {Looser}, Tobias J. and {Marconi}, Alessandro and {Maseda}, Michael V. and {Scholtz}, Jan and {Smit}, Renske and {{\"U}bler}, Hannah and {Wallace}, Imaan E.~B.},
        title = "{The chemical enrichment in the early Universe as probed by JWST via direct metallicity measurements at z {\ensuremath{\sim}} 8}",
      journal = {\mnras},
         year = 2023,
        month = jan,
       volume = {518},
       number = {1},
        pages = {425-438},
          doi = {10.1093/mnras/stac2737},
archivePrefix = {arXiv},
       eprint = {2207.12375},
 primaryClass = {astro-ph.GA},
       adsurl = {https://ui.adsabs.harvard.edu/abs/2023MNRAS.518..425C}
}

@ARTICLE{Brinchmann23,
       author = {{Brinchmann}, Jarle},
        title = "{High-z galaxies with JWST and local analogues - it is not only star formation}",
      journal = {\mnras},
         year = 2023,
        month = oct,
       volume = {525},
       number = {2},
        pages = {2087-2106},
          doi = {10.1093/mnras/stad1704},
archivePrefix = {arXiv},
       eprint = {2208.07467},
 primaryClass = {astro-ph.GA},
       adsurl = {https://ui.adsabs.harvard.edu/abs/2023MNRAS.525.2087B}
}

@ARTICLE{Silcock24,
       author = {{Silcock}, M.~S. and {Curtis-Lake}, E. and {Smith}, D.~J.~B. and {Wallace}, I.~E.~B. and {Vidal-Garc{\'\i}a}, A. and {Plat}, A. and {Hirschmann}, M. and {Feltre}, A. and {Chevallard}, J. and {Charlot}, S. and {Carniani}, S. and {Bunker}, A.~J.},
        title = "{Characterising the z {\ensuremath{\approx}} 7.66 Type-II AGN candidate SMACS S06355 using BEAGLE-AGN and JWST NIRSpec/NIRCam}",
      journal = {\mnras},
       volume = 541,
	   number = 4,
	   pages = {3822--3836},
         year = 2025,
        month = aug,
          doi = {10.1093/mnras/staf1087},
archivePrefix = {arXiv},
       eprint = {2410.18193},
 primaryClass = {astro-ph.GA},
       adsurl = {https://ui.adsabs.harvard.edu/abs/2025MNRAS.tmp.1082S}
}

@article{Koudmani21,
	title = {{A little FABLE: exploring AGN feedback in dwarf galaxies with cosmological simulations}},
	author = {{Koudmani}, Sophie and {Henden}, Nicholas A. and {Sijacki}, Debora},
	year = 2021,
	month = may,
	journal = {\mnras},
	volume = 503,
	number = 3,
	pages = {3568--3591},
	doi = {10.1093/mnras/stab677},
	archiveprefix = {arXiv},
	eprint = {2007.10342},
	primaryclass = {astro-ph.GA},
	adsurl = {https://ui.adsabs.harvard.edu/abs/2021MNRAS.503.3568K}
}

@ARTICLE{Trump2023,
       author = {{Trump}, Jonathan R. and {Arrabal Haro}, Pablo and {Simons}, Raymond C. and {Backhaus}, Bren E. and {Amor{\'\i}n}, Ricardo O. and {Dickinson}, Mark and {Fern{\'a}ndez}, Vital and {Papovich}, Casey and {Nicholls}, David C. and {Kewley}, Lisa J. and {Brunker}, Samantha W. and {Salzer}, John J. and {Wilkins}, Stephen M. and {Almaini}, Omar and {Bagley}, Micaela B. and {Berg}, Danielle A. and {Bhatawdekar}, Rachana and {Bisigello}, Laura and {Buat}, V{\'e}ronique and {Burgarella}, Denis and {Calabr{\`o}}, Antonello and {Casey}, Caitlin M. and {Ciesla}, Laure and {Cleri}, Nikko J. and {Cole}, Justin W. and {Cooper}, M.~C. and {Cooray}, Asantha R. and {Costantin}, Luca and {Croton}, Darren and {Ferguson}, Henry C. and {Finkelstein}, Steven L. and {Fujimoto}, Seiji and {Gardner}, Jonathan P. and {Gawiser}, Eric and {Giavalisco}, Mauro and {Grazian}, Andrea and {Grogin}, Norman A. and {Hathi}, Nimish P. and {Hirschmann}, Michaela and {Holwerda}, Benne W. and {Huertas-Company}, Marc and {Hutchison}, Taylor A. and {Jogee}, Shardha and {Juneau}, St{\'e}phanie and {Jung}, Intae and {Kartaltepe}, Jeyhan S. and {Kirkpatrick}, Allison and {Kocevski}, Dale D. and {Koekemoer}, Anton M. and {Lotz}, Jennifer M. and {Lucas}, Ray A. and {Magnelli}, Benjamin and {Matharu}, Jasleen and {P{\'e}rez-Gonz{\'a}lez}, Pablo G. and {Pirzkal}, Nor and {Rafelski}, Marc and {Rose}, Caitlin and {Seill{\'e}}, Lise-Marie and {Somerville}, Rachel S. and {Straughn}, Amber N. and {Tacchella}, Sandro and {Vanderhoof}, Brittany N. and {Weiner}, Benjamin J. and {Wuyts}, Stijn and {Yung}, L.~Y. Aaron and {Zavala}, Jorge A.},
        title = "{The Physical Conditions of Emission-line Galaxies at Cosmic Dawn from JWST/NIRSpec Spectroscopy in the SMACS 0723 Early Release Observations}",
      journal = {\apj},
         year = 2023,
        month = mar,
       volume = {945},
       number = {1},
          eid = {35},
        pages = {35},
          doi = {10.3847/1538-4357/acba8a},
archivePrefix = {arXiv},
       eprint = {2207.12388},
 primaryClass = {astro-ph.GA},
       adsurl = {https://ui.adsabs.harvard.edu/abs/2023ApJ...945...35T}
}

@ARTICLE{Pontoppidan2022,
       author = {{Pontoppidan}, Klaus M. and {Barrientes}, Jaclyn and {Blome}, Claire and {Braun}, Hannah and {Brown}, Matthew and {Carruthers}, Margaret and {Coe}, Dan and {DePasquale}, Joseph and {Espinoza}, N{\'e}stor and {Marin}, Macarena Garcia and {Gordon}, Karl D. and {Henry}, Alaina and {Hustak}, Leah and {James}, Andi and {Jenkins}, Ann and {Koekemoer}, Anton M. and {LaMassa}, Stephanie and {Law}, David and {Lockwood}, Alexandra and {Moro-Martin}, Amaya and {Mullally}, Susan E. and {Pagan}, Alyssa and {Player}, Dani and {Proffitt}, Charles and {Pulliam}, Christine and {Ramsay}, Leah and {Ravindranath}, Swara and {Reid}, Neill and {Robberto}, Massimo and {Sabbi}, Elena and {Ubeda}, Leonardo and {Balogh}, Michael and {Flanagan}, Kathryn and {Gardner}, Jonathan and {Hasan}, Hashima and {Meinke}, Bonnie and {Nota}, Antonella},
        title = "{The JWST Early Release Observations}",
      journal = {\apjl},
         year = 2022,
        month = sep,
       volume = {936},
       number = {1},
          eid = {L14},
        pages = {L14},
          doi = {10.3847/2041-8213/ac8a4e},
archivePrefix = {arXiv},
       eprint = {2207.13067},
 primaryClass = {astro-ph.IM},
       adsurl = {https://ui.adsabs.harvard.edu/abs/2022ApJ...936L..14P}
}

@ARTICLE{Kakkad2020,
       author = {{Kakkad}, D. and {Mainieri}, V. and {Vietri}, G. and {Carniani}, S. and {Harrison}, C.~M. and {Perna}, M. and {Scholtz}, J. and {Circosta}, C. and {Cresci}, G. and {Husemann}, B. and {Bischetti}, M. and {Feruglio}, C. and {Fiore}, F. and {Marconi}, A. and {Padovani}, P. and {Brusa}, M. and {Cicone}, C. and {Comastri}, A. and {Lanzuisi}, G. and {Mannucci}, F. and {Menci}, N. and {Netzer}, H. and {Piconcelli}, E. and {Puglisi}, A. and {Salvato}, M. and {Schramm}, M. and {Silverman}, J. and {Vignali}, C. and {Zamorani}, G. and {Zappacosta}, L.},
        title = "{SUPER. II. Spatially resolved ionised gas kinematics and scaling relations in z {\ensuremath{\sim}} 2 AGN host galaxies}",
      journal = {\aap},
         year = 2020,
        month = oct,
       volume = {642},
          eid = {A147},
        pages = {A147},
          doi = {10.1051/0004-6361/202038551},
archivePrefix = {arXiv},
       eprint = {2008.01728},
 primaryClass = {astro-ph.GA},
       adsurl = {https://ui.adsabs.harvard.edu/abs/2020A&A...642A.147K}
}

@BOOK{OsterbrockFerland2006,
       author = {{Osterbrock}, Donald E. and {Ferland}, Gary J.},
        title = "{Astrophysics of Gaseous Nebulae and Active Galactic Nuclei}",
         year = 2006,
       adsurl = {https://ui.adsabs.harvard.edu/abs/2006agna.book},
        publisher = {(CA: University Science Books)}
}

@ARTICLE{Shivaei2020,
       author = {{Shivaei}, Irene and {Reddy}, Naveen and {Rieke}, George and {Shapley}, Alice and {Kriek}, Mariska and {Battisti}, Andrew and {Mobasher}, Bahram and {Sanders}, Ryan and {Fetherolf}, Tara and {Azadi}, Mojegan and {Coil}, Alison L. and {Freeman}, William R. and {de Groot}, Laura and {Leung}, Gene and {Price}, Sedona H. and {Siana}, Brian and {Zick}, Tom},
        title = "{The MOSDEF Survey: The Variation of the Dust Attenuation Curve with Metallicity}",
      journal = {\apj},
         year = 2020,
        month = aug,
       volume = {899},
       number = {2},
          eid = {117},
        pages = {117},
          doi = {10.3847/1538-4357/aba35e},
archivePrefix = {arXiv},
       eprint = {2005.01742},
 primaryClass = {astro-ph.GA},
       adsurl = {https://ui.adsabs.harvard.edu/abs/2020ApJ...899..117S}
}

@ARTICLE{Venturi2024,
       author = {{Venturi}, G. and {Carniani}, S. and {Parlanti}, E. and {Kohandel}, M. and {Curti}, M. and {Pallottini}, A. and {Vallini}, L. and {Arribas}, S. and {Bunker}, A.~J. and {Cameron}, A.~J. and {Castellano}, M. and {Ferrara}, A. and {Fontana}, A. and {Gallerani}, S. and {Gelli}, V. and {Maiolino}, R. and {Ntormousi}, E. and {Pacifici}, C. and {Pentericci}, L. and {Salvadori}, S. and {Vanzella}, E.},
        title = "{Gas-phase metallicity gradients in galaxies at z {\ensuremath{\sim}} 6{\textendash}8}",
      journal = {\aap},
         year = 2024,
        month = nov,
       volume = {691},
          eid = {A19},
        pages = {A19},
          doi = {10.1051/0004-6361/202449855},
archivePrefix = {arXiv},
       eprint = {2403.03977},
 primaryClass = {astro-ph.GA},
       adsurl = {https://ui.adsabs.harvard.edu/abs/2024A&A...691A..19V}
}

@ARTICLE{Wisnioski2015,
       author = {{Wisnioski}, E. and {F{\"o}rster Schreiber}, N.~M. and {Wuyts}, S. and {Wuyts}, E. and {Bandara}, K. and {Wilman}, D. and {Genzel}, R. and {Bender}, R. and {Davies}, R. and {Fossati}, M. and {Lang}, P. and {Mendel}, J.~T. and {Beifiori}, A. and {Brammer}, G. and {Chan}, J. and {Fabricius}, M. and {Fudamoto}, Y. and {Kulkarni}, S. and {Kurk}, J. and {Lutz}, D. and {Nelson}, E.~J. and {Momcheva}, I. and {Rosario}, D. and {Saglia}, R. and {Seitz}, S. and {Tacconi}, L.~J. and {van Dokkum}, P.~G.},
        title = "{The KMOS$^{3D}$ Survey: Design, First Results, and the Evolution of Galaxy Kinematics from 0.7 <= z <= 2.7}",
      journal = {\apj},
         year = 2015,
        month = feb,
       volume = {799},
       number = {2},
          eid = {209},
        pages = {209},
          doi = {10.1088/0004-637X/799/2/209},
archivePrefix = {arXiv},
       eprint = {1409.6791},
 primaryClass = {astro-ph.GA},
       adsurl = {https://ui.adsabs.harvard.edu/abs/2015ApJ...799..209W}
}

@ARTICLE{Pillepich2019,
       author = {{Pillepich}, Annalisa and {Nelson}, Dylan and {Springel}, Volker and {Pakmor}, R{\"u}diger and {Torrey}, Paul and {Weinberger}, Rainer and {Vogelsberger}, Mark and {Marinacci}, Federico and {Genel}, Shy and {van der Wel}, Arjen and {Hernquist}, Lars},
        title = "{First results from the TNG50 simulation: the evolution of stellar and gaseous discs across cosmic time}",
      journal = {\mnras},
         year = 2019,
        month = dec,
       volume = {490},
       number = {3},
        pages = {3196-3233},
          doi = {10.1093/mnras/stz2338},
archivePrefix = {arXiv},
       eprint = {1902.05553},
 primaryClass = {astro-ph.GA},
       adsurl = {https://ui.adsabs.harvard.edu/abs/2019MNRAS.490.3196P}
}

@ARTICLE{Kohandel2024,
       author = {{Kohandel}, M. and {Pallottini}, A. and {Ferrara}, A. and {Zanella}, A. and {Rizzo}, F. and {Carniani}, S.},
        title = "{Dynamically cold disks in the early Universe: Myth or reality?}",
      journal = {\aap},
         year = 2024,
        month = may,
       volume = {685},
          eid = {A72},
        pages = {A72},
          doi = {10.1051/0004-6361/202348209},
archivePrefix = {arXiv},
       eprint = {2311.05832},
 primaryClass = {astro-ph.GA},
       adsurl = {https://ui.adsabs.harvard.edu/abs/2024A&A...685A..72K}
}

@ARTICLE{Rowland2024,
       author = {{Rowland}, Lucie E. and {Hodge}, Jacqueline and {Bouwens}, Rychard and {Mancera Pi{\~n}a}, Pavel E. and {Hygate}, Alexander and {Algera}, Hiddo and {Aravena}, Manuel and {Bowler}, Rebecca and {da Cunha}, Elisabete and {Dayal}, Pratika and {Ferrara}, Andrea and {Herard-Demanche}, Thomas and {Inami}, Hanae and {van Leeuwen}, Ivana and {de Looze}, Ilse and {Oesch}, Pascal and {Pallottini}, Andrea and {Phillips}, Si{\^a}n and {Rybak}, Matus and {Schouws}, Sander and {Smit}, Renske and {Sommovigo}, Laura and {Stefanon}, Mauro and {van der Werf}, Paul},
        title = "{REBELS-25: discovery of a dynamically cold disc galaxy at z = 7.31}",
      journal = {\mnras},
         year = 2024,
        month = dec,
       volume = {535},
       number = {3},
        pages = {2068-2091},
          doi = {10.1093/mnras/stae2217},
archivePrefix = {arXiv},
       eprint = {2405.06025},
 primaryClass = {astro-ph.GA},
       adsurl = {https://ui.adsabs.harvard.edu/abs/2024MNRAS.535.2068R}
}

@ARTICLE{deGraaff2024,
       author = {{de Graaff}, Anna and {Rix}, Hans-Walter and {Carniani}, Stefano and {Suess}, Katherine A. and {Charlot}, St{\'e}phane and {Curtis-Lake}, Emma and {Arribas}, Santiago and {Baker}, William M. and {Boyett}, Kristan and {Bunker}, Andrew J. and {Cameron}, Alex J. and {Chevallard}, Jacopo and {Curti}, Mirko and {Eisenstein}, Daniel J. and {Franx}, Marijn and {Hainline}, Kevin and {Hausen}, Ryan and {Ji}, Zhiyuan and {Johnson}, Benjamin D. and {Jones}, Gareth C. and {Maiolino}, Roberto and {Maseda}, Michael V. and {Nelson}, Erica and {Parlanti}, Eleonora and {Rawle}, Tim and {Robertson}, Brant and {Tacchella}, Sandro and {{\"U}bler}, Hannah and {Williams}, Christina C. and {Willmer}, Christopher N.~A. and {Willott}, Chris},
        title = "{Ionised gas kinematics and dynamical masses of z {\ensuremath{\gtrsim}} 6 galaxies from JADES/NIRSpec high-resolution spectroscopy}",
      journal = {\aap},
         year = 2024,
        month = apr,
       volume = {684},
          eid = {A87},
        pages = {A87},
          doi = {10.1051/0004-6361/202347755},
archivePrefix = {arXiv},
       eprint = {2308.09742},
 primaryClass = {astro-ph.GA},
       adsurl = {https://ui.adsabs.harvard.edu/abs/2024A&A...684A..87D}
}

@ARTICLE{ForsterSchreiber2018,
       author = {{F{\"o}rster Schreiber}, N.~M. and {Renzini}, A. and {Mancini}, C. and {Genzel}, R. and {Bouch{\'e}}, N. and {Cresci}, G. and {Hicks}, E.~K.~S. and {Lilly}, S.~J. and {Peng}, Y. and {Burkert}, A. and {Carollo}, C.~M. and {Cimatti}, A. and {Daddi}, E. and {Davies}, R.~I. and {Genel}, S. and {Kurk}, J.~D. and {Lang}, P. and {Lutz}, D. and {Mainieri}, V. and {McCracken}, H.~J. and {Mignoli}, M. and {Naab}, T. and {Oesch}, P. and {Pozzetti}, L. and {Scodeggio}, M. and {Shapiro Griffin}, K. and {Shapley}, A.~E. and {Sternberg}, A. and {Tacchella}, S. and {Tacconi}, L.~J. and {Wuyts}, S. and {Zamorani}, G.},
        title = "{The SINS/zC-SINF Survey of z {\ensuremath{\sim}} 2 Galaxy Kinematics: SINFONI Adaptive Optics-assisted Data and Kiloparsec-scale Emission-line Properties}",
      journal = {\apjs},
         year = 2018,
        month = oct,
       volume = {238},
       number = {2},
          eid = {21},
        pages = {21},
          doi = {10.3847/1538-4365/aadd49},
archivePrefix = {arXiv},
       eprint = {1802.07276},
 primaryClass = {astro-ph.GA},
       adsurl = {https://ui.adsabs.harvard.edu/abs/2018ApJS..238...21F}
}

@ARTICLE{Parlanti2023,
       author = {{Parlanti}, E. and {Carniani}, S. and {Pallottini}, A. and {Cignoni}, M. and {Cresci}, G. and {Kohandel}, M. and {Mannucci}, F. and {Marconi}, A.},
        title = "{ALMA hints at the presence of turbulent disk galaxies at z > 5}",
      journal = {\aap},
         year = 2023,
        month = may,
       volume = {673},
          eid = {A153},
        pages = {A153},
          doi = {10.1051/0004-6361/202245603},
archivePrefix = {arXiv},
       eprint = {2304.00036},
 primaryClass = {astro-ph.GA},
       adsurl = {https://ui.adsabs.harvard.edu/abs/2023A&A...673A.153P}
}

@ARTICLE{Bertola2025,
       author = {{Bertola}, E. and {Cresci}, G. and {Venturi}, G. and {Perna}, M. and {Circosta}, C. and {Tozzi}, G. and {Lamperti}, I. and {Vignali}, C. and {Arribas}, S. and {Bunker}, A.~J. and {Charlot}, S. and {Carniani}, S. and {Maiolino}, R. and {Rodr{\'\i}guez Del Pino}, B. and {{\"U}bler}, H. and {Willott}, C.~J. and {B{\"o}ker}, T. and {Marshall}, M.~A. and {Parlanti}, E. and {Scholtz}, J.},
        title = "{GA-NIFS: Mapping z ≃ 3.5 AGN-driven ionized outflows in the COSMOS field}",
      journal = {\aap},
         year = 2025,
        month = jul,
       volume = {699},
          eid = {A220},
        pages = {A220},
          doi = {10.1051/0004-6361/202554281},
archivePrefix = {arXiv},
       eprint = {2505.08867},
 primaryClass = {astro-ph.GA},
       adsurl = {https://ui.adsabs.harvard.edu/abs/2025A&A...699A.220B}
}

@ARTICLE{Parlanti2025,
       author = {{Parlanti}, Eleonora and {Carniani}, Stefano and {Venturi}, Giacomo and {Herrera-Camus}, Rodrigo and {Arribas}, Santiago and {Bunker}, Andrew J. and {Charlot}, St{\'e}phane and {D'Eugenio}, Francesco and {Maiolino}, Roberto and {Perna}, Michele and {{\"U}bler}, Hannah and {B{\"o}ker}, Torsten and {Cresci}, Giovanni and {Curti}, Mirko and {Jones}, Gareth C. and {Lamperti}, Isabella and {P{\'e}rez-Gonz{\'a}lez}, Pablo G. and {Del Pino}, Bruno Rodr{\'\i}guez and {Zamora}, Sandra},
        title = "{GA-NIFS: Multiphase analysis of a star-forming galaxy at z {\ensuremath{\sim}} 5.5}",
      journal = {\aap},
         year = 2025,
        month = mar,
       volume = {695},
          eid = {A6},
        pages = {A6},
          doi = {10.1051/0004-6361/202451692},
archivePrefix = {arXiv},
       eprint = {2407.19008},
 primaryClass = {astro-ph.GA},
       adsurl = {https://ui.adsabs.harvard.edu/abs/2025A&A...695A...6P}
}

@ARTICLE{Lutz2020,
       author = {{Lutz}, D. and {Sturm}, E. and {Janssen}, A. and {Veilleux}, S. and {Aalto}, S. and {Cicone}, C. and {Contursi}, A. and {Davies}, R.~I. and {Feruglio}, C. and {Fischer}, J. and {Fluetsch}, A. and {Garcia-Burillo}, S. and {Genzel}, R. and {Gonz{\'a}lez-Alfonso}, E. and {Graci{\'a}-Carpio}, J. and {Herrera-Camus}, R. and {Maiolino}, R. and {Schruba}, A. and {Shimizu}, T. and {Sternberg}, A. and {Tacconi}, L.~J. and {Wei{\ss}}, A.},
        title = "{Molecular outflows in local galaxies: Method comparison and a role of intermittent AGN driving}",
      journal = {\aap},
         year = 2020,
        month = jan,
       volume = {633},
          eid = {A134},
        pages = {A134},
          doi = {10.1051/0004-6361/201936803},
archivePrefix = {arXiv},
       eprint = {1911.05608},
 primaryClass = {astro-ph.GA},
       adsurl = {https://ui.adsabs.harvard.edu/abs/2020A&A...633A.134L}
}

@ARTICLE{Perna2025,
       author = {{Perna}, Michele and {Arribas}, Santiago and {Ji}, Xihan and {Marconcini}, Cosimo and {Lamperti}, Isabella and {Bertola}, Elena and {Circosta}, Chiara and {D'Eugenio}, Francesco and {{\"U}bler}, Hannah and {B{\"o}ker}, Torsten and {Maiolino}, Roberto and {Bunker}, Andrew J. and {Carniani}, Stefano and {Charlot}, St{\'e}phane and {Willott}, Chris J. and {Cresci}, Giovanni and {Marconi}, Alessandro and {Parlanti}, Eleonora and {Rodr{\'\i}guez Del Pino}, Bruno and {Scholtz}, Jan and {Venturi}, Giacomo},
        title = "{GA-NIFS: A galaxy-wide outflow in a Compton-thick mini-broad-absorption-line quasar at z = 3.5 probed in emission and absorption}",
      journal = {\aap},
         year = 2025,
        month = feb,
       volume = {694},
          eid = {A170},
        pages = {A170},
          doi = {10.1051/0004-6361/202453090},
archivePrefix = {arXiv},
       eprint = {2411.13698},
 primaryClass = {astro-ph.GA},
       adsurl = {https://ui.adsabs.harvard.edu/abs/2025A&A...694A.170P}
}

@ARTICLE{Zamora2024,
       author = {{Zamora}, Sandra and {Venturi}, Giacomo and {Carniani}, Stefano and {Bertola}, Elena and {Parlanti}, Eleonora and {Perna}, Michele and {Arribas}, Santiago and {B{\"o}ker}, Torsten and {Bunker}, Andrew J. and {Charlot}, St{\'e}phane and {D'Eugenio}, Francesco and {Maiolino}, Roberto and {Del Pino}, Bruno Rodr{\'\i}guez and {{\"U}bler}, Hannah and {Cresci}, Giovanni and {Jones}, Gareth C. and {Lamperti}, Isabella},
        title = "{GA-NIFS: The highly overdense system BR1202-0725 at z {\ensuremath{\sim}} 4.7: A double active galactic nucleus with fast outflows plus eight companion galaxies}",
      journal = {\aap},
         year = 2025,
        month = oct,
       volume = {702},
          eid = {A102},
        pages = {A102},
          doi = {10.1051/0004-6361/202453236},
archivePrefix = {arXiv},
       eprint = {2412.02751},
 primaryClass = {astro-ph.GA},
       adsurl = {https://ui.adsabs.harvard.edu/abs/2025A&A...702A.102Z}
}

@ARTICLE{Lamperti2024,
       author = {{Lamperti}, Isabella and {Arribas}, Santiago and {Perna}, Michele and {Rodr{\'\i}guez Del Pino}, Bruno and {Circosta}, Chiara and {P{\'e}rez-Gonz{\'a}lez}, Pablo G. and {Bunker}, Andrew J. and {Carniani}, Stefano and {Charlot}, St{\'e}phane and {D'Eugenio}, Francesco and {Maiolino}, Roberto and {{\"U}bler}, Hannah and {Willott}, Chris J. and {Bertola}, Elena and {B{\"o}ker}, Torsten and {Cresci}, Giovanni and {Curti}, Mirko and {Jones}, Gareth C. and {Kumari}, Nimisha and {Parlanti}, Eleonora and {Scholtz}, Jan and {Venturi}, Giacomo},
        title = "{GA-NIFS: JWST/NIRSpec IFS view of the z {\ensuremath{\sim}} 3.5 galaxy GS5001 and its close environment at the core of a large-scale overdensity}",
      journal = {\aap},
         year = 2024,
        month = nov,
       volume = {691},
          eid = {A153},
        pages = {A153},
          doi = {10.1051/0004-6361/202451021},
archivePrefix = {arXiv},
       eprint = {2406.10348},
 primaryClass = {astro-ph.GA},
       adsurl = {https://ui.adsabs.harvard.edu/abs/2024A&A...691A.153L}
}

@ARTICLE{delPino2024,
       author = {{Rodr{\'\i}guez Del Pino}, B. and {Perna}, M. and {Arribas}, S. and {D'Eugenio}, F. and {Lamperti}, I. and {P{\'e}rez-Gonz{\'a}lez}, P.~G. and {{\"U}bler}, H. and {Bunker}, A. and {Carniani}, S. and {Charlot}, S. and {Maiolino}, R. and {Willott}, C.~J. and {B{\"o}ker}, T. and {Chevallard}, J. and {Cresci}, G. and {Curti}, M. and {Jones}, G.~C. and {Parlanti}, E. and {Scholtz}, J. and {Venturi}, G.},
        title = "{GA-NIFS: Co-evolution within a highly star-forming galaxy group at z {\ensuremath{\sim}} 3.7 witnessed by JWST/NIRSpec IFS}",
      journal = {\aap},
         year = 2024,
        month = apr,
       volume = {684},
          eid = {A187},
        pages = {A187},
          doi = {10.1051/0004-6361/202348057},
archivePrefix = {arXiv},
       eprint = {2309.14431},
 primaryClass = {astro-ph.GA},
       adsurl = {https://ui.adsabs.harvard.edu/abs/2024A&A...684A.187R}
}

@ARTICLE{Parlanti2024,
       author = {{Parlanti}, Eleonora and {Carniani}, Stefano and {{\"U}bler}, Hannah and {Venturi}, Giacomo and {Circosta}, Chiara and {D'Eugenio}, Francesco and {Arribas}, Santiago and {Bunker}, Andrew J. and {Charlot}, St{\'e}phane and {L{\"u}tzgendorf}, Nora and {Maiolino}, Roberto and {Perna}, Michele and {Rodr{\'\i}guez Del Pino}, Bruno and {Willott}, Chris J. and {B{\"o}ker}, Torsten and {Cameron}, Alex J. and {Chevallard}, Jacopo and {Cresci}, Giovanni and {Jones}, Gareth C. and {Kumari}, Nimisha and {Lamperti}, Isabella and {Scholtz}, Jan},
        title = "{GA-NIFS: Early-stage feedback in a heavily obscured active galactic nucleus at z = 4.76}",
      journal = {\aap},
         year = 2024,
        month = apr,
       volume = {684},
          eid = {A24},
        pages = {A24},
          doi = {10.1051/0004-6361/202347914},
archivePrefix = {arXiv},
       eprint = {2309.05713},
 primaryClass = {astro-ph.GA},
       adsurl = {https://ui.adsabs.harvard.edu/abs/2024A&A...684A..24P}
}

@ARTICLE{Marshall2023,
       author = {{Marshall}, Madeline A. and {Perna}, Michele and {Willott}, Chris J. and {Maiolino}, Roberto and {Scholtz}, Jan and {{\"U}bler}, Hannah and {Carniani}, Stefano and {Arribas}, Santiago and {L{\"u}tzgendorf}, Nora and {Bunker}, Andrew J. and {Charlot}, Stephane and {Ferruit}, Pierre and {Jakobsen}, Peter and {Rix}, Hans-Walter and {Rodr{\'\i}guez Del Pino}, Bruno and {B{\"o}ker}, Torsten and {Cameron}, Alex J. and {Cresci}, Giovanni and {Curtis-Lake}, Emma and {Jones}, Gareth C. and {Kumari}, Nimisha and {P{\'e}rez-Gonz{\'a}lez}, Pablo G. and {Reed}, Sophie L.},
        title = "{GA-NIFS: Black hole and host galaxy properties of two z ≃ 6.8 quasars from the NIRSpec IFU}",
      journal = {\aap},
         year = 2023,
        month = oct,
       volume = {678},
          eid = {A191},
        pages = {A191},
          doi = {10.1051/0004-6361/202346113},
archivePrefix = {arXiv},
       eprint = {2302.04795},
 primaryClass = {astro-ph.GA},
       adsurl = {https://ui.adsabs.harvard.edu/abs/2023A&A...678A.191M}
}

@ARTICLE{Curti2020,
       author = {{Curti}, Mirko and {Mannucci}, Filippo and {Cresci}, Giovanni and {Maiolino}, Roberto},
        title = "{The mass-metallicity and the fundamental metallicity relation revisited on a fully T$_{e}$-based abundance scale for galaxies}",
      journal = {\mnras},
         year = 2020,
        month = jan,
       volume = {491},
       number = {1},
        pages = {944-964},
          doi = {10.1093/mnras/stz2910},
archivePrefix = {arXiv},
       eprint = {1910.00597},
 primaryClass = {astro-ph.GA},
       adsurl = {https://ui.adsabs.harvard.edu/abs/2020MNRAS.491..944C}
}

@ARTICLE{Jakobsen22,
       author = {{Jakobsen}, P. and {Ferruit}, P. and {Alves de Oliveira}, C. and {Arribas}, S. and {Bagnasco}, G. and {Barho}, R. and {Beck}, T.~L. and {Birkmann}, S. and {B{\"o}ker}, T. and {Bunker}, A.~J. and {Charlot}, S. and {de Jong}, P. and {de Marchi}, G. and {Ehrenwinkler}, R. and {Falcolini}, M. and {Fels}, R. and {Franx}, M. and {Franz}, D. and {Funke}, M. and {Giardino}, G. and {Gnata}, X. and {Holota}, W. and {Honnen}, K. and {Jensen}, P.~L. and {Jentsch}, M. and {Johnson}, T. and {Jollet}, D. and {Karl}, H. and {Kling}, G. and {K{\"o}hler}, J. and {Kolm}, M. -G. and {Kumari}, N. and {Lander}, M.~E. and {Lemke}, R. and {L{\'o}pez-Caniego}, M. and {L{\"u}tzgendorf}, N. and {Maiolino}, R. and {Manjavacas}, E. and {Marston}, A. and {Maschmann}, M. and {Maurer}, R. and {Messerschmidt}, B. and {Moseley}, S.~H. and {Mosner}, P. and {Mott}, D.~B. and {Muzerolle}, J. and {Pirzkal}, N. and {Pittet}, J. -F. and {Plitzke}, A. and {Posselt}, W. and {Rapp}, B. and {Rauscher}, B.~J. and {Rawle}, T. and {Rix}, H. -W. and {R{\"o}del}, A. and {Rumler}, P. and {Sabbi}, E. and {Salvignol}, J. -C. and {Schmid}, T. and {Sirianni}, M. and {Smith}, C. and {Strada}, P. and {te Plate}, M. and {Valenti}, J. and {Wettemann}, T. and {Wiehe}, T. and {Wiesmayer}, M. and {Willott}, C.~J. and {Wright}, R. and {Zeidler}, P. and {Zincke}, C.},
        title = "{The Near-Infrared Spectrograph (NIRSpec) on the James Webb Space Telescope. I. Overview of the instrument and its capabilities}",
      journal = {\aap},
         year = 2022,
        month = may,
       volume = {661},
          eid = {A80},
        pages = {A80},
          doi = {10.1051/0004-6361/202142663},
archivePrefix = {arXiv},
       eprint = {2202.03305},
 primaryClass = {astro-ph.IM},
       adsurl = {https://ui.adsabs.harvard.edu/abs/2022A&A...661A..80J}
}

@ARTICLE{Boker22,
       author = {{B{\"o}ker}, T. and {Arribas}, S. and {L{\"u}tzgendorf}, N. and {Alves de Oliveira}, C. and {Beck}, T.~L. and {Birkmann}, S. and {Bunker}, A.~J. and {Charlot}, S. and {de Marchi}, G. and {Ferruit}, P. and {Giardino}, G. and {Jakobsen}, P. and {Kumari}, N. and {L{\'o}pez-Caniego}, M. and {Maiolino}, R. and {Manjavacas}, E. and {Marston}, A. and {Moseley}, S.~H. and {Muzerolle}, J. and {Ogle}, P. and {Pirzkal}, N. and {Rauscher}, B. and {Rawle}, T. and {Rix}, H. -W. and {Sabbi}, E. and {Sargent}, B. and {Sirianni}, M. and {te Plate}, M. and {Valenti}, J. and {Willott}, C.~J. and {Zeidler}, P.},
        title = "{The Near-Infrared Spectrograph (NIRSpec) on the James Webb Space Telescope. III. Integral-field spectroscopy}",
      journal = {\aap},
         year = 2022,
        month = may,
       volume = {661},
          eid = {A82},
        pages = {A82},
          doi = {10.1051/0004-6361/202142589},
archivePrefix = {arXiv},
       eprint = {2202.03308},
 primaryClass = {astro-ph.IM},
       adsurl = {https://ui.adsabs.harvard.edu/abs/2022A&A...661A..82B}
}

@ARTICLE{vDokkum01,
       author = {{van Dokkum}, Pieter G.},
        title = "{Cosmic-Ray Rejection by Laplacian Edge Detection}",
      journal = {\pasp},
         year = 2001,
        month = nov,
       volume = {113},
       number = {789},
        pages = {1420-1427},
          doi = {10.1086/323894},
archivePrefix = {arXiv},
       eprint = {astro-ph/0108003},
 primaryClass = {astro-ph},
       adsurl = {https://ui.adsabs.harvard.edu/abs/2001PASP..113.1420V}
}

@ARTICLE{Perna23,
       author = {{Perna}, M. and {Arribas}, S. and {Marshall}, M. and {D'Eugenio}, F. and {{\"U}bler}, H. and {Bunker}, A. and {Charlot}, S. and {Carniani}, S. and {Jakobsen}, P. and {Maiolino}, R. and {Rodr{\'\i}guez Del Pino}, B. and {Willott}, C.~J. and {B{\"o}ker}, T. and {Circosta}, C. and {Cresci}, G. and {Curti}, M. and {Husemann}, B. and {Kumari}, N. and {Lamperti}, I. and {P{\'e}rez-Gonz{\'a}lez}, P.~G. and {Scholtz}, J.},
        title = "{GA-NIFS: The ultra-dense, interacting environment of a dual AGN at z {\ensuremath{\sim}} 3.3 revealed by JWST/NIRSpec IFS}",
      journal = {\aap},
         year = 2023,
        month = nov,
       volume = {679},
          eid = {A89},
        pages = {A89},
          doi = {10.1051/0004-6361/202346649},
archivePrefix = {arXiv},
       eprint = {2304.06756},
 primaryClass = {astro-ph.GA},
       adsurl = {https://ui.adsabs.harvard.edu/abs/2023A&A...679A..89P}
}

@ARTICLE{DEugenio23ifs,
       author = {{D'Eugenio}, Francesco and {P{\'e}rez-Gonz{\'a}lez}, Pablo G. and {Maiolino}, Roberto and {Scholtz}, Jan and {Perna}, Michele and {Circosta}, Chiara and {{\"U}bler}, Hannah and {Arribas}, Santiago and {B{\"o}ker}, Torsten and {Bunker}, Andrew J. and {Carniani}, Stefano and {Charlot}, Stephane and {Chevallard}, Jacopo and {Cresci}, Giovanni and {Curtis-Lake}, Emma and {Jones}, Gareth C. and {Kumari}, Nimisha and {Lamperti}, Isabella and {Looser}, Tobias J. and {Parlanti}, Eleonora and {Rix}, Hans-Walter and {Robertson}, Brant and {Rodr{\'\i}guez Del Pino}, Bruno and {Tacchella}, Sandro and {Venturi}, Giacomo and {Willott}, Chris J.},
        title = "{A fast-rotator post-starburst galaxy quenched by supermassive black-hole feedback at z = 3}",
      journal = {Nature Astronomy},
         year = 2024,
        month = nov,
       volume = {8},
        pages = {1443-1456},
          doi = {10.1038/s41550-024-02345-1},
archivePrefix = {arXiv},
       eprint = {2308.06317},
 primaryClass = {astro-ph.GA},
       adsurl = {https://ui.adsabs.harvard.edu/abs/2024NatAs...8.1443D}
}

@article{Nakajima22,
	title = {{Diagnostics for PopIII galaxies and direct collapse black holes in the early universe}},
	author = {{Nakajima}, K. and {Maiolino}, R.},
	year = 2022,
	month = jul,
	journal = {\mnras},
	volume = 513,
	number = 4,
	pages = {5134--5147},
	doi = {10.1093/mnras/stac1242},
	archiveprefix = {arXiv},
	eprint = {2204.11870},
	primaryclass = {astro-ph.GA},
	adsurl = {https://ui.adsabs.harvard.edu/abs/2022MNRAS.513.5134N}
}

@ARTICLE{Sanders23,
       author = {{Sanders}, Ryan L. and {Shapley}, Alice E. and {Topping}, Michael W. and {Reddy}, Naveen A. and {Brammer}, Gabriel B.},
        title = "{Excitation and Ionization Properties of Star-forming Galaxies at z = 2.0-9.3 with JWST/NIRSpec}",
      journal = {\apj},
         year = 2023,
        month = sep,
       volume = {955},
       number = {1},
          eid = {54},
        pages = {54},
          doi = {10.3847/1538-4357/acedad},
archivePrefix = {arXiv},
       eprint = {2301.06696},
 primaryClass = {astro-ph.GA},
       adsurl = {https://ui.adsabs.harvard.edu/abs/2023ApJ...955...54S}
}

@ARTICLE{Wang2022,
       author = {{Wang}, Xin and {Jones}, Tucker and {Vulcani}, Benedetta and {Treu}, Tommaso and {Morishita}, Takahiro and {Roberts-Borsani}, Guido and {Malkan}, Matthew A. and {Henry}, Alaina and {Brammer}, Gabriel and {Strait}, Victoria and {Brada{\v{c}}}, Maru{\v{s}}a and {Boyett}, Kristan and {Calabr{\`o}}, Antonello and {Castellano}, Marco and {Fontana}, Adriano and {Glazebrook}, Karl and {Kelly}, Patrick L. and {Leethochawalit}, Nicha and {Marchesini}, Danilo and {Santini}, P. and {Trenti}, M. and {Yang}, Lilan},
        title = "{Early Results from GLASS-JWST. IV. Spatially Resolved Metallicity in a Low-mass z   3 Galaxy with NIRISS}",
      journal = {\apjl},
         year = 2022,
        month = oct,
       volume = {938},
       number = {2},
          eid = {L16},
        pages = {L16},
          doi = {10.3847/2041-8213/ac959e},
archivePrefix = {arXiv},
       eprint = {2207.13113},
 primaryClass = {astro-ph.GA},
       adsurl = {https://ui.adsabs.harvard.edu/abs/2022ApJ...938L..16W}
}

@ARTICLE{Garcia25,
       author = {{Garcia}, Alex M. and {Torrey}, Paul and {Bhagwat}, Aniket and {Wright}, Ruby J. and {Chen}, Qian-Hui and {Grasha}, Kathryn and {Ridolfo}, Sophia and {Hemler}, Z.~S. and {Sarkar}, Arnab and {Chakraborty}, Priyanka and {Nelson}, Erica J. and {Sanders}, Ryan L. and {Costa}, Tiago and {Vogelsberger}, Mark and {Kewley}, Lisa J. and {Ellison}, Sara L. and {Hernquist}, Lars},
        title = "{Metallicity Gradients in Modern Cosmological Simulations. I. Tension between Smooth Stellar Feedback Models and Observations}",
      journal = {\apj},
         year = 2025,
        month = aug,
       volume = {989},
       number = {2},
          eid = {147},
        pages = {147},
          doi = {10.3847/1538-4357/adea51},
archivePrefix = {arXiv},
       eprint = {2503.03804},
 primaryClass = {astro-ph.GA},
       adsurl = {https://ui.adsabs.harvard.edu/abs/2025ApJ...989..147G}
}

@ARTICLE{PyNeb2015,
       author = {{Luridiana}, V. and {Morisset}, C. and {Shaw}, R.~A.},
        title = "{PyNeb: a new tool for analyzing emission lines. I. Code description and validation of results}",
      journal = {\aap},
         year = 2015,
        month = jan,
       volume = {573},
          eid = {A42},
        pages = {A42},
          doi = {10.1051/0004-6361/201323152},
archivePrefix = {arXiv},
       eprint = {1410.6662},
 primaryClass = {astro-ph.IM},
       adsurl = {https://ui.adsabs.harvard.edu/abs/2015A&A...573A..42L}
}

@ARTICLE{KennicuttEvans2012,
       author = {{Kennicutt}, Robert C. and {Evans}, Neal J.},
        title = "{Star Formation in the Milky Way and Nearby Galaxies}",
      journal = {\araa},
         year = 2012,
        month = sep,
       volume = {50},
        pages = {531-608},
          doi = {10.1146/annurev-astro-081811-125610},
archivePrefix = {arXiv},
       eprint = {1204.3552},
 primaryClass = {astro-ph.GA},
       adsurl = {https://ui.adsabs.harvard.edu/abs/2012ARA&A..50..531K}
}

@ARTICLE{Kennicutt1998,
       author = {{Kennicutt}, Jr., Robert C.},
        title = "{The Global Schmidt Law in Star-forming Galaxies}",
      journal = {\apj},
         year = 1998,
        month = may,
       volume = {498},
       number = {2},
        pages = {541-552},
          doi = {10.1086/305588},
archivePrefix = {arXiv},
       eprint = {astro-ph/9712213},
 primaryClass = {astro-ph},
       adsurl = {https://ui.adsabs.harvard.edu/abs/1998ApJ...498..541K}
}

@ARTICLE{Asplund2009,
       author = {{Asplund}, Martin and {Grevesse}, Nicolas and {Sauval}, A. Jacques and {Scott}, Pat},
        title = "{The Chemical Composition of the Sun}",
      journal = {\araa},
         year = 2009,
        month = sep,
       volume = {47},
       number = {1},
        pages = {481-522},
          doi = {10.1146/annurev.astro.46.060407.145222},
archivePrefix = {arXiv},
       eprint = {0909.0948},
 primaryClass = {astro-ph.SR},
       adsurl = {https://ui.adsabs.harvard.edu/abs/2009ARA&A..47..481A}
}

@ARTICLE{curtis-lake_2023,
       author = {{Curtis-Lake}, Emma and {Carniani}, Stefano and {Cameron}, Alex and {Charlot}, Stephane and {Jakobsen}, Peter and {Maiolino}, Roberto and {Bunker}, Andrew and {Witstok}, Joris and {Smit}, Renske and {Chevallard}, Jacopo and {Willott}, Chris and {Ferruit}, Pierre and {Arribas}, Santiago and {Bonaventura}, Nina and {Curti}, Mirko and {D'Eugenio}, Francesco and {Franx}, Marijn and {Giardino}, Giovanna and {Looser}, Tobias J. and {L{\"u}tzgendorf}, Nora and {Maseda}, Michael V. and {Rawle}, Tim and {Rix}, Hans-Walter and {Rodr{\'\i}guez del Pino}, Bruno and {{\"U}bler}, Hannah and {Sirianni}, Marco and {Dressler}, Alan and {Egami}, Eiichi and {Eisenstein}, Daniel J. and {Endsley}, Ryan and {Hainline}, Kevin and {Hausen}, Ryan and {Johnson}, Benjamin D. and {Rieke}, Marcia and {Robertson}, Brant and {Shivaei}, Irene and {Stark}, Daniel P. and {Tacchella}, Sandro and {Williams}, Christina C. and {Willmer}, Christopher N.~A. and {Bhatawdekar}, Rachana and {Bowler}, Rebecca and {Boyett}, Kristan and {Chen}, Zuyi and {de Graaff}, Anna and {Helton}, Jakob M. and {Hviding}, Raphael E. and {Jones}, Gareth C. and {Kumari}, Nimisha and {Lyu}, Jianwei and {Nelson}, Erica and {Perna}, Michele and {Sandles}, Lester and {Saxena}, Aayush and {Suess}, Katherine A. and {Sun}, Fengwu and {Topping}, Michael W. and {Wallace}, Imaan E.~B. and {Whitler}, Lily},
        title = "{Spectroscopic confirmation of four metal-poor galaxies at z = 10.3-13.2}",
      journal = {Nature Astronomy},
         year = 2023,
        month = may,
       volume = {7},
        pages = {622-632},
          doi = {10.1038/s41550-023-01918-w},
archivePrefix = {arXiv},
       eprint = {2212.04568},
 primaryClass = {astro-ph.GA},
       adsurl = {https://ui.adsabs.harvard.edu/abs/2023NatAs...7..622C}
}

@ARTICLE{Hsiao23,
       author = {{Hsiao}, Tiger Yu-Yang and {Coe}, Dan and {Abdurro'uf} and {Whitler}, Lily and {Jung}, Intae and {Khullar}, Gourav and {Meena}, Ashish Kumar and {Dayal}, Pratika and {Barrow}, Kirk S.~S. and {Santos-Olmsted}, Lillian and {Casselman}, Adam and {Vanzella}, Eros and {Nonino}, Mario and {Jim{\'e}nez-Teja}, Yolanda and {Oguri}, Masamune and {Stark}, Daniel P. and {Furtak}, Lukas J. and {Zitrin}, Adi and {Adamo}, Angela and {Brammer}, Gabriel and {Bradley}, Larry and {Diego}, Jose M. and {Zackrisson}, Erik and {Finkelstein}, Steven L. and {Windhorst}, Rogier A. and {Bhatawdekar}, Rachana and {Hutchison}, Taylor A. and {Broadhurst}, Tom and {Dimauro}, Paola and {Andrade-Santos}, Felipe and {Eldridge}, Jan J. and {Acebron}, Ana and {Avila}, Roberto J. and {Bayliss}, Matthew B. and {Ben{\'\i}tez}, Alex and {Binggeli}, Christian and {Bolan}, Patricia and {Brada{\v{c}}}, Maru{\v{s}}a and {Carnall}, Adam C. and {Conselice}, Christopher J. and {Donahue}, Megan and {Frye}, Brenda and {Fujimoto}, Seiji and {Henry}, Alaina and {James}, Bethan L. and {Kassin}, Susan A. and {Kewley}, Lisa and {Larson}, Rebecca L. and {Lauer}, Tod and {Law}, David and {Mahler}, Guillaume and {Mainali}, Ramesh and {McCandliss}, Stephan and {Nicholls}, David and {Pirzkal}, Norbert and {Postman}, Marc and {Rigby}, Jane R. and {Ryan}, Russell and {Senchyna}, Peter and {Sharon}, Keren and {Shimizu}, Ikko and {Strait}, Victoria and {Tang}, Mengtao and {Trenti}, Michele and {Vikaeus}, Anton and {Welch}, Brian},
        title = "{JWST Reveals a Possible z {\ensuremath{\sim}} 11 Galaxy Merger in Triply Lensed MACS0647{\textendash}JD}",
      journal = {\apjl},
         year = 2023,
        month = jun,
       volume = {949},
       number = {2},
          eid = {L34},
        pages = {L34},
          doi = {10.3847/2041-8213/acc94b},
archivePrefix = {arXiv},
       eprint = {2210.14123},
 primaryClass = {astro-ph.GA},
       adsurl = {https://ui.adsabs.harvard.edu/abs/2023ApJ...949L..34H}
}

@ARTICLE{maiolino_gnz11_2023,
       author = {{Maiolino}, Roberto and {Scholtz}, Jan and {Witstok}, Joris and {Carniani}, Stefano and {D'Eugenio}, Francesco and {de Graaff}, Anna and {{\"U}bler}, Hannah and {Tacchella}, Sandro and {Curtis-Lake}, Emma and {Arribas}, Santiago and {Bunker}, Andrew and {Charlot}, St{\'e}phane and {Chevallard}, Jacopo and {Curti}, Mirko and {Looser}, Tobias J. and {Maseda}, Michael V. and {Rawle}, Timothy D. and {Rodr{\'\i}guez del Pino}, Bruno and {Willott}, Chris J. and {Egami}, Eiichi and {Eisenstein}, Daniel J. and {Hainline}, Kevin N. and {Robertson}, Brant and {Williams}, Christina C. and {Willmer}, Christopher N.~A. and {Baker}, William M. and {Boyett}, Kristan and {DeCoursey}, Christa and {Fabian}, Andrew C. and {Helton}, Jakob M. and {Ji}, Zhiyuan and {Jones}, Gareth C. and {Kumari}, Nimisha and {Laporte}, Nicolas and {Nelson}, Erica J. and {Perna}, Michele and {Sandles}, Lester and {Shivaei}, Irene and {Sun}, Fengwu},
        title = "{A small and vigorous black hole in the early Universe}",
      journal = {\nat},
         year = 2024,
        month = mar,
       volume = {627},
       number = {8002},
        pages = {59-63},
          doi = {10.1038/s41586-024-07052-5},
archivePrefix = {arXiv},
       eprint = {2305.12492},
 primaryClass = {astro-ph.GA},
       adsurl = {https://ui.adsabs.harvard.edu/abs/2024Natur.627...59M}
}

@ARTICLE{Scholtz+2025_disk,
       author = {{Scholtz}, J. and {Parlanti}, E. and {Carniani}, S. and {Kohandel}, M. and {Sun}, F. and {Danhaive}, A.~L. and {Maiolino}, R. and {Arribas}, S. and {Bhatawdekar}, R. and {Bunker}, A.~J. and {Charlot}, S. and {D'Eugenio}, F. and {Ferrara}, A. and {Ji}, Z. and {Jones}, Gareth C. and {Rinaldi}, P. and {Robertson}, B. and {Pallottini}, A. and {Shivaei}, I. and {Sun}, Y. and {Tacchella}, S. and {{\"U}bler}, H. and {Venturi}, G.},
        title = "{Tentative rotation in a galaxy at z\raisebox{-0.5ex}\textasciitilde14 with ALMA}",
      journal = {\mnras},
         year = 2025,
        month = oct,
          doi = {10.1093/mnrasl/slaf109},
archivePrefix = {arXiv},
       eprint = {2503.10751},
 primaryClass = {astro-ph.GA},
       adsurl = {https://ui.adsabs.harvard.edu/abs/2025MNRAS.tmpL.102S}
}

@ARTICLE{Carnall23a,
       author = {{Carnall}, Adam C. and {McLure}, Ross J. and {Dunlop}, James S. and {McLeod}, Derek J. and {Wild}, Vivienne and {Cullen}, Fergus and {Magee}, Dan and {Begley}, Ryan and {Cimatti}, Andrea and {Donnan}, Callum T. and {Hamadouche}, Massissilia L. and {Jewell}, Sophie M. and {Walker}, Sam},
        title = "{A massive quiescent galaxy at redshift 4.658}",
      journal = {\nat},
         year = 2023,
        month = jul,
       volume = {619},
       number = {7971},
        pages = {716-719},
          doi = {10.1038/s41586-023-06158-6},
archivePrefix = {arXiv},
       eprint = {2301.11413},
 primaryClass = {astro-ph.GA},
       adsurl = {https://ui.adsabs.harvard.edu/abs/2023Natur.619..716C}
}

@ARTICLE{Carnall23b,
       author = {{Carnall}, A.~C. and {McLeod}, D.~J. and {McLure}, R.~J. and {Dunlop}, J.~S. and {Begley}, R. and {Cullen}, F. and {Donnan}, C.~T. and {Hamadouche}, M.~L. and {Jewell}, S.~M. and {Jones}, E.~W. and {Pollock}, C.~L. and {Wild}, V.},
        title = "{A surprising abundance of massive quiescent galaxies at 3 < z < 5 in the first data from JWST CEERS}",
      journal = {\mnras},
         year = 2023,
        month = apr,
       volume = {520},
       number = {3},
        pages = {3974-3985},
          doi = {10.1093/mnras/stad369},
archivePrefix = {arXiv},
       eprint = {2208.00986},
 primaryClass = {astro-ph.GA},
       adsurl = {https://ui.adsabs.harvard.edu/abs/2023MNRAS.520.3974C}
}

@ARTICLE{Looser23SFH,
       author = {{Looser}, Tobias J. and {D'Eugenio}, Francesco and {Maiolino}, Roberto and {Tacchella}, Sandro and {Curti}, Mirko and {Arribas}, Santiago and {Baker}, William M. and {Baum}, Stefi and {Bonaventura}, Nina and {Boyett}, Kristan and {Bunker}, Andrew J. and {Carniani}, Stefano and {Charlot}, Stephane and {Chevallard}, Jacopo and {Curtis-Lake}, Emma and {Lola Danhaive}, A. and {Eisenstein}, Daniel J. and {de Graaff}, Anna and {Hainline}, Kevin and {Ji}, Zhiyuan and {Johnson}, Benjamin D. and {Kumari}, Nimisha and {Nelson}, Erica and {Parlanti}, Eleonora and {Rix}, Hans-Walter and {Robertson}, Brant and {Del Pino}, Bruno Rodr{\'\i}guez and {Sandles}, Lester and {Scholtz}, Jan and {Smit}, Renske and {Stark}, Daniel P. and {{\"U}bler}, Hannah and {Williams}, Christina C. and {Willott}, Chris and {Witstok}, Joris},
        title = "{JADES: Differing assembly histories of galaxies: Observational evidence for bursty star formation histories and (mini-)quenching in the first billion years of the Universe}",
      journal = {\aap},
         year = 2025,
        month = may,
       volume = {697},
          eid = {A88},
        pages = {A88},
          doi = {10.1051/0004-6361/202347102},
archivePrefix = {arXiv},
       eprint = {2306.02470},
 primaryClass = {astro-ph.GA},
       adsurl = {https://ui.adsabs.harvard.edu/abs/2025A&A...697A..88L}
}

@ARTICLE{Gelli24,
       author = {{Gelli}, Viola and {Salvadori}, Stefania and {Ferrara}, Andrea and {Pallottini}, Andrea},
        title = "{Can Supernovae Quench Star Formation in High-z Galaxies?}",
      journal = {\apj},
         year = 2024,
        month = mar,
       volume = {964},
       number = {1},
          eid = {76},
        pages = {76},
          doi = {10.3847/1538-4357/ad23ec},
archivePrefix = {arXiv},
       eprint = {2310.03065},
 primaryClass = {astro-ph.GA},
       adsurl = {https://ui.adsabs.harvard.edu/abs/2024ApJ...964...76G}
}

@ARTICLE{Arrabal_Haro_nature_2023,
       author = {{Arrabal Haro}, Pablo and {Dickinson}, Mark and {Finkelstein}, Steven L. and {Kartaltepe}, Jeyhan S. and {Donnan}, Callum T. and {Burgarella}, Denis and {Carnall}, Adam C. and {Cullen}, Fergus and {Dunlop}, James S. and {Fern{\'a}ndez}, Vital and {Fujimoto}, Seiji and {Jung}, Intae and {Krips}, Melanie and {Larson}, Rebecca L. and {Papovich}, Casey and {P{\'e}rez-Gonz{\'a}lez}, Pablo G. and {Amor{\'\i}n}, Ricardo O. and {Bagley}, Micaela B. and {Buat}, V{\'e}ronique and {Casey}, Caitlin M. and {Chworowsky}, Katherine and {Cohen}, Seth H. and {Ferguson}, Henry C. and {Giavalisco}, Mauro and {Huertas-Company}, Marc and {Hutchison}, Taylor A. and {Kocevski}, Dale D. and {Koekemoer}, Anton M. and {Lucas}, Ray A. and {McLeod}, Derek J. and {McLure}, Ross J. and {Pirzkal}, Norbert and {Seill{\'e}}, Lise-Marie and {Trump}, Jonathan R. and {Weiner}, Benjamin J. and {Wilkins}, Stephen M. and {Zavala}, Jorge A.},
        title = "{Confirmation and refutation of very luminous galaxies in the early Universe}",
      journal = {\nat},
         year = 2023,
        month = oct,
       volume = {622},
       number = {7984},
        pages = {707-711},
          doi = {10.1038/s41586-023-06521-7},
archivePrefix = {arXiv},
       eprint = {2303.15431},
 primaryClass = {astro-ph.GA},
       adsurl = {https://ui.adsabs.harvard.edu/abs/2023Natur.622..707A}
}

@ARTICLE{Cameron23,
       author = {{Cameron}, Alex J. and {Saxena}, Aayush and {Bunker}, Andrew J. and {D'Eugenio}, Francesco and {Carniani}, Stefano and {Maiolino}, Roberto and {Curtis-Lake}, Emma and {Ferruit}, Pierre and {Jakobsen}, Peter and {Arribas}, Santiago and {Bonaventura}, Nina and {Charlot}, Stephane and {Chevallard}, Jacopo and {Curti}, Mirko and {Looser}, Tobias J. and {Maseda}, Michael V. and {Rawle}, Tim and {Rodr{\'\i}guez Del Pino}, Bruno and {Smit}, Renske and {{\"U}bler}, Hannah and {Willott}, Chris and {Witstok}, Joris and {Egami}, Eiichi and {Eisenstein}, Daniel J. and {Johnson}, Benjamin D. and {Hainline}, Kevin and {Rieke}, Marcia and {Robertson}, Brant E. and {Stark}, Daniel P. and {Tacchella}, Sandro and {Williams}, Christina C. and {Willmer}, Christopher N.~A. and {Bhatawdekar}, Rachana and {Bowler}, Rebecca and {Boyett}, Kristan and {Circosta}, Chiara and {Helton}, Jakob M. and {Jones}, Gareth C. and {Kumari}, Nimisha and {Ji}, Zhiyuan and {Nelson}, Erica and {Parlanti}, Eleonora and {Sandles}, Lester and {Scholtz}, Jan and {Sun}, Fengwu},
        title = "{JADES: Probing interstellar medium conditions at z {\ensuremath{\sim}} 5.5-9.5 with ultra-deep JWST/NIRSpec spectroscopy}",
      journal = {\aap},
         year = 2023,
        month = sep,
       volume = {677},
          eid = {A115},
        pages = {A115},
          doi = {10.1051/0004-6361/202346107},
archivePrefix = {arXiv},
       eprint = {2302.04298},
 primaryClass = {astro-ph.GA},
       adsurl = {https://ui.adsabs.harvard.edu/abs/2023A&A...677A.115C}
}

@ARTICLE{Scholtz2025_AGN,
       author = {{Scholtz}, Jan and {Maiolino}, Roberto and {D'Eugenio}, Francesco and {Curtis-Lake}, Emma and {Carniani}, Stefano and {Charlot}, Stephane and {Curti}, Mirko and {Silcock}, Maddie S. and {Arribas}, Santiago and {Baker}, William and {Bhatawdekar}, Rachana and {Boyett}, Kristan and {Bunker}, Andrew J. and {Chevallard}, Jacopo and {Circosta}, Chiara and {Eisenstein}, Daniel J. and {Hainline}, Kevin and {Hausen}, Ryan and {Ji}, Xihan and {Ji}, Zhiyuan and {Johnson}, Benjamin D. and {Kumari}, Nimisha and {Looser}, Tobias J. and {Lyu}, Jianwei and {Maseda}, Michael V. and {Parlanti}, Eleonora and {Perna}, Michele and {Rieke}, Marcia and {Robertson}, Brant and {Del Pino}, Bruno Rodr{\'\i}guez and {Sun}, Fengwu and {Tacchella}, Sandro and {{\"U}bler}, Hannah and {Venturi}, Giacomo and {Williams}, Christina C. and {Willmer}, Christopher N.~A. and {Willott}, Chris and {Witstok}, Joris},
        title = "{JADES: A large population of obscured, narrow-line active galactic nuclei at high redshift}",
      journal = {\aap},
         year = 2025,
        month = may,
       volume = {697},
          eid = {A175},
        pages = {A175},
          doi = {10.1051/0004-6361/202348804},
archivePrefix = {arXiv},
       eprint = {2311.18731},
 primaryClass = {astro-ph.GA},
       adsurl = {https://ui.adsabs.harvard.edu/abs/2025A&A...697A.175S}
}

@ARTICLE{furtak23,
       author = {{Furtak}, Lukas J. and {Labb{\'e}}, Ivo and {Zitrin}, Adi and {Greene}, Jenny E. and {Dayal}, Pratika and {Chemerynska}, Iryna and {Kokorev}, Vasily and {Miller}, Tim B. and {Goulding}, Andy D. and {de Graaff}, Anna and {Bezanson}, Rachel and {Brammer}, Gabriel B. and {Cutler}, Sam E. and {Leja}, Joel and {Pan}, Richard and {Price}, Sedona H. and {Wang}, Bingjie and {Weaver}, John R. and {Whitaker}, Katherine E. and {Atek}, Hakim and {Bogd{\'a}n}, {\'A}kos and {Charlot}, St{\'e}phane and {Curtis-Lake}, Emma and {van Dokkum}, Pieter and {Endsley}, Ryan and {Feldmann}, Robert and {Fudamoto}, Yoshinobu and {Fujimoto}, Seiji and {Glazebrook}, Karl and {Juneau}, St{\'e}phanie and {Marchesini}, Danilo and {Maseda}, Micheal V. and {Nelson}, Erica and {Oesch}, Pascal A. and {Plat}, Ad{\`e}le and {Setton}, David J. and {Stark}, Daniel P. and {Williams}, Christina C.},
        title = "{A high black-hole-to-host mass ratio in a lensed AGN in the early Universe}",
      journal = {\nat},
         year = 2024,
        month = apr,
       volume = {628},
       number = {8006},
        pages = {57-61},
          doi = {10.1038/s41586-024-07184-8},
archivePrefix = {arXiv},
       eprint = {2308.05735},
 primaryClass = {astro-ph.GA},
       adsurl = {https://ui.adsabs.harvard.edu/abs/2024Natur.628...57F}
}

@ARTICLE{greene23,
       author = {{Greene}, Jenny E. and {Labbe}, Ivo and {Goulding}, Andy D. and {Furtak}, Lukas J. and {Chemerynska}, Iryna and {Kokorev}, Vasily and {Dayal}, Pratika and {Volonteri}, Marta and {Williams}, Christina C. and {Wang}, Bingjie and {Setton}, David J. and {Burgasser}, Adam J. and {Bezanson}, Rachel and {Atek}, Hakim and {Brammer}, Gabriel and {Cutler}, Sam E. and {Feldmann}, Robert and {Fujimoto}, Seiji and {Glazebrook}, Karl and {de Graaff}, Anna and {Khullar}, Gourav and {Leja}, Joel and {Marchesini}, Danilo and {Maseda}, Michael V. and {Matthee}, Jorryt and {Miller}, Tim B. and {Naidu}, Rohan P. and {Nanayakkara}, Themiya and {Oesch}, Pascal A. and {Pan}, Richard and {Papovich}, Casey and {Price}, Sedona H. and {van Dokkum}, Pieter and {Weaver}, John R. and {Whitaker}, Katherine E. and {Zitrin}, Adi},
        title = "{UNCOVER Spectroscopy Confirms the Surprising Ubiquity of Active Galactic Nuclei in Red Sources at z > 5}",
      journal = {\apj},
         year = 2024,
        month = mar,
       volume = {964},
       number = {1},
          eid = {39},
        pages = {39},
          doi = {10.3847/1538-4357/ad1e5f},
archivePrefix = {arXiv},
       eprint = {2309.05714},
 primaryClass = {astro-ph.GA},
       adsurl = {https://ui.adsabs.harvard.edu/abs/2024ApJ...964...39G}
}

@ARTICLE{Harikane23,
       author = {{Harikane}, Yuichi and {Zhang}, Yechi and {Nakajima}, Kimihiko and {Ouchi}, Masami and {Isobe}, Yuki and {Ono}, Yoshiaki and {Hatano}, Shun and {Xu}, Yi and {Umeda}, Hiroya},
        title = "{A JWST/NIRSpec First Census of Broad-line AGNs at z = 4-7: Detection of 10 Faint AGNs with M $_{BH}$ {}10$^{6}$-{}10$^{8}$ M $_{{\ensuremath{\odot}}}$ and Their Host Galaxy Properties}",
      journal = {\apj},
         year = 2023,
        month = dec,
       volume = {959},
       number = {1},
          eid = {39},
        pages = {39},
          doi = {10.3847/1538-4357/ad029e},
archivePrefix = {arXiv},
       eprint = {2303.11946},
 primaryClass = {astro-ph.GA},
       adsurl = {https://ui.adsabs.harvard.edu/abs/2023ApJ...959...39H}
}

@ARTICLE{Curti_Jades_24,
       author = {{Curti}, Mirko and {Maiolino}, Roberto and {Curtis-Lake}, Emma and {Chevallard}, Jacopo and {Carniani}, Stefano and {D'Eugenio}, Francesco and {Looser}, Tobias J. and {Scholtz}, Jan and {Charlot}, Stephane and {Cameron}, Alex and {{\"U}bler}, Hannah and {Witstok}, Joris and {Boyett}, Kristian and {Laseter}, Isaac and {Sandles}, Lester and {Arribas}, Santiago and {Bunker}, Andrew and {Giardino}, Giovanna and {Maseda}, Michael V. and {Rawle}, Tim and {Rodr{\'\i}guez Del Pino}, Bruno and {Smit}, Renske and {Willott}, Chris J. and {Eisenstein}, Daniel J. and {Hausen}, Ryan and {Johnson}, Benjamin and {Rieke}, Marcia and {Robertson}, Brant and {Tacchella}, Sandro and {Williams}, Christina C. and {Willmer}, Christopher and {Baker}, William M. and {Bhatawdekar}, Rachana and {Egami}, Eiichi and {Helton}, Jakob M. and {Ji}, Zhiyuan and {Kumari}, Nimisha and {Perna}, Michele and {Shivaei}, Irene and {Sun}, Fengwu},
        title = "{JADES: Insights into the low-mass end of the mass-metallicity-SFR relation at 3 < z < 10 from deep JWST/NIRSpec spectroscopy}",
      journal = {\aap},
         year = 2024,
        month = apr,
       volume = {684},
          eid = {A75},
        pages = {A75},
          doi = {10.1051/0004-6361/202346698},
archivePrefix = {arXiv},
       eprint = {2304.08516},
 primaryClass = {astro-ph.GA},
       adsurl = {https://ui.adsabs.harvard.edu/abs/2024A&A...684A..75C}
}

@ARTICLE{reddy_ism_2023,
       author = {{Reddy}, Naveen A. and {Topping}, Michael W. and {Sanders}, Ryan L. and {Shapley}, Alice E. and {Brammer}, Gabriel},
        title = "{A JWST/NIRSpec Exploration of the Connection between Ionization Parameter, Electron Density, and Star-formation-rate Surface Density in z = 2.7-6.3 Galaxies}",
      journal = {\apj},
         year = 2023,
        month = aug,
       volume = {952},
       number = {2},
          eid = {167},
        pages = {167},
          doi = {10.3847/1538-4357/acd754},
archivePrefix = {arXiv},
       eprint = {2303.11397},
 primaryClass = {astro-ph.GA},
       adsurl = {https://ui.adsabs.harvard.edu/abs/2023ApJ...952..167R}
}

@ARTICLE{Schaerer22,
       author = {{Schaerer}, D. and {Marques-Chaves}, R. and {Barrufet}, L. and {Oesch}, P. and {Izotov}, Y.~I. and {Naidu}, R. and {Guseva}, N.~G. and {Brammer}, G.},
        title = "{First look with JWST spectroscopy: Resemblance among z {\ensuremath{\sim}} 8 galaxies and local analogs}",
      journal = {\aap},
         year = 2022,
        month = sep,
       volume = {665},
          eid = {L4},
        pages = {L4},
          doi = {10.1051/0004-6361/202244556},
archivePrefix = {arXiv},
       eprint = {2207.10034},
 primaryClass = {astro-ph.GA},
       adsurl = {https://ui.adsabs.harvard.edu/abs/2022A&A...665L...4S}
}

@ARTICLE{castellano_ghz12_2024,
       author = {{Castellano}, Marco and {Napolitano}, Lorenzo and {Fontana}, Adriano and {Roberts-Borsani}, Guido and {Treu}, Tommaso and {Vanzella}, Eros and {Zavala}, Jorge A. and {Arrabal Haro}, Pablo and {Calabr{\`o}}, Antonello and {Llerena}, Mario and {Mascia}, Sara and {Merlin}, Emiliano and {Paris}, Diego and {Pentericci}, Laura and {Santini}, Paola and {Bakx}, Tom J.~L.~C. and {Bergamini}, Pietro and {Cupani}, Guido and {Dickinson}, Mark and {Filippenko}, Alexei V. and {Glazebrook}, Karl and {Grillo}, Claudio and {Kelly}, Patrick L. and {Malkan}, Matthew A. and {Mason}, Charlotte A. and {Morishita}, Takahiro and {Nanayakkara}, Themiya and {Rosati}, Piero and {Sani}, Eleonora and {Wang}, Xin and {Yoon}, Ilsang},
        title = "{JWST NIRSpec Spectroscopy of the Remarkable Bright Galaxy GHZ2/GLASS-z12 at Redshift 12.34}",
      journal = {\apj},
         year = 2024,
        month = sep,
       volume = {972},
       number = {2},
          eid = {143},
        pages = {143},
          doi = {10.3847/1538-4357/ad5f88},
archivePrefix = {arXiv},
       eprint = {2403.10238},
 primaryClass = {astro-ph.GA},
       adsurl = {https://ui.adsabs.harvard.edu/abs/2024ApJ...972..143C}
}

@ARTICLE{Maiolino23JADES,
       author = {{Maiolino}, Roberto and {Scholtz}, Jan and {Curtis-Lake}, Emma and {Carniani}, Stefano and {Baker}, William and {de Graaff}, Anna and {Tacchella}, Sandro and {{\"U}bler}, Hannah and {D'Eugenio}, Francesco and {Witstok}, Joris and {Curti}, Mirko and {Arribas}, Santiago and {Bunker}, Andrew J. and {Charlot}, St{\'e}phane and {Chevallard}, Jacopo and {Eisenstein}, Daniel J. and {Egami}, Eiichi and {Ji}, Zhiyuan and {Jones}, Gareth C. and {Lyu}, Jianwei and {Rawle}, Tim and {Robertson}, Brant and {Rujopakarn}, Wiphu and {Perna}, Michele and {Sun}, Fengwu and {Venturi}, Giacomo and {Williams}, Christina C. and {Willott}, Chris},
        title = "{JADES: The diverse population of infant black holes at 4 < z < 11: Merging, tiny, poor, but mighty}",
      journal = {\aap},
         year = 2024,
        month = nov,
       volume = {691},
          eid = {A145},
        pages = {A145},
          doi = {10.1051/0004-6361/202347640},
archivePrefix = {arXiv},
       eprint = {2308.01230},
 primaryClass = {astro-ph.GA},
       adsurl = {https://ui.adsabs.harvard.edu/abs/2024A&A...691A.145M}
}

@article{Onoue23,
	title = {{A Candidate for the Least-massive Black Hole in the First 1.1 Billion Years of the Universe}},
	author = {{Onoue}, Masafusa and {Inayoshi}, Kohei and {Ding}, Xuheng and {Li}, Wenxiu and {Li}, Zhengrong and {Molina}, Juan and {Inoue}, Akio K. and {Jiang}, Linhua and {Ho}, Luis C.},
	year = 2023,
	month = jan,
	journal = {\apjl},
	volume = 942,
	number = 1,
	pages = {L17},
	doi = {10.3847/2041-8213/aca9d3},
	eid = {L17},
	archiveprefix = {arXiv},
	eprint = {2209.07325},
	primaryclass = {astro-ph.GA},
	adsurl = {https://ui.adsabs.harvard.edu/abs/2023ApJ...942L..17O}
}

@ARTICLE{Matthee23,
       author = {{Matthee}, Jorryt and {Naidu}, Rohan P. and {Brammer}, Gabriel and {Chisholm}, John and {Eilers}, Anna-Christina and {Goulding}, Andy and {Greene}, Jenny and {Kashino}, Daichi and {Labbe}, Ivo and {Lilly}, Simon J. and {Mackenzie}, Ruari and {Oesch}, Pascal A. and {Weibel}, Andrea and {Wuyts}, Stijn and {Xiao}, Mengyuan and {Bordoloi}, Rongmon and {Bouwens}, Rychard and {van Dokkum}, Pieter and {Illingworth}, Garth and {Kramarenko}, Ivan and {Maseda}, Michael V. and {Mason}, Charlotte and {Meyer}, Romain A. and {Nelson}, Erica J. and {Reddy}, Naveen A. and {Shivaei}, Irene and {Simcoe}, Robert A. and {Yue}, Minghao},
        title = "{Little Red Dots: An Abundant Population of Faint Active Galactic Nuclei at z {\ensuremath{\sim}} 5 Revealed by the EIGER and FRESCO JWST Surveys}",
      journal = {\apj},
         year = 2024,
        month = mar,
       volume = {963},
       number = {2},
          eid = {129},
        pages = {129},
          doi = {10.3847/1538-4357/ad2345},
archivePrefix = {arXiv},
       eprint = {2306.05448},
 primaryClass = {astro-ph.GA},
       adsurl = {https://ui.adsabs.harvard.edu/abs/2024ApJ...963..129M}
}

@ARTICLE{Ubler23,
       author = {{{\"U}bler}, Hannah and {Maiolino}, Roberto and {Curtis-Lake}, Emma and {P{\'e}rez-Gonz{\'a}lez}, Pablo G. and {Curti}, Mirko and {Perna}, Michele and {Arribas}, Santiago and {Charlot}, St{\'e}phane and {Marshall}, Madeline A. and {D'Eugenio}, Francesco and {Scholtz}, Jan and {Bunker}, Andrew and {Carniani}, Stefano and {Ferruit}, Pierre and {Jakobsen}, Peter and {Rix}, Hans-Walter and {Rodr{\'\i}guez Del Pino}, Bruno and {Willott}, Chris J. and {Boeker}, Torsten and {Cresci}, Giovanni and {Jones}, Gareth C. and {Kumari}, Nimisha and {Rawle}, Tim},
        title = "{GA-NIFS: A massive black hole in a low-metallicity AGN at z {\ensuremath{\sim}} 5.55 revealed by JWST/NIRSpec IFS}",
      journal = {\aap},
         year = 2023,
        month = sep,
       volume = {677},
          eid = {A145},
        pages = {A145},
          doi = {10.1051/0004-6361/202346137},
archivePrefix = {arXiv},
       eprint = {2302.06647},
 primaryClass = {astro-ph.GA},
       adsurl = {https://ui.adsabs.harvard.edu/abs/2023A&A...677A.145U}
}

@ARTICLE{Kocevski23,
       author = {{Kocevski}, Dale D. and {Onoue}, Masafusa and {Inayoshi}, Kohei and {Trump}, Jonathan R. and {Arrabal Haro}, Pablo and {Grazian}, Andrea and {Dickinson}, Mark and {Finkelstein}, Steven L. and {Kartaltepe}, Jeyhan S. and {Hirschmann}, Michaela and {Aird}, James and {Holwerda}, Benne W. and {Fujimoto}, Seiji and {Juneau}, St{\'e}phanie and {Amor{\'\i}n}, Ricardo O. and {Backhaus}, Bren E. and {Bagley}, Micaela B. and {Barro}, Guillermo and {Bell}, Eric F. and {Bisigello}, Laura and {Calabr{\`o}}, Antonello and {Cleri}, Nikko J. and {Cooper}, M.~C. and {Ding}, Xuheng and {Grogin}, Norman A. and {Ho}, Luis C. and {Hutchison}, Taylor A. and {Inoue}, Akio K. and {Jiang}, Linhua and {Jones}, Brenda and {Koekemoer}, Anton M. and {Li}, Wenxiu and {Li}, Zhengrong and {McGrath}, Elizabeth J. and {Molina}, Juan and {Papovich}, Casey and {P{\'e}rez-Gonz{\'a}lez}, Pablo G. and {Pirzkal}, Nor and {Wilkins}, Stephen M. and {Yang}, Guang and {Yung}, L.~Y. Aaron},
        title = "{Hidden Little Monsters: Spectroscopic Identification of Low-mass, Broad-line AGNs at z > 5 with CEERS}",
      journal = {\apjl},
         year = 2023,
        month = sep,
       volume = {954},
       number = {1},
          eid = {L4},
        pages = {L4},
          doi = {10.3847/2041-8213/ace5a0},
archivePrefix = {arXiv},
       eprint = {2302.00012},
 primaryClass = {astro-ph.GA},
       adsurl = {https://ui.adsabs.harvard.edu/abs/2023ApJ...954L...4K}
}

@ARTICLE{dressler24,
       author = {{Dressler}, Alan and {Rieke}, Marcia and {Eisenstein}, Daniel and {Stark}, Daniel P. and {Burns}, Chris and {Bhatawdekar}, Rachana and {Bonaventura}, Nina and {Boyett}, Kristan and {Bunker}, Andrew J. and {Carniani}, Stefano and {Charlot}, Stephane and {Hausen}, Ryan and {Misselt}, Karl and {Tacchella}, Sandro and {Willmer}, Christopher},
        title = "{Building the First Galaxies{\textemdash}Chapter 2. Starbursts Dominate the Star Formation Histories of 6 < z < 12 Galaxies}",
      journal = {\apj},
         year = 2024,
        month = apr,
       volume = {964},
       number = {2},
          eid = {150},
        pages = {150},
          doi = {10.3847/1538-4357/ad1923},
archivePrefix = {arXiv},
       eprint = {2306.02469},
 primaryClass = {astro-ph.GA},
       adsurl = {https://ui.adsabs.harvard.edu/abs/2024ApJ...964..150D}
}

@ARTICLE{endsley_sfh_2023,
       author = {{Endsley}, Ryan and {Stark}, Daniel P. and {Whitler}, Lily and {Topping}, Michael W. and {Chen}, Zuyi and {Plat}, Ad{\`e}le and {Chisholm}, John and {Charlot}, St{\'e}phane},
        title = "{A JWST/NIRCam study of key contributors to reionization: the star-forming and ionizing properties of UV-faint z   7-8 galaxies}",
      journal = {\mnras},
         year = 2023,
        month = sep,
       volume = {524},
       number = {2},
        pages = {2312-2330},
          doi = {10.1093/mnras/stad1919},
archivePrefix = {arXiv},
       eprint = {2208.14999},
 primaryClass = {astro-ph.GA},
       adsurl = {https://ui.adsabs.harvard.edu/abs/2023MNRAS.524.2312E}
}

@ARTICLE{Looser23a,
       author = {{Looser}, Tobias J. and {D'Eugenio}, Francesco and {Maiolino}, Roberto and {Witstok}, Joris and {Sandles}, Lester and {Curtis-Lake}, Emma and {Chevallard}, Jacopo and {Tacchella}, Sandro and {Johnson}, Benjamin D. and {Baker}, William M. and {Suess}, Katherine A. and {Carniani}, Stefano and {Ferruit}, Pierre and {Arribas}, Santiago and {Bonaventura}, Nina and {Bunker}, Andrew J. and {Cameron}, Alex J. and {Charlot}, Stephane and {Curti}, Mirko and {de Graaff}, Anna and {Maseda}, Michael V. and {Rawle}, Tim and {Rix}, Hans-Walter and {Del Pino}, Bruno Rodr{\'\i}guez and {Smit}, Renske and {{\"U}bler}, Hannah and {Willott}, Chris and {Alberts}, Stacey and {Egami}, Eiichi and {Eisenstein}, Daniel J. and {Endsley}, Ryan and {Hausen}, Ryan and {Rieke}, Marcia and {Robertson}, Brant and {Shivaei}, Irene and {Williams}, Christina C. and {Boyett}, Kristan and {Chen}, Zuyi and {Ji}, Zhiyuan and {Jones}, Gareth C. and {Kumari}, Nimisha and {Nelson}, Erica and {Perna}, Michele and {Saxena}, Aayush and {Scholtz}, Jan},
        title = "{A recently quenched galaxy 700 million years after the Big Bang}",
      journal = {\nat},
         year = 2024,
        month = may,
       volume = {629},
       number = {8010},
        pages = {53-57},
          doi = {10.1038/s41586-024-07227-0},
archivePrefix = {arXiv},
       eprint = {2302.14155},
 primaryClass = {astro-ph.GA},
       adsurl = {https://ui.adsabs.harvard.edu/abs/2024Natur.629...53L}
}

@ARTICLE{robertson_JOF_LF_2023,
       author = {{Robertson}, Brant and {Johnson}, Benjamin D. and {Tacchella}, Sandro and {Eisenstein}, Daniel J. and {Hainline}, Kevin and {Arribas}, Santiago and {Baker}, William M. and {Bunker}, Andrew J. and {Carniani}, Stefano and {Cargile}, Phillip A. and {Carreira}, Courtney and {Charlot}, Stephane and {Chevallard}, Jacopo and {Curti}, Mirko and {Curtis-Lake}, Emma and {D'Eugenio}, Francesco and {Egami}, Eiichi and {Hausen}, Ryan and {Helton}, Jakob M. and {Jakobsen}, Peter and {Ji}, Zhiyuan and {Jones}, Gareth C. and {Maiolino}, Roberto and {Maseda}, Michael V. and {Nelson}, Erica and {P{\'e}rez-Gonz{\'a}lez}, Pablo G. and {Pusk{\'a}s}, D{\'a}vid and {Rieke}, Marcia and {Smit}, Renske and {Sun}, Fengwu and {{\"U}bler}, Hannah and {Whitler}, Lily and {Williams}, Christina C. and {Willmer}, Christopher N.~A. and {Willott}, Chris and {Witstok}, Joris},
        title = "{Earliest Galaxies in the JADES Origins Field: Luminosity Function and Cosmic Star Formation Rate Density 300 Myr after the Big Bang}",
      journal = {\apj},
         year = 2024,
        month = jul,
       volume = {970},
       number = {1},
          eid = {31},
        pages = {31},
          doi = {10.3847/1538-4357/ad463d},
archivePrefix = {arXiv},
       eprint = {2312.10033},
 primaryClass = {astro-ph.GA},
       adsurl = {https://ui.adsabs.harvard.edu/abs/2024ApJ...970...31R}
}

@ARTICLE{carniani_z14_2024,
       author = {{Carniani}, Stefano and {Hainline}, Kevin and {D'Eugenio}, Francesco and {Eisenstein}, Daniel J. and {Jakobsen}, Peter and {Witstok}, Joris and {Johnson}, Benjamin D. and {Chevallard}, Jacopo and {Maiolino}, Roberto and {Helton}, Jakob M. and {Willott}, Chris and {Robertson}, Brant and {Alberts}, Stacey and {Arribas}, Santiago and {Baker}, William M. and {Bhatawdekar}, Rachana and {Boyett}, Kristan and {Bunker}, Andrew J. and {Cameron}, Alex J. and {Cargile}, Phillip A. and {Charlot}, St{\'e}phane and {Curti}, Mirko and {Curtis-Lake}, Emma and {Egami}, Eiichi and {Giardino}, Giovanna and {Isaak}, Kate and {Ji}, Zhiyuan and {Jones}, Gareth C. and {Kumari}, Nimisha and {Maseda}, Michael V. and {Parlanti}, Eleonora and {P{\'e}rez-Gonz{\'a}lez}, Pablo G. and {Rawle}, Tim and {Rieke}, George and {Rieke}, Marcia and {Del Pino}, Bruno Rodr{\'\i}guez and {Saxena}, Aayush and {Scholtz}, Jan and {Smit}, Renske and {Sun}, Fengwu and {Tacchella}, Sandro and {{\"U}bler}, Hannah and {Venturi}, Giacomo and {Williams}, Christina C. and {Willmer}, Christopher N.~A.},
        title = "{Spectroscopic confirmation of two luminous galaxies at a redshift of 14}",
      journal = {\nat},
         year = 2024,
        month = sep,
       volume = {633},
       number = {8029},
        pages = {318-322},
          doi = {10.1038/s41586-024-07860-9},
archivePrefix = {arXiv},
       eprint = {2405.18485},
 primaryClass = {astro-ph.GA},
       adsurl = {https://ui.adsabs.harvard.edu/abs/2024Natur.633..318C}
}

@ARTICLE{d_eugenio_gsz12_2023,
       author = {{D'Eugenio}, Francesco and {Maiolino}, Roberto and {Carniani}, Stefano and {Chevallard}, Jacopo and {Curtis-Lake}, Emma and {Witstok}, Joris and {Charlot}, Stephane and {Baker}, William M. and {Arribas}, Santiago and {Boyett}, Kristan and {Bunker}, Andrew J. and {Curti}, Mirko and {Eisenstein}, Daniel J. and {Hainline}, Kevin and {Ji}, Zhiyuan and {Johnson}, Benjamin D. and {Kumari}, Nimisha and {Looser}, Tobias J. and {Nakajima}, Kimihiko and {Nelson}, Erica and {Rieke}, Marcia and {Robertson}, Brant and {Scholtz}, Jan and {Smit}, Renske and {Sun}, Fengwu and {Venturi}, Giacomo and {Tacchella}, Sandro and {{\"U}bler}, Hannah and {Willmer}, Christopher N.~A. and {Willott}, Chris},
        title = "{JADES: Carbon enrichment 350 Myr after the Big Bang}",
      journal = {\aap},
         year = 2024,
        month = sep,
       volume = {689},
          eid = {A152},
        pages = {A152},
          doi = {10.1051/0004-6361/202348636},
archivePrefix = {arXiv},
       eprint = {2311.09908},
 primaryClass = {astro-ph.GA},
       adsurl = {https://ui.adsabs.harvard.edu/abs/2024A&A...689A.152D}
}

@ARTICLE{Glazebrook24,
       author = {{Glazebrook}, Karl and {Nanayakkara}, Themiya and {Schreiber}, Corentin and {Lagos}, Claudia and {Kawinwanichakij}, Lalitwadee and {Jacobs}, Colin and {Chittenden}, Harry and {Brammer}, Gabriel and {Kacprzak}, Glenn G. and {Labbe}, Ivo and {Marchesini}, Danilo and {Marsan}, Z. Cemile and {Oesch}, Pascal A. and {Papovich}, Casey and {Remus}, Rhea-Silvia and {Tran}, Kim-Vy H. and {Esdaile}, James and {Chandro-Gomez}, Angel},
        title = "{A massive galaxy that formed its stars at z {\ensuremath{\approx}} 11}",
      journal = {\nat},
         year = 2024,
        month = apr,
       volume = {628},
       number = {8007},
        pages = {277-281},
          doi = {10.1038/s41586-024-07191-9},
archivePrefix = {arXiv},
       eprint = {2308.05606},
 primaryClass = {astro-ph.GA},
       adsurl = {https://ui.adsabs.harvard.edu/abs/2024Natur.628..277G}
}

@ARTICLE{Taylor25,
       author = {{Taylor}, Anthony J. and {Kokorev}, Vasily and {Kocevski}, Dale D. and {Akins}, Hollis B. and {Cullen}, Fergus and {Dickinson}, Mark and {Finkelstein}, Steven L. and {Arrabal Haro}, Pablo and {Bromm}, Volker and {Giavalisco}, Mauro and {Inayoshi}, Kohei and {Juneau}, St{\'e}phanie and {Leung}, Gene C.~K. and {P{\'e}rez-Gonz{\'a}lez}, Pablo G. and {Somerville}, Rachel S. and {Trump}, Jonathan R. and {Amor{\'\i}n}, Ricardo O. and {Barro}, Guillermo and {Burgarella}, Denis and {Brooks}, Madisyn and {Carnall}, Adam C. and {Casey}, Caitlin M. and {Cheng}, Yingjie and {Chisholm}, John and {Chworowsky}, Katherine and {Davis}, Kelcey and {Donnan}, Callum T. and {Dunlop}, James S. and {Ellis}, Richard S. and {Fern{\'a}ndez}, Vital and {Fujimoto}, Seiji and {Grogin}, Norman A. and {Gupta}, Ansh R. and {Hathi}, Nimish P. and {Jung}, Intae and {Hirschmann}, Michaela and {Kartaltepe}, Jeyhan S. and {Koekemoer}, Anton M. and {Larson}, Rebecca L. and {Leung}, Ho-Hin and {Llerena}, Mario and {Lucas}, Ray A. and {McLeod}, Derek J. and {McLure}, Ross and {Napolitano}, Lorenzo and {Papovich}, Casey and {Stanton}, Thomas M. and {Tripodi}, Roberta and {Wang}, Xin and {Wilkins}, Stephen M. and {Yung}, L.~Y. Aaron and {Zavala}, Jorge A.},
        title = "{CAPERS-LRD-z9: A Gas-enshrouded Little Red Dot Hosting a Broad-line Active Galactic Nucleus at z = 9.288}",
      journal = {\apjl},
         year = 2025,
        month = aug,
       volume = {989},
       number = {1},
          eid = {L7},
        pages = {L7},
          doi = {10.3847/2041-8213/ade789},
archivePrefix = {arXiv},
       eprint = {2505.04609},
 primaryClass = {astro-ph.GA},
       adsurl = {https://ui.adsabs.harvard.edu/abs/2025ApJ...989L...7T}
}

@ARTICLE{Juodzbalis25,
       author = {{Juod{\v{z}}balis}, Ignas and {Maiolino}, Roberto and {Baker}, William M. and {Lake}, Emma Curtis and {Scholtz}, Jan and {D'Eugenio}, Francesco and {Trefoloni}, Bartolomeo and {Isobe}, Yuki and {Tacchella}, Sandro and {Bunker}, Andrew J. and {Carniani}, Stefano and {Charlot}, St{\'e}phane and {Jones}, Gareth C. and {Parlanti}, Eleonora and {Perna}, Michele and {Rinaldi}, Pierluigi and {Robertson}, Brant and {{\"U}bler}, Hannah and {Venturi}, Giacomo and {Willott}, Chris},
        title = "{JADES: comprehensive census of broad-line AGN from Reionization to Cosmic Noon revealed by JWST}",
      journal = {arXiv e-prints},
         year = 2025,
        month = apr,
          eid = {arXiv:2504.03551},
        pages = {arXiv:2504.03551},
          doi = {10.48550/arXiv.2504.03551},
archivePrefix = {arXiv},
       eprint = {2504.03551},
 primaryClass = {astro-ph.GA},
       adsurl = {https://ui.adsabs.harvard.edu/abs/2025arXiv250403551J}
}

@ARTICLE{Harrison16,
       author = {{Harrison}, C.~M. and {Alexander}, D.~M. and {Mullaney}, J.~R. and {Stott}, J.~P. and {Swinbank}, A.~M. and {Arumugam}, V. and {Bauer}, F.~E. and {Bower}, R.~G. and {Bunker}, A.~J. and {Sharples}, R.~M.},
        title = "{The KMOS AGN Survey at High redshift (KASHz): the prevalence and drivers of ionized outflows in the host galaxies of X-ray AGN}",
      journal = {\mnras},
         year = 2016,
        month = feb,
       volume = {456},
       number = {2},
        pages = {1195-1220},
          doi = {10.1093/mnras/stv2727},
archivePrefix = {arXiv},
       eprint = {1511.00008},
 primaryClass = {astro-ph.GA},
       adsurl = {https://ui.adsabs.harvard.edu/abs/2016MNRAS.456.1195H}
}

@ARTICLE{Marshall24,
       author = {{Marshall}, Madeline A. and {Yue}, Minghao and {Eilers}, Anna-Christina and {Scholtz}, Jan and {Perna}, Michele and {Willott}, Chris J. and {Maiolino}, Roberto and {{\"U}bler}, Hannah and {Arribas}, Santiago and {Bunker}, Andrew J. and {Charlot}, Stephane and {Rodr{\'\i}guez Del Pino}, Bruno and {B{\"o}ker}, Torsten and {Carniani}, Stefano and {Circosta}, Chiara and {Cresci}, Giovanni and {D'Eugenio}, Francesco and {Jones}, Gareth C. and {Venturi}, Giacomo and {Bordoloi}, Rongmon and {Kashino}, Daichi and {Mackenzie}, Ruari and {Matthee}, Jorryt and {Naidu}, Rohan and {Simcoe}, Robert A.},
        title = "{GA-NIFS and EIGER: A merging quasar host at z = 7 with an overmassive black hole}",
      journal = {\aap},
         year = 2025,
        month = oct,
       volume = {702},
          eid = {A50},
        pages = {A50},
          doi = {10.1051/0004-6361/202452650},
archivePrefix = {arXiv},
       eprint = {2410.11035},
 primaryClass = {astro-ph.GA},
       adsurl = {https://ui.adsabs.harvard.edu/abs/2025A&A...702A..50M}
}

@ARTICLE{Cresci2023,
       author = {{Cresci}, G. and {Tozzi}, G. and {Perna}, M. and {Brusa}, M. and {Marconcini}, C. and {Marconi}, A. and {Carniani}, S. and {Brienza}, M. and {Giroletti}, M. and {Belfiore}, F. and {Ginolfi}, M. and {Mannucci}, F. and {Ulivi}, L. and {Scholtz}, J. and {Venturi}, G. and {Arribas}, S. and {{\"U}bler}, H. and {D'Eugenio}, F. and {Mingozzi}, M. and {Balmaverde}, B. and {Capetti}, A. and {Parlanti}, E. and {Zana}, T.},
        title = "{Bubbles and outflows: The novel JWST/NIRSpec view of the z = 1.59 obscured quasar XID2028}",
      journal = {\aap},
         year = 2023,
        month = apr,
       volume = {672},
          eid = {A128},
        pages = {A128},
          doi = {10.1051/0004-6361/202346001},
archivePrefix = {arXiv},
       eprint = {2301.11060},
 primaryClass = {astro-ph.GA},
       adsurl = {https://ui.adsabs.harvard.edu/abs/2023A&A...672A.128C}
}

@ARTICLE{Bischetti2017,
       author = {{Bischetti}, M. and {Piconcelli}, E. and {Vietri}, G. and {Bongiorno}, A. and {Fiore}, F. and {Sani}, E. and {Marconi}, A. and {Duras}, F. and {Zappacosta}, L. and {Brusa}, M. and {Comastri}, A. and {Cresci}, G. and {Feruglio}, C. and {Giallongo}, E. and {La Franca}, F. and {Mainieri}, V. and {Mannucci}, F. and {Martocchia}, S. and {Ricci}, F. and {Schneider}, R. and {Testa}, V. and {Vignali}, C.},
        title = "{The WISSH quasars project. I. Powerful ionised outflows in hyper-luminous quasars}",
      journal = {\aap},
         year = 2017,
        month = feb,
       volume = {598},
          eid = {A122},
        pages = {A122},
          doi = {10.1051/0004-6361/201629301},
archivePrefix = {arXiv},
       eprint = {1612.03728},
 primaryClass = {astro-ph.GA},
       adsurl = {https://ui.adsabs.harvard.edu/abs/2017A&A...598A.122B}
}

@ARTICLE{Scholtz25-COS3018,
       author = {{Scholtz}, J. and {Curti}, M. and {D'Eugenio}, F. and {{\"U}bler}, H. and {Maiolino}, R. and {Marconcini}, C. and {Smit}, R. and {Perna}, M. and {Witstok}, J. and {Arribas}, S. and {B{\"o}ker}, T. and {Bunker}, A.~J. and {Carniani}, S. and {Charlot}, S. and {Cresci}, G. and {Lamperti}, I. and {Parlanti}, E. and {P{\'e}rez-Gonz{\'a}lez}, P.~G. and {Rodr{\'\i}guez Del Pino}, B. and {Venturi}, G.},
        title = "{GA-NIFS: ISM properties and metal enrichment in a merger-driven starburst during the epoch of reionization probed with JWST and ALMA}",
      journal = {\mnras},
         year = 2025,
        month = may,
       volume = {539},
       number = {3},
        pages = {2463-2484},
          doi = {10.1093/mnras/staf518},
archivePrefix = {arXiv},
       eprint = {2411.07695},
 primaryClass = {astro-ph.GA},
       adsurl = {https://ui.adsabs.harvard.edu/abs/2025MNRAS.539.2463S}
}

@ARTICLE{Ji2024ganifs,
       author = {{Ji}, Xihan and {{\"U}bler}, Hannah and {Maiolino}, Roberto and {D'Eugenio}, Francesco and {Arribas}, Santiago and {Bunker}, Andrew J. and {Charlot}, St{\'e}phane and {Perna}, Michele and {Rodr{\'\i}guez Del Pino}, Bruno and {B{\"o}ker}, Torsten and {Cresci}, Giovanni and {Curti}, Mirko and {Kumari}, Nimisha and {Lamperti}, Isabella},
        title = "{GA-NIFS: an extremely nitrogen-loud and chemically stratified galaxy at z   5.55}",
      journal = {\mnras},
         year = 2024,
        month = nov,
       volume = {535},
       number = {1},
        pages = {881-908},
          doi = {10.1093/mnras/stae2375},
archivePrefix = {arXiv},
       eprint = {2404.04148},
 primaryClass = {astro-ph.GA},
       adsurl = {https://ui.adsabs.harvard.edu/abs/2024MNRAS.535..881J}
}

@ARTICLE{Arribas2024,
       author = {{Arribas}, Santiago and {Perna}, Michele and {Rodr{\'\i}guez Del Pino}, Bruno and {Lamperti}, Isabella and {D'Eugenio}, Francesco and {P{\'e}rez-Gonz{\'a}lez}, Pablo G. and {Jones}, Gareth C. and {Crespo G{\'o}mez}, Alejandro and {Curti}, Mirko and {Lim}, Seunghwan and {{\'A}lvarez-M{\'a}rquez}, Javier and {Bunker}, Andrew J. and {Carniani}, Stefano and {Charlot}, St{\'e}phane and {Jakobsen}, Peter and {Maiolino}, Roberto and {{\"U}bler}, Hannah and {Willott}, Chris J. and {B{\"o}ker}, Torsten and {Chevallard}, Jacopo and {Circosta}, Chiara and {Cresci}, Giovanni and {Kumari}, Nimisha and {Parlanti}, Eleonora and {Scholtz}, Jan and {Venturi}, Giacomo and {Witstok}, Joris},
        title = "{GA-NIFS: The core of an extremely massive protocluster at the epoch of reionisation probed with JWST/NIRSpec}",
      journal = {\aap},
         year = 2024,
        month = aug,
       volume = {688},
          eid = {A146},
        pages = {A146},
          doi = {10.1051/0004-6361/202348824},
archivePrefix = {arXiv},
       eprint = {2312.00899},
 primaryClass = {astro-ph.GA},
       adsurl = {https://ui.adsabs.harvard.edu/abs/2024A&A...688A.146A}
}

@ARTICLE{perez-gonzalez-jekyll,
       author = {{P{\'e}rez-Gonz{\'a}lez}, Pablo G. and {D'Eugenio}, Francesco and {Rodr{\'\i}guez del Pino}, Bruno and {Perna}, Michele and {{\"U}bler}, Hannah and {Maiolino}, Roberto and {Arribas}, Santiago and {Cresci}, Giovanni and {Lamperti}, Isabella and {Bunker}, Andrew J. and {Carniani}, Stefano and {Charlot}, Stephane and {Willott}, Chris J. and {B{\"o}ker}, Torsten and {Parlanti}, Eleonora and {Scholtz}, Jan and {Venturi}, Giacomo and {Barro}, Guillermo and {Costantin}, Luca and {Mart{\'\i}n-Navarro}, Ignacio and {Dunlop}, James S. and {Magee}, Daniel},
        title = "{Accelerated quenching and chemical enhancement of massive galaxies in a z {\ensuremath{\approx}} 4 gas-rich halo}",
      journal = {Nature Astronomy},
         year = 2025,
        month = aug,
       volume = {9},
        pages = {1240-1255},
          doi = {10.1038/s41550-025-02586-8},
archivePrefix = {arXiv},
       eprint = {2405.03744},
 primaryClass = {astro-ph.GA},
       adsurl = {https://ui.adsabs.harvard.edu/abs/2025NatAs...9.1240P}
}

@ARTICLE{Ubler24,
       author = {{{\"U}bler}, Hannah and {Maiolino}, Roberto and {P{\'e}rez-Gonz{\'a}lez}, Pablo G. and {D'Eugenio}, Francesco and {Perna}, Michele and {Curti}, Mirko and {Arribas}, Santiago and {Bunker}, Andrew and {Carniani}, Stefano and {Charlot}, St{\'e}phane and {Rodr{\'\i}guez Del Pino}, Bruno and {Baker}, William and {B{\"o}ker}, Torsten and {Cresci}, Giovanni and {Dunlop}, James and {Grogin}, Norman A. and {Jones}, Gareth C. and {Kumari}, Nimisha and {Lamperti}, Isabella and {Laporte}, Nicolas and {Marshall}, Madeline A. and {Mazzolari}, Giovanni and {Parlanti}, Eleonora and {Rawle}, Tim and {Scholtz}, Jan and {Venturi}, Giacomo and {Witstok}, Joris},
        title = "{GA-NIFS: JWST discovers an offset AGN 740 million years after the big bang}",
      journal = {\mnras},
         year = 2024,
        month = jun,
       volume = {531},
       number = {1},
        pages = {355-365},
          doi = {10.1093/mnras/stae943},
archivePrefix = {arXiv},
       eprint = {2312.03589},
 primaryClass = {astro-ph.GA},
       adsurl = {https://ui.adsabs.harvard.edu/abs/2024MNRAS.531..355U}
}

@ARTICLE{Pascale2022,
       author = {{Pascale}, Massimo and {Frye}, Brenda L. and {Diego}, Jose and {Furtak}, Lukas J. and {Zitrin}, Adi and {Broadhurst}, Tom and {Conselice}, Christopher J. and {Dai}, Liang and {Ferreira}, Leonardo and {Adams}, Nathan J. and {Kamieneski}, Patrick and {Foo}, Nicholas and {Kelly}, Patrick and {Chen}, Wenlei and {Lim}, Jeremy and {Meena}, Ashish K. and {Wilkins}, Stephen M. and {Bhatawdekar}, Rachana and {Windhorst}, Rogier A.},
        title = "{Unscrambling the Lensed Galaxies in JWST Images behind SMACS 0723}",
      journal = {\apjl},
         year = 2022,
        month = oct,
       volume = {938},
       number = {1},
          eid = {L6},
        pages = {L6},
          doi = {10.3847/2041-8213/ac9316},
archivePrefix = {arXiv},
       eprint = {2207.07102},
 primaryClass = {astro-ph.GA},
       adsurl = {https://ui.adsabs.harvard.edu/abs/2022ApJ...938L...6P}
}

@ARTICLE{Caminha22,
       author = {{Caminha}, G.~B. and {Suyu}, S.~H. and {Mercurio}, A. and {Brammer}, G. and {Bergamini}, P. and {Acebron}, A. and {Vanzella}, E.},
        title = "{First JWST observations of a gravitational lens. Mass model from new multiple images with near-infrared observations of SMACS J0723.3{\ensuremath{-}}7327}",
      journal = {\aap},
         year = 2022,
        month = oct,
       volume = {666},
          eid = {L9},
        pages = {L9},
          doi = {10.1051/0004-6361/202244517},
archivePrefix = {arXiv},
       eprint = {2207.07567},
 primaryClass = {astro-ph.GA},
       adsurl = {https://ui.adsabs.harvard.edu/abs/2022A&A...666L...9C}
}

@ARTICLE{Davies2020,
       author = {{Davies}, Rebecca L. and {F{\"o}rster Schreiber}, N.~M. and {Lutz}, D. and {Genzel}, R. and {Belli}, S. and {Shimizu}, T.~T. and {Contursi}, A. and {Davies}, R.~I. and {Herrera-Camus}, R. and {Lee}, M.~M. and {Naab}, T. and {Price}, S.~H. and {Renzini}, A. and {Schruba}, A. and {Sternberg}, A. and {Tacconi}, L.~J. and {{\"U}bler}, H. and {Wisnioski}, E. and {Wuyts}, S.},
        title = "{From Nuclear to Circumgalactic: Zooming in on AGN-driven Outflows at z {\ensuremath{\sim}} 2.2 with SINFONI}",
      journal = {\apj},
         year = 2020,
        month = may,
       volume = {894},
       number = {1},
          eid = {28},
        pages = {28},
          doi = {10.3847/1538-4357/ab86ad},
archivePrefix = {arXiv},
       eprint = {2004.02891},
 primaryClass = {astro-ph.GA},
       adsurl = {https://ui.adsabs.harvard.edu/abs/2020ApJ...894...28D}
}

@ARTICLE{Pu2014,
       author = {{Du}, Pu and {Wang}, Jian-Min and {Hu}, Chen and {Valls-Gabaud}, David and {Baldwin}, Jack A. and {Ge}, Jun-Qiang and {Xue}, Sui-Jian},
        title = "{Outflows from active galactic nuclei: the BLR-NLR metallicity correlation}",
      journal = {\mnras},
         year = 2014,
        month = mar,
       volume = {438},
       number = {4},
        pages = {2828-2838},
          doi = {10.1093/mnras/stt2386},
archivePrefix = {arXiv},
       eprint = {1312.3212},
 primaryClass = {astro-ph.GA},
       adsurl = {https://ui.adsabs.harvard.edu/abs/2014MNRAS.438.2828D}
}

@ARTICLE{Cooper2025,
       author = {{Cooper}, Ryan A. and {Caputi}, Karina I. and {Iani}, Edoardo and {Rinaldi}, Pierluigi and {Desprez}, Guillaume and {Navarro-Carrera}, Rafael},
        title = "{High-velocity outflows in [OIII] emitters at z=2.5-9 from JWST NIRSpec medium-resolution spectroscopy}",
      journal = {arXiv e-prints},
         year = 2025,
        month = feb,
          eid = {arXiv:2502.18310},
        pages = {arXiv:2502.18310},
          doi = {10.48550/arXiv.2502.18310},
archivePrefix = {arXiv},
       eprint = {2502.18310},
 primaryClass = {astro-ph.GA},
       adsurl = {https://ui.adsabs.harvard.edu/abs/2025arXiv250218310C}
}

@ARTICLE{Xu2025,
       author = {{Xu}, Yi and {Ouchi}, Masami and {Nakajima}, Kimihiko and {Harikane}, Yuichi and {Isobe}, Yuki and {Ono}, Yoshiaki and {Umeda}, Hiroya and {Zhang}, Yechi},
        title = "{Stellar- and AGN-driven Outflows in JWST Galaxies at z = 3{\textendash}9: More Frequent, Wider Opening Angles, and Mostly Bounded}",
      journal = {\apj},
         year = 2025,
        month = may,
       volume = {984},
       number = {2},
          eid = {182},
        pages = {182},
          doi = {10.3847/1538-4357/adc733},
archivePrefix = {arXiv},
       eprint = {2310.06614},
 primaryClass = {astro-ph.GA},
       adsurl = {https://ui.adsabs.harvard.edu/abs/2025ApJ...984..182X}
}

@ARTICLE{Marconcini24,
       author = {{Marconcini}, C. and {D'Eugenio}, F. and {Maiolino}, R. and {Arribas}, S. and {Bunker}, A. and {Carniani}, S. and {Charlot}, S. and {Perna}, M. and {Rodr{\'\i}guez Del Pino}, B. and {{\"U}bler}, H. and {Willott}, C.~J. and {B{\"o}ker}, T. and {Cresci}, G. and {Curti}, M. and {Jones}, G.~C. and {Lamperti}, I. and {Parlanti}, E. and {Venturi}, G.},
        title = "{GA-NIFS: the interplay between merger, star formation, and chemical enrichment in MACS1149-JD1 at z = 9.11 with JWST/NIRSpec}",
      journal = {\mnras},
         year = 2024,
        month = sep,
       volume = {533},
       number = {2},
        pages = {2488-2501},
          doi = {10.1093/mnras/stae1971},
archivePrefix = {arXiv},
       eprint = {2407.08616},
 primaryClass = {astro-ph.GA},
       adsurl = {https://ui.adsabs.harvard.edu/abs/2024MNRAS.533.2488M}
}

@ARTICLE{Qiu21,
       author = {{Qiu}, Yu and {McNamara}, Brian R. and {Bogdanovi{\'c}}, Tamara and {Inayoshi}, Kohei and {Ho}, Luis C.},
        title = "{On the Mass Loading of AGN-driven Outflows in Elliptical Galaxies and Clusters}",
      journal = {\apj},
         year = 2021,
        month = dec,
       volume = {923},
       number = {2},
          eid = {256},
        pages = {256},
          doi = {10.3847/1538-4357/ac2ede},
archivePrefix = {arXiv},
       eprint = {2103.06505},
 primaryClass = {astro-ph.GA},
       adsurl = {https://ui.adsabs.harvard.edu/abs/2021ApJ...923..256Q}
}

@ARTICLE{Bennett2024,
       author = {{Bennett}, Jake S. and {Sijacki}, Debora and {Costa}, Tiago and {Laporte}, Nicolas and {Witten}, Callum},
        title = "{The growth of the gargantuan black holes powering high-redshift quasars and their impact on the formation of early galaxies and protoclusters}",
      journal = {\mnras},
         year = 2024,
        month = jan,
       volume = {527},
       number = {1},
        pages = {1033-1054},
          doi = {10.1093/mnras/stad3179},
archivePrefix = {arXiv},
       eprint = {2305.11932},
 primaryClass = {astro-ph.GA},
       adsurl = {https://ui.adsabs.harvard.edu/abs/2024MNRAS.527.1033B}
}

@ARTICLE{MazzolariCEERS24,
       author = {{Mazzolari}, Giovanni and {Scholtz}, Jan and {Maiolino}, Roberto and {Gilli}, Roberto and {Traina}, Alberto and {L{\'o}pez}, Ivan E. and {{\"U}bler}, Hannah and {Trefoloni}, Bartolomeo and {D'Eugenio}, Francesco and {Ji}, Xihan and {Mignoli}, Marco and {Vito}, Fabio and {Vignali}, Cristian and {Brusa}, Marcella},
        title = "{Narrow-line AGN selection in CEERS: Spectroscopic selection, physical properties, and X-ray and radio analysis}",
      journal = {\aap},
         year = 2025,
        month = aug,
       volume = {700},
          eid = {A12},
        pages = {A12},
          doi = {10.1051/0004-6361/202451860},
archivePrefix = {arXiv},
       eprint = {2408.15615},
 primaryClass = {astro-ph.GA},
       adsurl = {https://ui.adsabs.harvard.edu/abs/2025A&A...700A..12M}
}

@ARTICLE{Liu2020dwarfAGN,
       author = {{Liu}, Weizhe and {Veilleux}, Sylvain and {Canalizo}, Gabriela and {Rupke}, David S.~N. and {Manzano-King}, Christina M. and {Bohn}, Thomas and {U}, Vivian},
        title = "{Integral Field Spectroscopy of Fast Outflows in Dwarf Galaxies with AGNs}",
      journal = {\apj},
         year = 2020,
        month = dec,
       volume = {905},
       number = {2},
          eid = {166},
        pages = {166},
          doi = {10.3847/1538-4357/abc269},
archivePrefix = {arXiv},
       eprint = {2010.09008},
 primaryClass = {astro-ph.GA},
       adsurl = {https://ui.adsabs.harvard.edu/abs/2020ApJ...905..166L}
}

@ARTICLE{RM25dwarfagn,
       author = {{Rodr{\'\i}guez Morales}, V. and {Mezcua}, M. and {Dom{\'\i}nguez S{\'a}nchez}, H. and {Audibert}, A. and {M{\"u}ller-S{\'a}nchez}, F. and {Siudek}, M. and {Er{\'o}stegui}, A.},
        title = "{MaNGA AGN dwarf galaxies (MAD): II. AGN outflows in dwarf galaxies}",
      journal = {\aap},
         year = 2025,
        month = may,
       volume = {697},
          eid = {A235},
        pages = {A235},
          doi = {10.1051/0004-6361/202453481},
archivePrefix = {arXiv},
       eprint = {2503.07779},
 primaryClass = {astro-ph.GA},
       adsurl = {https://ui.adsabs.harvard.edu/abs/2025A&A...697A.235R}
}

@ARTICLE{Rizzo20,
       author = {{Rizzo}, F. and {Vegetti}, S. and {Powell}, D. and {Fraternali}, F. and {McKean}, J.~P. and {Stacey}, H.~R. and {White}, S.~D.~M.},
        title = "{A dynamically cold disk galaxy in the early Universe}",
      journal = {\nat},
         year = 2020,
        month = aug,
       volume = {584},
       number = {7820},
        pages = {201-204},
          doi = {10.1038/s41586-020-2572-6},
archivePrefix = {arXiv},
       eprint = {2009.01251},
 primaryClass = {astro-ph.GA},
       adsurl = {https://ui.adsabs.harvard.edu/abs/2020Natur.584..201R}
}

@ARTICLE{Lelli21,
       author = {{Lelli}, Federico and {Di Teodoro}, Enrico M. and {Fraternali}, Filippo and {Man}, Allison W.~S. and {Zhang}, Zhi-Yu and {De Breuck}, Carlos and {Davis}, Timothy A. and {Maiolino}, Roberto},
        title = "{A massive stellar bulge in a regularly rotating galaxy 1.2 billion years after the Big Bang}",
      journal = {Science},
         year = 2021,
        month = feb,
       volume = {371},
       number = {6530},
        pages = {713-716},
          doi = {10.1126/science.abc1893},
archivePrefix = {arXiv},
       eprint = {2102.05957},
 primaryClass = {astro-ph.GA},
       adsurl = {https://ui.adsabs.harvard.edu/abs/2021Sci...371..713L}
}

@ARTICLE{Henden18,
       author = {{Henden}, Nicholas A. and {Puchwein}, Ewald and {Shen}, Sijing and {Sijacki}, Debora},
        title = "{The FABLE simulations: a feedback model for galaxies, groups, and clusters}",
      journal = {\mnras},
         year = 2018,
        month = oct,
       volume = {479},
       number = {4},
        pages = {5385-5412},
          doi = {10.1093/mnras/sty1780},
archivePrefix = {arXiv},
       eprint = {1804.05064},
 primaryClass = {astro-ph.GA},
       adsurl = {https://ui.adsabs.harvard.edu/abs/2018MNRAS.479.5385H}
}

@ARTICLE{Koudmani22,
       author = {{Koudmani}, Sophie and {Sijacki}, Debora and {Smith}, Matthew C.},
        title = "{Two can play at that game: constraining the role of supernova and AGN feedback in dwarf galaxies with cosmological zoom-in simulations}",
      journal = {\mnras},
         year = 2022,
        month = oct,
       volume = {516},
       number = {2},
        pages = {2112-2141},
          doi = {10.1093/mnras/stac2252},
archivePrefix = {arXiv},
       eprint = {2206.11274},
 primaryClass = {astro-ph.GA},
       adsurl = {https://ui.adsabs.harvard.edu/abs/2022MNRAS.516.2112K}
}

@ARTICLE{Piotrowska22,
       author = {{Piotrowska}, Joanna M. and {Bluck}, Asa F.~L. and {Maiolino}, Roberto and {Peng}, Yingjie},
        title = "{On the quenching of star formation in observed and simulated central galaxies: evidence for the role of integrated AGN feedback}",
      journal = {\mnras},
         year = 2022,
        month = may,
       volume = {512},
       number = {1},
        pages = {1052-1090},
          doi = {10.1093/mnras/stab3673},
archivePrefix = {arXiv},
       eprint = {2112.07672},
 primaryClass = {astro-ph.GA},
       adsurl = {https://ui.adsabs.harvard.edu/abs/2022MNRAS.512.1052P}
}

@ARTICLE{Scholtz2020,
       author = {{Scholtz}, J. and {Harrison}, C.~M. and {Rosario}, D.~J. and {Alexander}, D.~M. and {Chen}, C. -C. and {Kakkad}, D. and {Mainieri}, V. and {Tiley}, A.~L. and {Turner}, O. and {Cirasuolo}, M. and {Sharples}, R.~M. and {Stach}, S.},
        title = "{KASHz: No evidence for ionised outflows instantaneously suppressing star formation in moderate luminosity AGN at z {\ensuremath{\sim}} 1.4-2.6}",
      journal = {\mnras},
         year = 2020,
        month = mar,
       volume = {492},
       number = {3},
        pages = {3194-3216},
          doi = {10.1093/mnras/staa030},
archivePrefix = {arXiv},
       eprint = {2001.02242},
 primaryClass = {astro-ph.GA},
       adsurl = {https://ui.adsabs.harvard.edu/abs/2020MNRAS.492.3194S}
}

@ARTICLE{Lamperti21,
       author = {{Lamperti}, I. and {Harrison}, C.~M. and {Mainieri}, V. and {Kakkad}, D. and {Perna}, M. and {Circosta}, C. and {Scholtz}, J. and {Carniani}, S. and {Cicone}, C. and {Alexander}, D.~M. and {Bischetti}, M. and {Calistro Rivera}, G. and {Chen}, C. -C. and {Cresci}, G. and {Feruglio}, C. and {Fiore}, F. and {Mannucci}, F. and {Marconi}, A. and {Mart{\'\i}nez-Ram{\'\i}rez}, L.~N. and {Netzer}, H. and {Piconcelli}, E. and {Puglisi}, A. and {Rosario}, D.~J. and {Schramm}, M. and {Vietri}, G. and {Vignali}, C. and {Zappacosta}, L.},
        title = "{SUPER. V. ALMA continuum observations of z {\ensuremath{\sim}} 2 AGN and the elusive evidence of outflows influencing star formation}",
      journal = {\aap},
         year = 2021,
        month = oct,
       volume = {654},
          eid = {A90},
        pages = {A90},
          doi = {10.1051/0004-6361/202141363},
archivePrefix = {arXiv},
       eprint = {2109.02674},
 primaryClass = {astro-ph.GA},
       adsurl = {https://ui.adsabs.harvard.edu/abs/2021A&A...654A..90L}
}

@ARTICLE{Langeroodi23,
       author = {{Langeroodi}, Danial and {Hjorth}, Jens and {Chen}, Wenlei and {Kelly}, Patrick L. and {Williams}, Hayley and {Lin}, Yu-Heng and {Scarlata}, Claudia and {Zitrin}, Adi and {Broadhurst}, Tom and {Diego}, Jose M. and {Huang}, Xiaosheng and {Filippenko}, Alexei V. and {Foley}, Ryan J. and {Jha}, Saurabh and {Koekemoer}, Anton M. and {Oguri}, Masamune and {Perez-Fournon}, Ismael and {Pierel}, Justin and {Poidevin}, Frederick and {Strolger}, Lou},
        title = "{Evolution of the Mass-Metallicity Relation from Redshift z {\ensuremath{\approx}} 8 to the Local Universe}",
      journal = {\apj},
         year = 2023,
        month = nov,
       volume = {957},
       number = {1},
          eid = {39},
        pages = {39},
          doi = {10.3847/1538-4357/acdbc1},
archivePrefix = {arXiv},
       eprint = {2212.02491},
 primaryClass = {astro-ph.GA},
       adsurl = {https://ui.adsabs.harvard.edu/abs/2023ApJ...957...39L}
}

@ARTICLE{Strait23,
       author = {{Strait}, Victoria and {Brammer}, Gabriel and {Muzzin}, Adam and {Desprez}, Guillaume and {Asada}, Yoshihisa and {Abraham}, Roberto and {Brada{\v{c}}}, Maru{\v{s}}a and {Iyer}, Kartheik G. and {Martis}, Nicholas and {Mowla}, Lamiya and {Noirot}, Ga{\"e}l and {Sarrouh}, Ghassan T.~E. and {Sawicki}, Marcin and {Willott}, Chris and {Gould}, Katriona and {Grindlay}, Tess and {Matharu}, Jasleen and {Rihtar{\v{s}}i{\v{c}}}, Gregor},
        title = "{An Extremely Compact, Low-mass Galaxy on its Way to Quiescence at z = 5.2}",
      journal = {\apjl},
         year = 2023,
        month = jun,
       volume = {949},
       number = {2},
          eid = {L23},
        pages = {L23},
          doi = {10.3847/2041-8213/acd457},
archivePrefix = {arXiv},
       eprint = {2303.11349},
 primaryClass = {astro-ph.GA},
       adsurl = {https://ui.adsabs.harvard.edu/abs/2023ApJ...949L..23S}
}

@ARTICLE{Baker25,
       author = {{Baker}, William M. and {D'Eugenio}, Francesco and {Maiolino}, Roberto and {Bunker}, Andrew J. and {Simmonds}, Charlotte and {Tacchella}, Sandro and {Witstok}, Joris and {Arribas}, Santiago and {Carniani}, Stefano and {Charlot}, St{\'e}phane and {Chevallard}, Jacopo and {Curti}, Mirko and {Curtis-Lake}, Emma and {Jones}, Gareth C. and {Kumari}, Nimisha and {Rinaldi}, Pierluigi and {Robertson}, Brant and {Williams}, Christina C. and {Willott}, Chris and {Zhu}, Yongda},
        title = "{Zapped then napped? A rapidly quenched remnant leaker candidate with a steep spectroscopic {\ensuremath{\beta}}$_{UV}$ slope at z = 8.5}",
      journal = {\aap},
         year = 2025,
        month = may,
       volume = {697},
          eid = {A90},
        pages = {A90},
          doi = {10.1051/0004-6361/202553766},
archivePrefix = {arXiv},
       eprint = {2501.09070},
 primaryClass = {astro-ph.GA},
       adsurl = {https://ui.adsabs.harvard.edu/abs/2025A&A...697A..90B}
}

@ARTICLE{Nanayakkara25,
       author = {{Nanayakkara}, Themiya and {Glazebrook}, Karl and {Schreiber}, Corentin and {Chittenden}, Harry and {Brammer}, Gabriel and {Esdaile}, James and {Jacobs}, Colin and {Kacprzak}, Glenn G. and {Kawinwanichakij}, Lalitwadee and {Kimmig}, Lucas C. and {Labbe}, Ivo and {Lagos}, Claudia and {Marchesini}, Danilo and {Mart{\`\i}nez-Mar{\`\i}n}, M. and {Marsan}, Z. Cemile and {Oesch}, Pascal A. and {Papovich}, Casey and {Remus}, Rhea-Silvia and {Tran}, Kim-Vy H.},
        title = "{The Formation Histories of Massive and Quiescent Galaxies in the 3 < z < 4.5 Universe}",
      journal = {\apj},
         year = 2025,
        month = mar,
       volume = {981},
       number = {1},
          eid = {78},
        pages = {78},
          doi = {10.3847/1538-4357/ada6ac},
archivePrefix = {arXiv},
       eprint = {2410.02076},
 primaryClass = {astro-ph.GA},
       adsurl = {https://ui.adsabs.harvard.edu/abs/2025ApJ...981...78N}
}

@ARTICLE{Baker25Flamingo,
       author = {{Baker}, William M. and {Lim}, Seunghwan and {D'Eugenio}, Francesco and {Maiolino}, Roberto and {Ji}, Zhiyuan and {Arribas}, Santiago and {Bunker}, Andrew J. and {Carniani}, Stefano and {Charlot}, Stephane and {de Graaff}, Anna and {Hainline}, Kevin and {Looser}, Tobias J. and {Lyu}, Jianwei and {Rinaldi}, Pierluigi and {Robertson}, Brant and {Schaller}, Matthieu and {Schaye}, Joop and {Scholtz}, Jan and {{\"U}bler}, Hannah and {Williams}, Christina C. and {Willmer}, Christopher N.~A. and {Willott}, Chris and {Zhu}, Yongda},
        title = "{The abundance and nature of high-redshift quiescent galaxies from JADES spectroscopy and the FLAMINGO simulations}",
      journal = {\mnras},
         year = 2025,
        month = may,
       volume = {539},
       number = {1},
        pages = {557-589},
          doi = {10.1093/mnras/staf475},
archivePrefix = {arXiv},
       eprint = {2410.14773},
 primaryClass = {astro-ph.GA},
       adsurl = {https://ui.adsabs.harvard.edu/abs/2025MNRAS.539..557B}
}

@ARTICLE{Tacchella25,
       author = {{Tacchella}, Sandro and {McClymont}, William and {Scholtz}, Jan and {Maiolino}, Roberto and {Ji}, Xihan and {Villanueva}, Natalia C. and {Charlot}, St{\'e}phane and {D'Eugenio}, Francesco and {Helton}, Jakob M. and {Williams}, Christina C. and {Witstok}, Joris and {Bhatawdekar}, Rachana and {Carniani}, Stefano and {Chevallard}, Jacopo and {Curti}, Mirko and {Hainline}, Kevin and {Ji}, Zhiyuan and {Johnson}, Benjamin D. and {Leja}, Joel and {Li}, Yijia and {Maseda}, Michael V. and {Pusk{\'a}s}, D{\'a}vid and {Rieke}, Marcia and {Robertson}, Brant and {Shivaei}, Irene and {Silcock}, Maddie S. and {Simmonds}, Charlotte and {{\"U}bler}, Hannah and {Willmer}, Christopher N.~A. and {Willott}, Chris},
        title = "{Resolving the nature and putative nebular emission of GS9422: an obscured AGN without exotic stars}",
      journal = {\mnras},
         year = 2025,
        month = jun,
       volume = {540},
       number = {1},
        pages = {851-870},
          doi = {10.1093/mnras/staf718},
archivePrefix = {arXiv},
       eprint = {2404.02194},
 primaryClass = {astro-ph.GA},
       adsurl = {https://ui.adsabs.harvard.edu/abs/2025MNRAS.540..851T}
}

@ARTICLE{Kartaltepe23,
       author = {{Kartaltepe}, Jeyhan S. and {Rose}, Caitlin and {Vanderhoof}, Brittany N. and {McGrath}, Elizabeth J. and {Costantin}, Luca and {Cox}, Isabella G. and {Yung}, L.~Y. Aaron and {Kocevski}, Dale D. and {Wuyts}, Stijn and {Ferguson}, Henry C. and {Bagley}, Micaela B. and {Finkelstein}, Steven L. and {Amor{\'\i}n}, Ricardo O. and {Andrews}, Brett H. and {Arrabal Haro}, Pablo and {Backhaus}, Bren E. and {Behroozi}, Peter and {Bisigello}, Laura and {Calabr{\`o}}, Antonello and {Casey}, Caitlin M. and {Coogan}, Rosemary T. and {Cooper}, M.~C. and {Croton}, Darren and {de la Vega}, Alexander and {Dickinson}, Mark and {Fontana}, Adriano and {Franco}, Maximilien and {Grazian}, Andrea and {Grogin}, Norman A. and {Hathi}, Nimish P. and {Holwerda}, Benne W. and {Huertas-Company}, Marc and {Iyer}, Kartheik G. and {Jogee}, Shardha and {Jung}, Intae and {Kewley}, Lisa J. and {Kirkpatrick}, Allison and {Koekemoer}, Anton M. and {Liu}, James and {Lotz}, Jennifer M. and {Lucas}, Ray A. and {Newman}, Jeffrey A. and {Pacifici}, Camilla and {Pandya}, Viraj and {Papovich}, Casey and {Pentericci}, Laura and {P{\'e}rez-Gonz{\'a}lez}, Pablo G. and {Petersen}, Jayse and {Pirzkal}, Nor and {Rafelski}, Marc and {Ravindranath}, Swara and {Simons}, Raymond C. and {Snyder}, Gregory F. and {Somerville}, Rachel S. and {Stanway}, Elizabeth R. and {Straughn}, Amber N. and {Tacchella}, Sandro and {Trump}, Jonathan R. and {Vega-Ferrero}, Jes{\'u}s and {Wilkins}, Stephen M. and {Yang}, Guang and {Zavala}, Jorge A.},
        title = "{CEERS Key Paper. III. The Diversity of Galaxy Structure and Morphology at z = 3-9 with JWST}",
      journal = {\apjl},
         year = 2023,
        month = mar,
       volume = {946},
       number = {1},
          eid = {L15},
        pages = {L15},
          doi = {10.3847/2041-8213/acad01},
archivePrefix = {arXiv},
       eprint = {2210.14713},
 primaryClass = {astro-ph.GA},
       adsurl = {https://ui.adsabs.harvard.edu/abs/2023ApJ...946L..15K}
}

@ARTICLE{Treu23,
       author = {{Treu}, T. and {Calabr{\`o}}, A. and {Castellano}, M. and {Leethochawalit}, N. and {Merlin}, E. and {Fontana}, A. and {Yang}, L. and {Morishita}, T. and {Trenti}, M. and {Dressler}, A. and {Mason}, C. and {Paris}, D. and {Pentericci}, L. and {Roberts-Borsani}, G. and {Vulcani}, B. and {Boyett}, K. and {Bradac}, M. and {Glazebrook}, K. and {Jones}, T. and {Marchesini}, D. and {Mascia}, S. and {Nanayakkara}, T. and {Santini}, P. and {Strait}, V. and {Vanzella}, E. and {Wang}, X.},
        title = "{Early Results From GLASS-JWST. XII. The Morphology of Galaxies at the Epoch of Reionization}",
      journal = {\apjl},
         year = 2023,
        month = jan,
       volume = {942},
       number = {2},
          eid = {L28},
        pages = {L28},
          doi = {10.3847/2041-8213/ac9283},
archivePrefix = {arXiv},
       eprint = {2207.13527},
 primaryClass = {astro-ph.GA},
       adsurl = {https://ui.adsabs.harvard.edu/abs/2023ApJ...942L..28T}
}

@ARTICLE{Li25metgrad,
       author = {{Li}, Zihao and {Cai}, Zheng and {Wang}, Xin and {Li}, Zhaozhou and {Dekel}, Avishai and {Sarkar}, Kartick C. and {Ba{\~n}ados}, Eduardo and {Bian}, Fuyan and {Bhowmick}, Aklant K. and {Blecha}, Laura and {Bosman}, Sarah E.~I. and {Champagne}, Jaclyn B. and {Fan}, Xiaohui and {Golden-Marx}, Emmet and {Jun}, Hyunsung D. and {Li}, Mingyu and {Lin}, Xiaojing and {Liu}, Weizhe and {Sun}, Fengwu and {Trebitsch}, Maxime and {Walter}, Fabian and {Wang}, Feige and {Wu}, Yunjing and {Yang}, Jinyi and {Zhang}, Huanian and {Zhang}, Shiwu and {Zhuang}, Mingyang and {Zou}, Siwei},
        title = "{A 13 Billion Year View of Galaxy Growth: Metallicity Gradient Evolution from the Local Universe to z = 9 with JWST and Archival Surveys}",
      journal = {\apjs},
         year = 2025,
        month = oct,
       volume = {280},
       number = {2},
          eid = {62},
        pages = {62},
          doi = {10.3847/1538-4365/adfa70},
archivePrefix = {arXiv},
       eprint = {2506.12129},
 primaryClass = {astro-ph.GA},
       adsurl = {https://ui.adsabs.harvard.edu/abs/2025ApJS..280...62L}
}

@ARTICLE{Langeroodi24,
       author = {{Langeroodi}, Danial and {Hjorth}, Jens},
        title = "{NIRSpec View of the Appearance and Evolution of Balmer Breaks and the Transition from Bursty to Smooth Star Formation Histories from Deep Within the Epoch of Reionization to Cosmic Noon}",
      journal = {arXiv e-prints},
         year = 2024,
        month = apr,
          eid = {arXiv:2404.13045},
        pages = {arXiv:2404.13045},
          doi = {10.48550/arXiv.2404.13045},
archivePrefix = {arXiv},
       eprint = {2404.13045},
 primaryClass = {astro-ph.GA},
       adsurl = {https://ui.adsabs.harvard.edu/abs/2024arXiv240413045L}
}

@ARTICLE{Weibel25,
       author = {{Weibel}, Andrea and {de Graaff}, Anna and {Setton}, David J. and {Miller}, Tim B. and {Oesch}, Pascal A. and {Brammer}, Gabriel and {Lagos}, Claudia D.~P. and {Whitaker}, Katherine E. and {Williams}, Christina C. and {Baggen}, Josephine F.~W. and {Bezanson}, Rachel and {Boogaard}, Leindert A. and {Cleri}, Nikko J. and {Greene}, Jenny E. and {Hirschmann}, Michaela and {Hviding}, Raphael E. and {Kuruvanthodi}, Adarsh and {Labb{\'e}}, Ivo and {Leja}, Joel and {Maseda}, Michael V. and {Matthee}, Jorryt and {McConachie}, Ian and {Naidu}, Rohan P. and {Roberts-Borsani}, Guido and {Schaerer}, Daniel and {Suess}, Katherine A. and {Valentino}, Francesco and {van Dokkum}, Pieter and {Wang}, Bingjie},
        title = "{RUBIES Reveals a Massive Quiescent Galaxy at z = 7.3}",
      journal = {\apj},
         year = 2025,
        month = apr,
       volume = {983},
       number = {1},
          eid = {11},
        pages = {11},
          doi = {10.3847/1538-4357/adab7a},
archivePrefix = {arXiv},
       eprint = {2409.03829},
 primaryClass = {astro-ph.GA},
       adsurl = {https://ui.adsabs.harvard.edu/abs/2025ApJ...983...11W}
}

@ARTICLE{Nanayakkara24Nature,
       author = {{Nanayakkara}, Themiya and {Glazebrook}, Karl and {Jacobs}, Colin and {Kawinwanichakij}, Lalitwadee and {Schreiber}, Corentin and {Brammer}, Gabriel and {Esdaile}, James and {Kacprzak}, Glenn G. and {Labbe}, Ivo and {Lagos}, Claudia and {Marchesini}, Danilo and {Marsan}, Z. Cemile and {Oesch}, Pascal A. and {Papovich}, Casey and {Remus}, Rhea-Silvia and {Tran}, Kim-Vy H.},
        title = "{A population of faint, old, and massive quiescent galaxies at 3 <z <4 revealed by JWST NIRSpec Spectroscopy}",
      journal = {Scientific Reports},
         year = 2024,
        month = feb,
       volume = {14},
          eid = {3724},
        pages = {3724},
          doi = {10.1038/s41598-024-52585-4},
archivePrefix = {arXiv},
       eprint = {2212.11638},
 primaryClass = {astro-ph.GA},
       adsurl = {https://ui.adsabs.harvard.edu/abs/2024NatSR..14.3724N}
}

@ARTICLE{Baker25quiescent,
       author = {{Baker}, William M. and {Valentino}, Francesco and {Lagos}, Claudia del P. and {Ito}, Kei and {Jespersen}, Christian Kragh and {Gottumukkala}, Rashmi and {Hjorth}, Jens and {Langeroodi}, Danial and {Sedgewick}, Aidan},
        title = "{Exploring over 700 massive quiescent galaxies at z = 2-7: Demographics and stellar mass functions}",
      journal = {arXiv e-prints},
         year = 2025,
        month = jun,
          eid = {arXiv:2506.04119},
        pages = {arXiv:2506.04119},
          doi = {10.48550/arXiv.2506.04119},
archivePrefix = {arXiv},
       eprint = {2506.04119},
 primaryClass = {astro-ph.GA},
       adsurl = {https://ui.adsabs.harvard.edu/abs/2025arXiv250604119B}
}

@ARTICLE{Laseter24,
       author = {{Laseter}, Isaac H. and {Maseda}, Michael V. and {Curti}, Mirko and {Maiolino}, Roberto and {D'Eugenio}, Francesco and {Cameron}, Alex J. and {Looser}, Tobias J. and {Arribas}, Santiago and {Baker}, William M. and {Bhatawdekar}, Rachana and {Boyett}, Kristan and {Bunker}, Andrew J. and {Carniani}, Stefano and {Charlot}, Stephane and {Chevallard}, Jacopo and {Curtis-lake}, Emma and {Egami}, Eiichi and {Eisenstein}, Daniel J. and {Hainline}, Kevin and {Hausen}, Ryan and {Ji}, Zhiyuan and {Kumari}, Nimisha and {Perna}, Michele and {Rawle}, Tim and {Rix}, Hans-Walter and {Robertson}, Brant and {Rodr{\'\i}guez Del Pino}, Bruno and {Sandles}, Lester and {Scholtz}, Jan and {Smit}, Renske and {Tacchella}, Sandro and {{\"U}bler}, Hannah and {Williams}, Christina C. and {Willott}, Chris and {Witstok}, Joris},
        title = "{JADES: Detecting [OIII]{\ensuremath{\lambda}}4363 emitters and testing strong line calibrations in the high-z Universe with ultra-deep JWST/NIRSpec spectroscopy up to z {\ensuremath{\sim}} 9.5}",
      journal = {\aap},
         year = 2024,
        month = jan,
       volume = {681},
          eid = {A70},
        pages = {A70},
          doi = {10.1051/0004-6361/202347133},
archivePrefix = {arXiv},
       eprint = {2306.03120},
 primaryClass = {astro-ph.GA},
       adsurl = {https://ui.adsabs.harvard.edu/abs/2024A&A...681A..70L}
}

@ARTICLE{Maiolino19,
       author = {{Maiolino}, R. and {Mannucci}, F.},
        title = "{De re metallica: the cosmic chemical evolution of galaxies}",
      journal = {\aapr},
         year = 2019,
        month = feb,
       volume = {27},
       number = {1},
          eid = {3},
        pages = {3},
          doi = {10.1007/s00159-018-0112-2},
archivePrefix = {arXiv},
       eprint = {1811.09642},
 primaryClass = {astro-ph.GA},
       adsurl = {https://ui.adsabs.harvard.edu/abs/2019A&ARv..27....3M}
}

@ARTICLE{Curti2017,
       author = {{Curti}, M. and {Cresci}, G. and {Mannucci}, F. and {Marconi}, A. and {Maiolino}, R. and {Esposito}, S.},
        title = "{New fully empirical calibrations of strong-line metallicity indicators in star-forming galaxies}",
      journal = {\mnras},
         year = 2017,
        month = feb,
       volume = {465},
       number = {2},
        pages = {1384-1400},
          doi = {10.1093/mnras/stw2766},
archivePrefix = {arXiv},
       eprint = {1610.06939},
 primaryClass = {astro-ph.GA},
       adsurl = {https://ui.adsabs.harvard.edu/abs/2017MNRAS.465.1384C}
}

@ARTICLE{Kumari21,
       author = {{Kumari}, Nimisha and {Maiolino}, Roberto and {Trussler}, James and {Mannucci}, Filippo and {Cresci}, Giovanni and {Curti}, Mirko and {Marconi}, Alessandro and {Belfiore}, Francesco},
        title = "{The extension of the fundamental metallicity relation beyond the BPT star-forming sequence: Evidence for both gas accretion and starvation}",
      journal = {\aap},
         year = 2021,
        month = dec,
       volume = {656},
          eid = {A140},
        pages = {A140},
          doi = {10.1051/0004-6361/202140757},
archivePrefix = {arXiv},
       eprint = {2108.12437},
 primaryClass = {astro-ph.GA},
       adsurl = {https://ui.adsabs.harvard.edu/abs/2021A&A...656A.140K}
}

@ARTICLE{Baker23metfundep,
       author = {{Baker}, William M. and {Maiolino}, Roberto and {Belfiore}, Francesco and {Curti}, Mirko and {Bluck}, Asa F.~L. and {Lin}, Lihwai and {Ellison}, Sara L. and {Thorp}, Mallory and {Pan}, Hsi-An},
        title = "{The metallicity's fundamental dependence on both local and global galactic quantities}",
      journal = {\mnras},
         year = 2023,
        month = feb,
       volume = {519},
       number = {1},
        pages = {1149-1170},
          doi = {10.1093/mnras/stac3594},
archivePrefix = {arXiv},
       eprint = {2210.03755},
 primaryClass = {astro-ph.GA},
       adsurl = {https://ui.adsabs.harvard.edu/abs/2023MNRAS.519.1149B}
}

@ARTICLE{Chisholm2018,
       author = {{Chisholm}, J. and {Tremonti}, C. and {Leitherer}, C.},
        title = "{Metal-enriched galactic outflows shape the mass-metallicity relationship}",
      journal = {\mnras},
         year = 2018,
        month = dec,
       volume = {481},
       number = {2},
        pages = {1690-1706},
          doi = {10.1093/mnras/sty2380},
archivePrefix = {arXiv},
       eprint = {1808.10453},
 primaryClass = {astro-ph.GA},
       adsurl = {https://ui.adsabs.harvard.edu/abs/2018MNRAS.481.1690C}
}

@ARTICLE{Cresci10,
       author = {{Cresci}, G. and {Mannucci}, F. and {Maiolino}, R. and {Marconi}, A. and {Gnerucci}, A. and {Magrini}, L.},
        title = "{Gas accretion as the origin of chemical abundance gradients in distant galaxies}",
      journal = {\nat},
         year = 2010,
        month = oct,
       volume = {467},
       number = {7317},
        pages = {811-813},
          doi = {10.1038/nature09451},
archivePrefix = {arXiv},
       eprint = {1010.2534},
 primaryClass = {astro-ph.CO},
       adsurl = {https://ui.adsabs.harvard.edu/abs/2010Natur.467..811C}
}

%\begin{thebibliography}{99}
%\end{thebibliography}

%%%%%%%%%%%%%%%%%%%%%%%%%%%%%%%%%%%%%%%%%%%%%%%%%%

%%%%%%%%%%%%%%%%% APPENDICES %%%%%%%%%%%%%%%%%%%%%

\appendix

\section{PSF Determination}\label{sec:star}

\begin{figure}
        \centering
	\includegraphics[width=\linewidth]{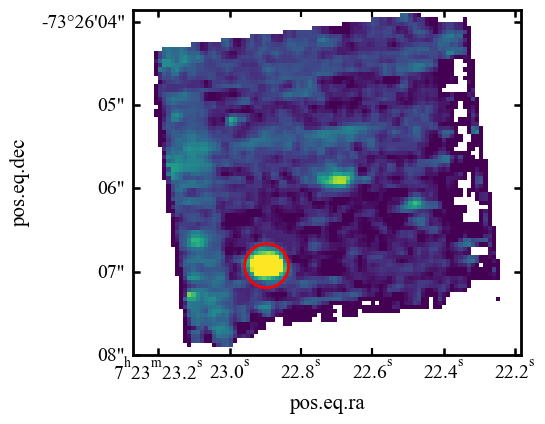}
    \caption{Mock F356W NIRCam image created from the NIRSpec/IFU observation of ID6355, showing the full IFU FoV. The point source present in the FoV used for PSF determination is shown circled in red. The large blob of emission at the centre of the figure corresponds to ID6355. The IFU data shown here has been realigned to match the NIRCam observations.}
    \label{fig:star_fov}
\end{figure}

\begin{figure}
        \centering
	\includegraphics[width=\linewidth]{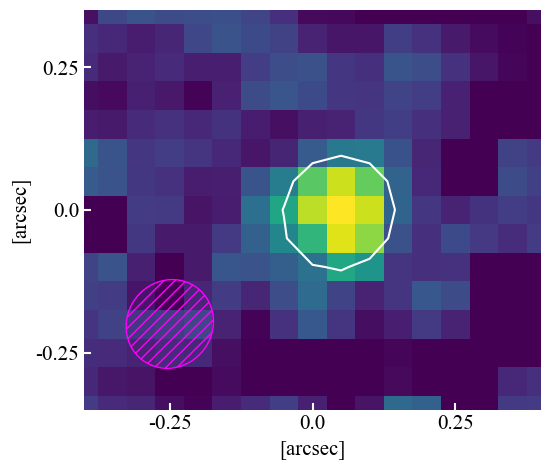}
    \caption{Image of the star visible in the FoV of the ID6355 observation, collapsed over the $\OIII\lambda$5007 emission line. The image is fitted with a 2D Gaussian to calculate the associated PSF, with a white contour plotted at the 1$\sigma$ confidence interval. The magenta hatched ellipse in the bottom left illustrates the fitted PSF for comparison.} 
    \label{fig:PSF_Gauss}
\end{figure}

In this work, we derive the instrumental PSF of the NIRSpec/IFU (shown, for example, in Figs.~\ref{fig:6355intfluxmaps}, \ref{fig:10612intfluxmaps}, \ref{fig:6355_kinem} and \ref{fig:10612_kinem}) based on a point source present in the FoV of our observation of ID6355 (see Fig. \ref{fig:star_fov}). The PSF corresponding to a particular emission line is calculated by fitting an image of the IFU cube, collapsed over a narrow wavelength range centred on that line, with a 2D Gaussian (see Fig. \ref{fig:PSF_Gauss}). Noting that the observations of ID6355 and ID10612 were taken with only a short interval of time between them, we assume that same PSFs applied to our observations of ID10612. This star was also used as a test of our astrometry realignment (described in Sec.~\ref{sec:Realignment}).


\section{MSA and IFU Aperture Comparison}\label{sec:msavifu}
\begin{figure}
    \centering
	\includegraphics[width=\linewidth, trim = {0 0 0 0}]{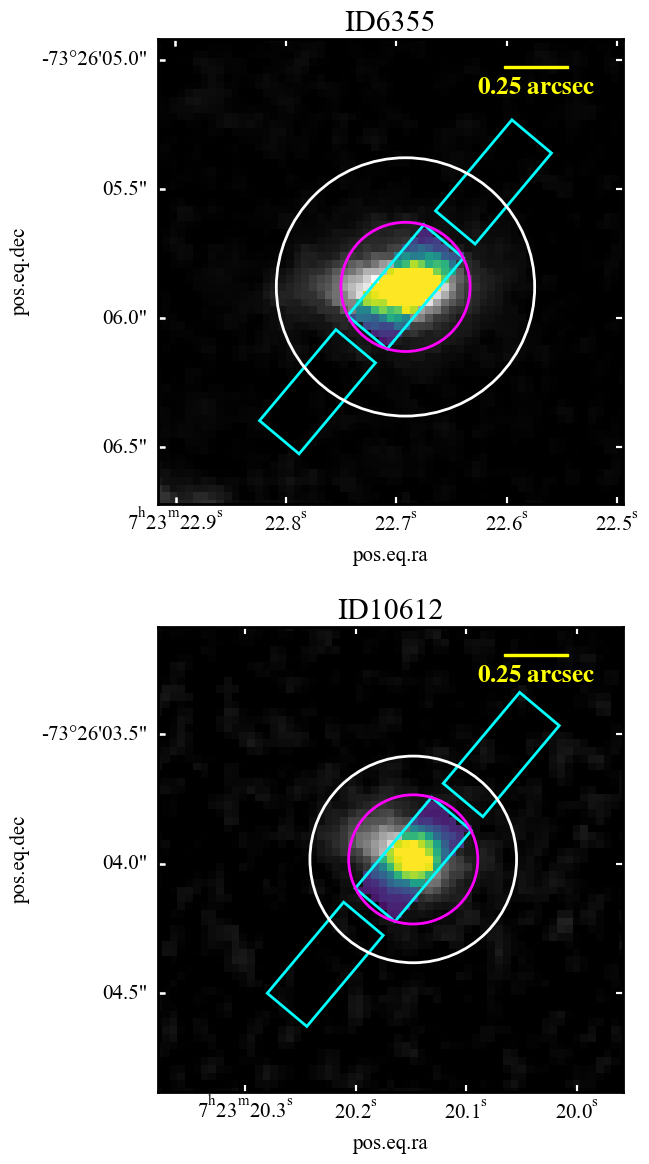}
    \caption{NIRCam F444W cutouts of each galaxy (\textbf{Top panel:} ID6355. \textbf{Bottom panel:} ID10612), illustrating the sizes of the MSA and IFU apertures. The MSA apertures are shown as the cyan rectangles, with each central rectangle showing while the region of galaxy flux extraction while the surrounding two rectangles show the background detections. The IFU apertures used to calculate galaxy-integrated properties in this study, circular apertures of radius $0.25''$ for both galaxies, are shown in magenta.   (galaxy-integrated properties) and 0.5 (star formation rates) are shown as the magenta and white circles, respectively.}
    \label{fig:ifuvmsa}
\end{figure}

As discussed in Sec.~\ref{sec:Comparison}, we compare our results to earlier results obtained from the MSA observations. In Fig.~\ref{fig:ifuvmsa}, we highlight the difference in aperture size between the IFU and MSA observations of our targets. This illustrates that the $0.25''$ radius apertures chosen to derive integrated galaxy properties from IFU data capture the bulk of each galaxy's \OIIIall flux, including galaxy regions not captured by the MSA observations. Furthermore, the $0.5''$ and $0.4''$ radius apertures fully capture any star formation in ID6355 and ID10612 respectively, offering a far more complete picture of the total star formation rate than provided by the MSA observations.

\section{PySersic Fitting}\label{sec:pysersic}
\begin{figure*}
    \centering
	\includegraphics[width=0.85\paperwidth, trim = {5cm 0 0 0}]{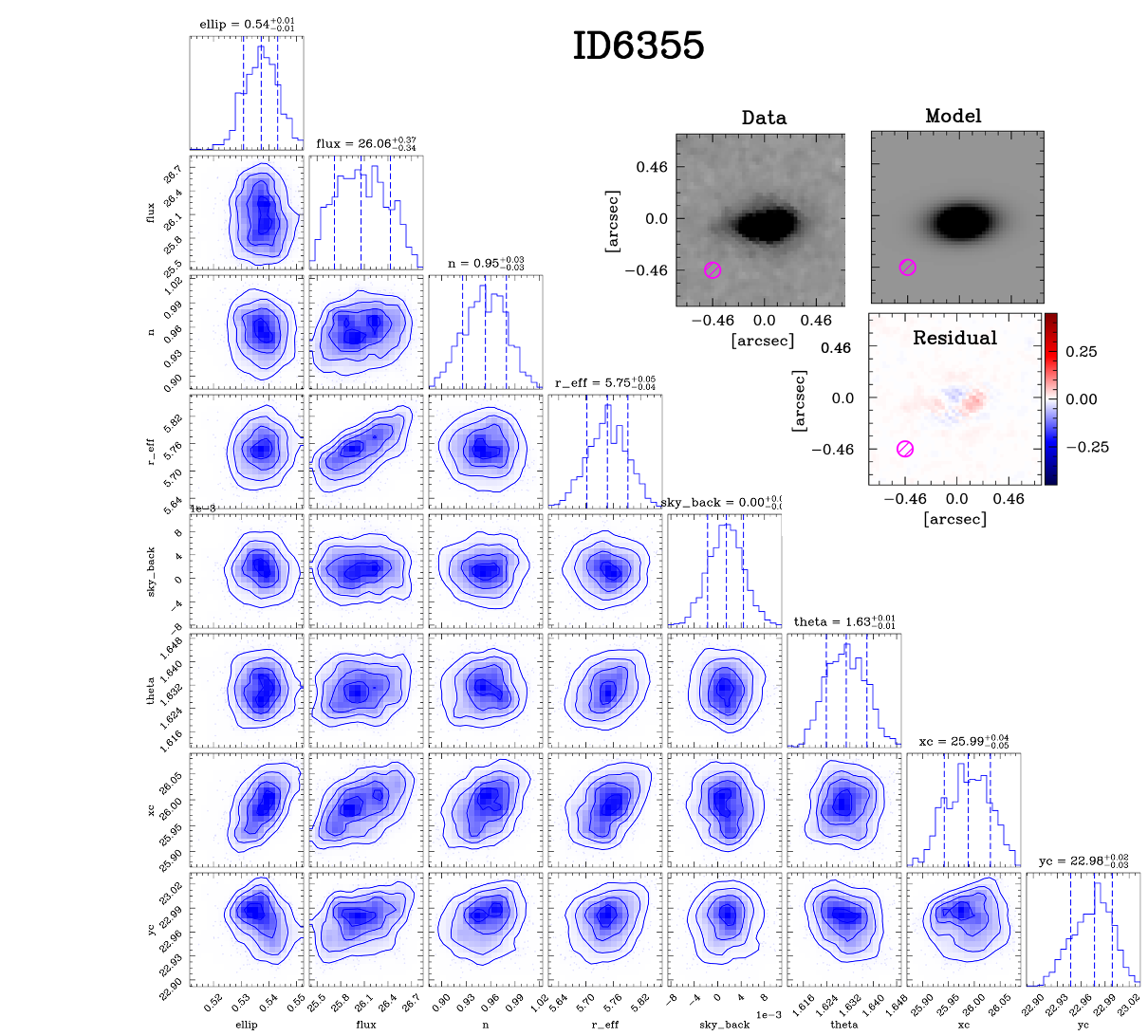}
    \caption{A visual summary of results from fitting ID6355 with \textsc{PySersic}. The bottom left of the figure shows the posterior distributions of the fitted parameters, along with best fit values with their associated errors. The top right corner contains the data, model and residual plots. The magenta hatched ellipse in each of the model/data/residual plots shows the NIRCam PSF.}
    \label{fig:6355posterior}
\end{figure*}

\begin{figure*}
    \centering
	\includegraphics[width=0.85\paperwidth, trim = {5cm 0 0 0}]{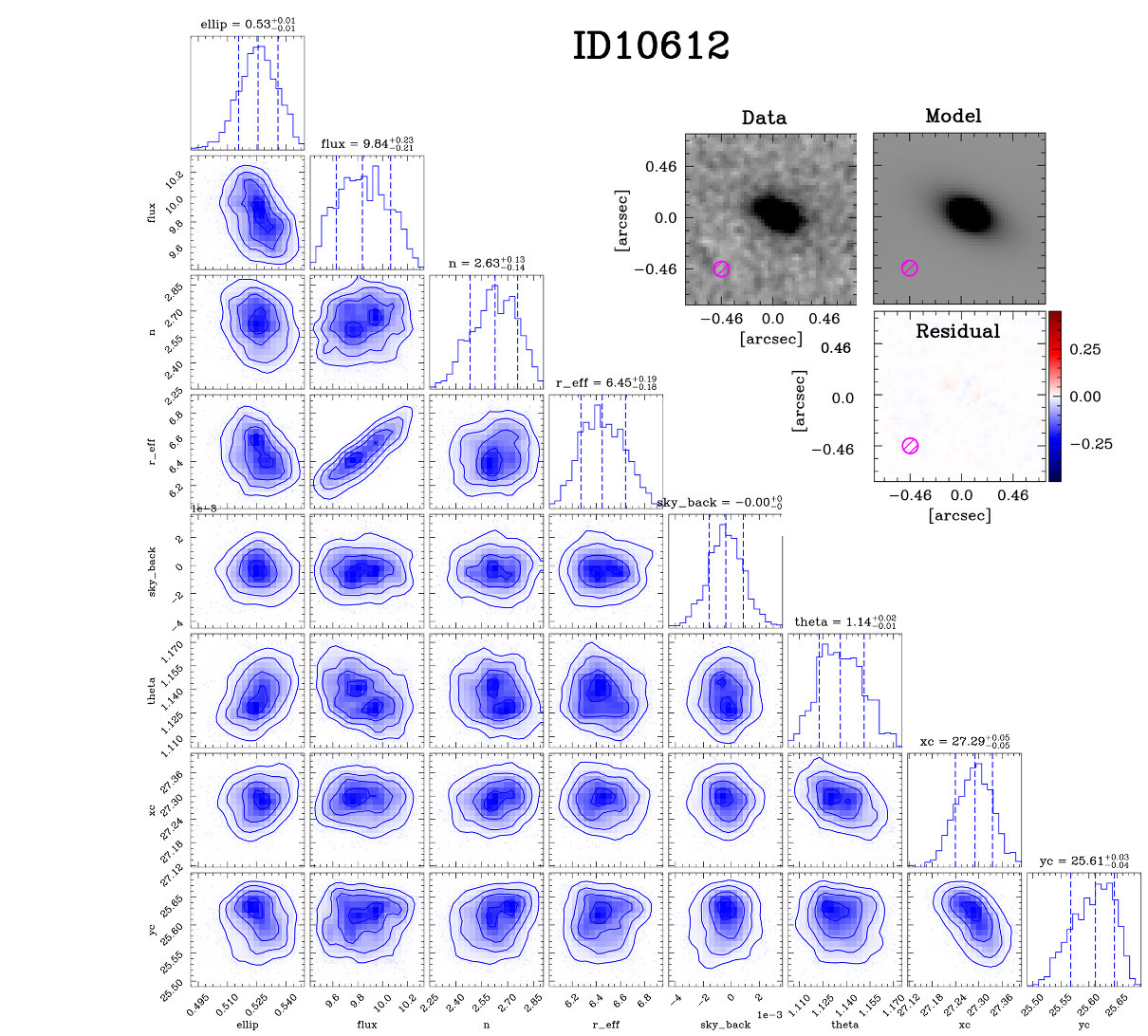}
    \caption{A visual summary of results from fitting ID10612 with \textsc{PySersic}. The bottom left of the figure shows the posterior distributions of the fitted parameters, along with the best fit values and their associated errors. The top right corner contains the data, model and residual plots. The magenta hatched ellipse in each of the model/data/residual plots shows the NIRCam PSF.}
    \label{fig:10612posterior}
\end{figure*}

As briefly introduced in Sec.~\ref{sec:morph-fit}, we model the F444W NIRCam images of our two galaxies using a Sérsic profile in order to estimate parameters, such as the inclination, which are needed to de-project and interpret the measured kinematics. This modelling also adds insight into the morphology of the two systems. We use the Bayesian code \textsc{PySersic} with a one-component model, and additionally fit a for a flat sky background, using the relevant NIRCam F444W PSF, derived following the methodology of \citet{Ji2024jades}. This is done to ensure that any residual background flux does not bias the modelling of the source itself. No masking is needed as there as there are no neighbouring sources in the cutouts. The results are shown in Figs. \ref{fig:6355posterior} and \ref{fig:10612posterior}. As seen in the residuals, the single Sérsic model does not reproduce the entire light profile, which is expected in galaxies at high-redshift which are in earlier stages of their assembly history and hence tend to be clumpier. However, the fits remain satisfactory and are representative of the galaxy size and inclination.

\section{Results from Simulations}\label{sec:simchoice}
\begin{figure*}
    \centering
	\includegraphics[width=0.85\paperwidth]{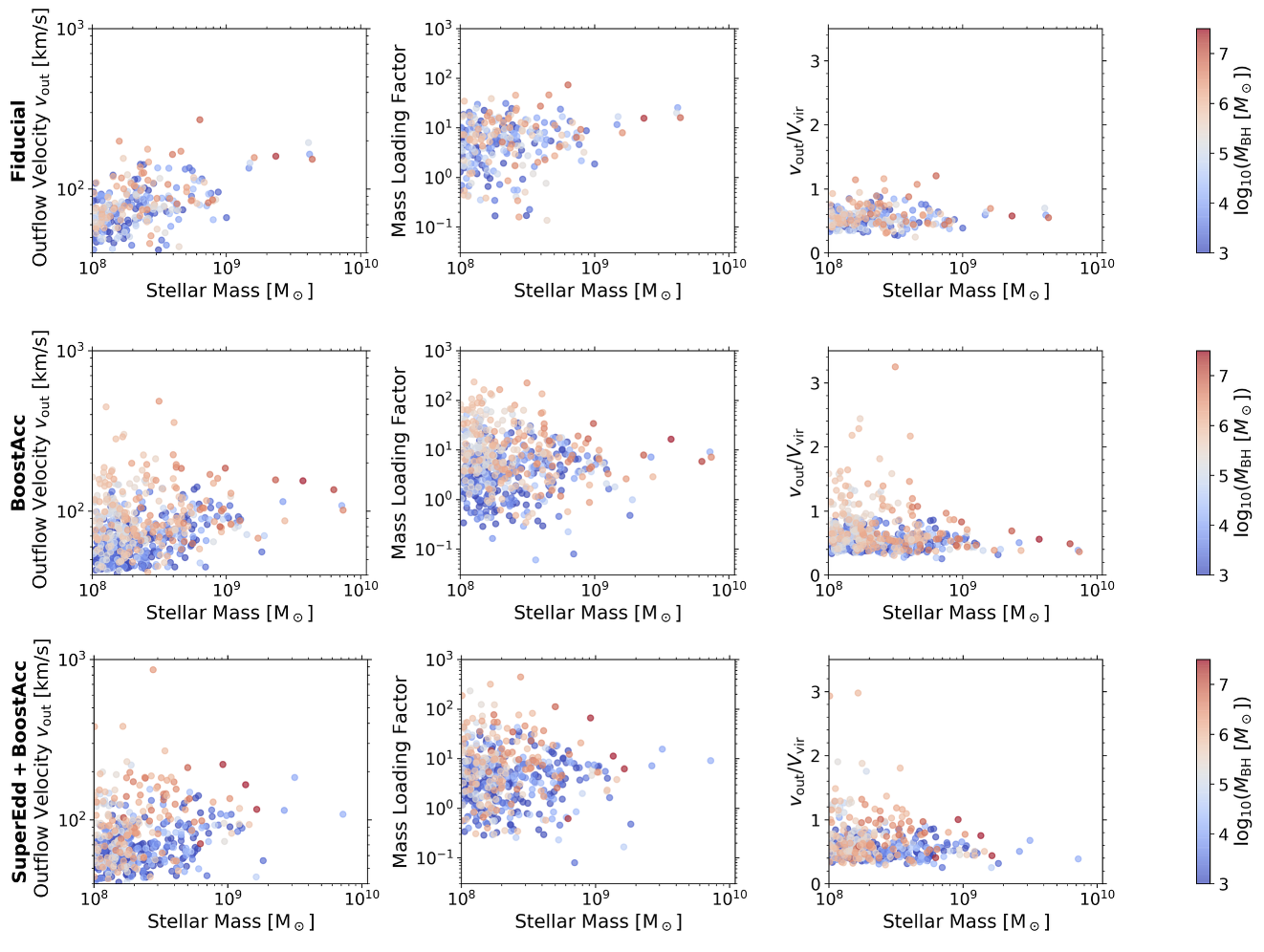}
    \caption{A summary of simulated outflow properties at $z\sim7.6$ from \textsc{Aesopica}. The points on each plot are colour-coded according to the black hole mass of the simulated galaxy, with more massive black holes indicated by the red points. \textbf{Rows, from top to bottom}: Results from \textit{Fiducial}, \textit{BoostAcc} and \textit{SuperEdd+BoostAcc} variations on the main \textsc{Aesopica} simulations. \textbf{Columns, from left to right}: The simulated outflow velocity, mass loading factor, and $v_\mathrm{out}/v_\mathrm{vir}$. Note that following \citet{Koudmani22}, we use the virial velocity as proxy for the escape velocity. Boosted AGN accretion leads to significantly stronger outflows in low-mass galaxies.}
    \label{fig:aesopicasims}
\end{figure*}

As discussed in Sec.~\ref{sec:impact}, we compare our results to three variations within the \textsc{Aesopica} simulation suite (\textit{Fiducial}, \textit{BoostAcc} and \textit{SuperEdd+BoostAcc}). In Fig.~\ref{fig:aesopicasims}, we summarise the simulated outflow properties of each variant as functions of stellar mass. Across all three variations (but most notably for the models with boosted AGN accretion), for a given stellar mass, higher outflow velocities and mass loading factors appear to be associated with overmassive black holes ($M_\mathrm{BH} \sim 10^{6}~\mathrm{M_\odot}$ for our stellar mass range of interest).

%%%%%%%%%%%%%%%%%%%%%%%%%%%%%%%%%%%%%%%%%%%%%%%%%%

% Don't change these lines
\bsp	% typesetting comment
\label{lastpage}
\end{document}